\documentclass[onecolumn,showpacs,amsmath,amssymb,prd,floatfix,preprintnumbers]{revtex4}
\usepackage[dvipdfmx]{graphicx}
\usepackage{bm}
\usepackage{amsmath}
\usepackage{color}
\usepackage{hyperref}
\usepackage{comment}
\usepackage{multirow}
\usepackage[varg]{txfonts}

\hypersetup{
  bookmarksnumbered=true,
  linkcolor=red,
  citecolor=green,
  urlcolor=cyan
}

\newcommand{\bx}{{\bf x}}

\newcommand{\bs}{{\bf s}}
\newcommand{\bk}{{\bf k}}

\newcommand{\hMpci}{$h\,$Mpc$^{-1}$}
\newcommand{\hiMpc}{$h^{-1}\,$Mpc}

\newcommand{\Om}{\Omega_{\rm m}}

\renewcommand{\L}{\mathcal{L}}

\def\avrg#1{\left\langle #1 \right\rangle}

\newcommand{\simgt}{\lower.5ex\hbox{$\; \buildrel > \over \sim \;$}}
\newcommand{\simlt}{\lower.5ex\hbox{$\; \buildrel < \over \sim \;$}}

\begin{document}

\title[]{Cosmological information content in redshift-space power spectrum of SDSS-like galaxies in the quasi-nonlinear regime up to $k=0.3$~\hMpci }

\author{Yosuke Kobayashi${}^{1,2}$}
\email{yosuke.kobayashi@ipmu.jp}
\author{Takahiro Nishimichi${}^{3,1}$}
\author{Masahiro Takada${}^{1}$}
\author{Ryuichi Takahashi${}^{4}$}
\affiliation{
${}^{1}$Kavli Institute for the Physics and Mathematics of the Universe
(WPI), The University of Tokyo Institutes for Advanced Study (UTIAS),
The University of Tokyo, Chiba 277-8583, Japan\\
${}^{2}$ Physics Department,
The University of Tokyo, Bunkyo, Tokyo 113-0031, Japan\\
${}^3$Center for Gravitational Physics, Yukawa Institute for Theoretical Physics, Kyoto University, Kyoto 606-8502, Japan\\
${}^4$Faculty of Science and Technology, Hirosaki University, 3 Bunkyo-cho, Hirosaki, Aomori 036-8561, Japan
}

\date{\today}

\begin{abstract}
Clustering properties and peculiar velocities of halos in large-scale structure carry a wealth of cosmological information over a wide range of scales from linear to nonlinear scales.
We use halo catalogs in a suite of high-resolution $N$-body simulations to construct mock catalogs of galaxies that resemble the SDSS-like luminous early-type galaxies at three redshift bins in the range $0.15\le z\le 0.7$.
To do this we include 10 nuisance parameters to model variations in halo-galaxy connections for each redshift bin; the halo occupation distribution, and the spatial and velocity distributions of galaxies in the host halos.
We evaluate the Fisher information matrix for the redshift-space power spectrum of SDSS-like galaxies using different sets of the mock catalogs that are generated from changes in each of model parameters; cosmological parameters ($\sigma_8$ and $\Omega_{\rm m}$), the halo-galaxy connection parameters, and the cosmological distances ($D_{\rm A}$ and $H$ parameters at each redshift bin) for modeling an apparent geometrical distortion of the redshift-space power spectrum (the Alcock-Paczynski effect).
We show that combining the monopole and quadrupole power spectra of galaxies allows for
precise
estimations of the cosmological parameters and the cosmological distances, even after marginalization over the halo-galaxy parameters, by lifting the parameter degeneracies that are otherwise inevitable if either of the two spectra alone is used.
When including the galaxy power spectrum information up to $k=0.3$~\hMpci, we find about factor of 6 gain in the cosmological information content of ($\sigma_8,\Om,$ $D_{\rm A}$'s and $H$'s)
compared to $k=0.2$~\hMpci.
We also discuss the use of redshift-space galaxy power spectrum for a model-independent measurement of redshift-space distortion strength and a possible impact of the assembly bias on the cosmological parameters.
\end{abstract}

\preprint{YITP-19-64}
\maketitle

\section{Introduction}
\label{sec:intro}

The cosmic large-scale structure (LSS), which originates from the gravitational amplification of primordial fluctuations over the cosmic time in an expanding universe, is one of the most fundamental observables of cosmology.
In particular redshift-space clustering statistics of galaxies, measurable from a wide-area spectroscopic survey of galaxies, is one of the most powerful probes for estimating cosmological parameters including dark energy parameters, as well as for testing properties of gravity on cosmological scales.
To aim at addressing these fundamental questions in cosmology, various wide-area spectroscopic surveys are ongoing or being planned: these include the
Sloan Digital Sky Survey (SDSS) \footnote{\url{https://www.sdss.org}}, the SDSS-III Baryon Oscillation Spectroscopic Survey (BOSS) \cite{2013AJ....145...10D}, the SDSS-IV extended BOSS survey (eBOSS) \cite{2016AJ....151...44D}, the Dark Energy Spectroscopic Instrument (DESI) \cite{Aghamousa:2016zmz}, the Subaru Prime Focus Spectrograph (PFS) \cite{2014PASJ...66R...1T}, the ESA Euclid mission \cite{Laureijs:2011gra} and the NASA Wide Field Infrared Survey Telescope (WFIRST) mission \cite{Gehrels:2014spa}.

Galaxies are biased tracers of the underlying matter distribution in the large-scale structure, and the relation between the spatial distributions of galaxies and dark matter leaves an uncertainty -- the so-called galaxy bias uncertainty.
On the other hand
their peculiar velocities are believed to follow the the underlying velocity field of matter (mostly dark matter), at least on large scales, driven by the gravity due to inhomogenieties.
The peculiar velocities then cause a modulation in redshifts of individual galaxies observed in a spectroscopic survey.
Hence the observed galaxy distribution in redshift space \cite{1977ApJ...212L...3S} is apparently distorted along the line-of-sight direction -- the so-called
redshift-space distortion (RSD) effect \cite{kaiser84,Hamilton00}.
Even if the intrinsic (real-space) galaxy distribution is statistically homogeneous and isotropic, the RSD effect induces an apparent anisotropic pattern in the galaxy distribution in redshift space.
For instance, the two-point correlation function of galaxies becomes a function of two-dimensional separations between pairs of galaxies that are parallel and perpendicular to the line-of-sight direction.

Measurements of the redshift-space clustering of galaxies have been used to constrain cosmological parameters as well as test gravity theory \citep{peacock01,verde00,beutler12,2004PhRvD..69j3501T,Eisenstein:2005su,reid09,blake11a,reid12,2013A&A...557A..54D,2016PASJ...68...38O,2017MNRAS.470.2617A}.
Whilst these results look promising, a main challenge in extracting the cosmological information from the redshift-space clustering is inherent in uncertainties in modeling galaxy bias as well as nonlinearities in galaxy clustering and RSD effects.
In particular physics inherent in galaxy formation and evolution is impossible or at least very difficult to model from the first principles, and an empirical approach, e.g. introducing parameters to model the effects of galaxy bias, needs to be employed and then the parameters have to be marginalized over when deriving cosmological constraints.
Although a model employing the Kaiser formula plus the linear bias model is accurate and robust in the linear regime, it ceases to be accurate in the quasi-nonlinear regime, especially more quickly at $k\gtrsim 0.05~$\hMpci for the RSD effect compared to $\gtrsim 0.1$~\hMpci for the real-space clustering.
Hence, to include the clustering information up to the quasi-nonlinear regime, which has a more statistical power, the previous works usually employ the perturbation theory approach
including the perturbative expansion of galaxy bias \citep{scoccimarro04,2011MNRAS.417.1913R,taruya10,nishimichi11,2013MNRAS.429.1674C} \citep[also see][for a review]{bernardeau02,Desjacques18}
or the extension including the corrections of UV-sensitive calculations in the standard perturbation theory \citep{baumann12}. The cosmological information extractable from the RSD effect or bias parameters within these perturbative frameworks were also investigated \citep[e.g.,][]{taruya11,2012MNRAS.427..327D,ishikawa:2013lr}.
Although the perturbation theory approach gives a well-defined framework of nonlinear structure formation, it breaks down in the nonlinear regime, and the validation range and flexibility in models of the redshift-space clustering, especially the nonlinear RSD effect and nonlinear galaxy bias, are not yet clear or have not been fully explored.

An alternative approach to model the nonlinear redshift-space clustering is a method using simulations. Hydrodynamical simulations including galaxy formation physics are ideal, but are still computationally expensive \citep[e.g. see][for the recent developtment]{2018MNRAS.475..676S}.
Instead $N$-body simulations are tractable, and allow for running many realizations for different cosmological models \citep{Coyote1,MiraTitan1}.
In particular, we can identify dark matter halos from a concentration of particles in phase space in $N$-body simulation output as places where galaxies are likely to be formed.
With currently available numerical resources we can construct an ensemble of halo catalog realizations for different cosmological models, and build a database or ``emulation tool'' that allows for a fast, accurate computation of clustering statistics of halos as a function of cosmological models, halo masses, redshifts and separations (wavenumbers) \citep{Nishimichi:2018etk} \citep[also see][]{2015ApJ...810...35K,2018arXiv180405867Z,2018MNRAS.tmp.2206W}.
For example, the SDSS BOSS provides the large spectroscopic sample of luminous early-type galaxies that are believed to reside in halos with a typical mass of $10^{13}~h^{-1}M_\odot$, and such massive halos are relatively easy to simulate for a sufficiently large cosmological volume ($\sim 1~h^{-1}$Gpc).

Hence the purpose of this paper is to assess cosmological information content in the redshift-space power spectrum of galaxies assuming that we can accurately model the nonlinear clustering, nonlinear biasing and RSD effects of halos hosting the galaxies.
For this purpose, we use a suite of halo catalogs, constructed from high-resolution $N$-body simulations, to build mocks of galaxies that resemble the SDSS BOSS galaxies, using the halo occupation distribution (HOD) method \citep{1998ApJ...494....1J,seljak:2000uq,peacock:2000qy,2005ApJ...633..791Z}.
To create realistic mock catalogs, we include a sufficient number of nuisance parameters that model the halo-galaxy connection; the population of central and satellite galaxies, the off-centering effects of central galaxies, and the spatial and velocity distributions of satellite galaxies in the host halos.
We then measure the monopole and quadrupole moments of the redshift-space galaxy power spectrum as our hypothetical observables of the redshift-space galaxy clustering.
We evaluate the sensitivity of redshift-space galaxy power spectrum to each of model parameters using different mock catalogs of galaxies that are generated from changes in each of model parameters (cosmological parameters and the halo-galaxy connection parameters).
In doing this we also include apparent anisotropic clustering patterns in the redshift-space galaxy distribution that are caused if an assumed cosmological model, which needs to be employed to convert the observed angular separations and redshifts to the comoving coordinates in the clustering analysis, is different from the underlying true cosmology -- the so-called Alcock-Paczynski (AP) effect \cite{alcock79}.
Thus we use the redshift-space power spectra of galaxies, measured from these mocks, to numerically evaluate the Fisher information matrix for the SDSS-like galaxies, and then assess the cosmological information content up to the quasi-nonlinear regime
($k_{\rm max}=0.3~$\hMpci in our exercise), after marginalization over the halo-galaxy connection parameters.
Our study is somewhat based on the similar motivation to the previous works \cite{2015PhRvD..92j3516O,2017JCAP...10..009H}, but our work is different in the sense that we purely rely on the $N$-body simulations to assess cosmological information contents inherent in the redshift-space power spectrum of host halos up to the nonlinear regime, while the previous works used the perturbation theory based model.

The structure of this paper is as follows:
In \S~\ref{sec:preliminaries} we briefly describe the redshift-space galaxy power spectrum and the halo model formalism, and define various parameters we will use in the subsequent sections.
In \S~\ref{sec:simulation} we give details of the $N$-body simulations and the procedures to mock catalogs for the SDSS-like galaxies, as well as to measure the redshift-space power spectrum from the mock catalog.
In this section we also study how a change in each of the model parameters alter the galaxy power spectrum.
\S~\ref{sec:results} gives the main results of this paper.
We show forecasts of cosmological parameters and the cosmological distances expected from the redshift-space power spectrum of SDSS-like galaxies.
In \S~\ref{sec:discussion} we give discussion of how the redshift-space power spectrum can be used to robustly measure the strength of RSD effect in the presence of uncertainties in the halo-galaxy connection, and discuss a possible impact of the assembly bias on the cosmological parameter estimation from the redshift-space galaxy power spectrum.
In \S~\ref{sec:conclusion} we give a conclusion.

\section{Preliminaries}
\label{sec:preliminaries}

We extensively use a sufficiently large number of realizations of the mock catalogs for a hypothetical galaxy survey, which are constructed from halo catalogs of high-resolution $N$-body simulations, and then perform
a Fisher matrix analysis using the galaxy power spectra measured from the mocks. In this section, we describe basics of redshift-space distortion effect on the galaxy power spectrum and the halo occupation distribution (HOD) that gives a prescription of connecting galaxies to halos in the $N$-body simulation realization.

\subsection{Redshift-space galaxy power spectrum}

In redshift space the positions of galaxies appear to be modulated by their peculiar velocities along the line-of-sight direction as
\begin{align}
 \bs = \bx + \frac{v_{\rm LoS}(\bx)}{aH(z)}\bf{\hat{e}}_{\rm LoS},
 \label{eq:rsd_def}
\end{align}
where $\bs$ or $\bx$ is the redshift- or real-space position of a galaxy, respectively, and $v_{\rm LoS}$ is the line-of-sight component of the peculiar velocity, $a$ is the scale factor, $H(z)$ is the Hubble expansion rate at redshift $z$, and $\bf{\hat{e}}_{\rm LoS}$ is the unit vector along the line-of-sight direction.
The density fluctuation field of galaxies in redshift space is related to the real-space density field via the number conservation of galaxies as
\begin{align}
\label{eq:massconserve}
 [1 + \delta^{\rm S}\!(\bs)] {\rm d}^3\bs = [1+ \delta(\bx)] {\rm d}^3\bx,
\end{align}
where $\delta(\bx)$ is the density fluctuation field at the position $\bx$, and quantities with superscript ``S'' throughout this paper denote their redshift-space quantities.
Throughout this paper we employ the distant observer approximation or equivalently the plane-parallel approximation.
Under this approximation, we can take the line-of-sight direction to all the galaxies to be along the $z$-axis ($\bf{\hat{e}}_{\rm LoS} = \bf{\hat{z}}$) without loss of generality.
Fourier-transforming Eq.~(\ref{eq:massconserve}) yields
\begin{align}
\delta_{\rm D}(\bk) + \delta^{\rm S}\!(\bk) = \int {\rm d}^3\bx e^{i \bk\cdot \left[\bx + \frac{v_{z}(\bx)}{a H(z)} \bf{\hat{z}}\right]} [1+ \delta(\bx)],
\end{align}
where $\delta_{\rm D}\!(\bk)$ is the Dirac's delta function.
The redshift-space distortion breaks the statistical isotropy of galaxy distribution.
The redshift-space two-point correlation function of galaxies in Fourier space, i.e. the redshift-space power spectrum, is defined as
\begin{align}
 \left \langle \delta^{\rm S}\!(\bk) \delta^{\rm S}\!(\bk') \right \rangle = (2\pi)^3 \delta_{\rm D}(\bk + \bk') P^{\rm S}\!(\bk).
\end{align}
Even if the real-space power spectrum depends only on the length of wavevector $k=|{\bf k}|$ due to the statistical isotropy, the redshift-space distortion
leads the redshift-space power spectrum to depend on the direction of ${\bf k}$ with respect to
the line-of-sight direction.
Since the statistical isotropy retains in the two-dimensional plane perpendicular to the line-of-sight direction, denoted as ${\bf k}_\perp$, the ${\bf k}$-direction dependence of redshift-space power spectrum is fully characterized by the direction cosine $\mu\equiv k_z/k$.
We can then decompose the $\mu$-dependence of the
redshift-space power spectrum into multipole moments based on the Legendre polynomial expansion as
\begin{align}
 P^{\rm S}_\ell(k) \equiv \frac{2\ell+1}{2} \int_{-1}^{1}\!{\rm d}\mu~ P^{\rm S}\!(k,\mu) ~ {\cal L}_{\ell}(\mu),
\end{align}
where ${\cal L}_{\ell}(\mu)$ is the $\ell$-th order Legendre polynomial function.
The monopole ($\ell = 0$) and the quadrupole ($\ell = 2$) moments are dominant terms of the anisotropic power spectrum at least on scales up to the quasi-nonlinear regime we are interested in.
The current-generation galaxy redshift surveys such as the SDSS survey enable a significant detection of these two dominant moments \cite{Beutler:2016arn,Beutler:2016ixs}.

\subsection{Halo model prescription of galaxy power spectrum}
\label{subsec:halomodel}

A cosmological analysis of galaxy clustering often employs theoretical models based on linear perturbation theory or its extensions including higher-order corrections.
Galaxies, selected under some criteria, are a biased tracer of the underlying matter distribution in the large-scale structure,
which introduces uncertainty in their relation to the underlying matter distribution -- the so-called galaxy bias.
Since physical processes of galaxy formation and evolution are complicated and difficult to accurately model from the first principles, our understanding of galaxy bias is limited.
Hence, a phenomenological parametrized model of galaxy bias needs to be used for a cosmological analysis of galaxy clustering.
In this paper we employ the halo model approach \citep{peacock:2000qy,seljak:2000uq,ma:2000lr,scoccimarro:2001fj}
\citep[also see][for a review]{Cooray02}
to model galaxy bias, and then address how we can extract cosmological information from the redshift-space galaxy power spectrum after marginalizing over a large number of nuisance parameters to model galaxy bias uncertainty.

In the halo model, we assume that all matter is associated with halos, and then the matter auto correlation function is described by the contribution from pairs of mass elements in the same halo and those in two distinct halos, which are referred to as the 1-halo and 2-halo terms, respectively.
Likewise, the galaxy correlation function is decomposed into the 1- and the 2-halo terms.
We adopt the halo occupation distribution \citep[HOD;][]{1998ApJ...494....1J,scoccimarro:2001fj} to model how many galaxies reside in halos of a given mass.
We consider an HOD model in which the mean number of central and satellite galaxies are given separately:
\begin{align}
\avrg{N}\!(M)&=\avrg{N_{\rm c}}\!(M)+\avrg{N_{\rm s}}\!(M),
\end{align}
where the notation $\avrg{\hspace{1em}}(M)$ denotes the average of a quantity for halos of a given mass $M$.
We employ the mean HOD for central galaxies given by the following form:
\begin{align}
\avrg{N_{\rm c}}\!(M)=\frac{1}{2}\left[1+{\rm erf}\left(\frac{\log M-\log M_{\rm min}}{\sigma_{\log M}}\right)\right],
\label{eq:Nc}
\end{align}
where ${\rm erf}(x)$ is the error function and $M_{\rm min}$ and $\sigma_{\log M}$ are model parameters. Note $\avrg{N_{\rm c}}\le 1$.
The mean central HOD, $\avrg{N_{\rm c}}\!(M)$, can be interpreted as the probability that a halo with mass $M$ hosts a central galaxy.
The mean central HOD considered here has properties that $\avrg{N_{\rm c}} \rightarrow 0$ for halos with $M\ll M_{\rm min}$, while $\avrg{N_{\rm c}}\rightarrow 1$ for halos
with $M\gg M_{\rm min}$.
In our fiducial model we assume that halos host central galaxies following a Bernoulli process with the probability solely determined by the halo mass,
or equivalently we ignore a possible extra dependence of the central HOD, often referred to as the \textit{halo assembly bias}, on other physical properties such as large-scale environments or internal structures of halos (density profile, formation epoch, etc.)
for our default choice. We will come back to the impact of assembly bias effect.

For the mean HOD of satellite galaxies, we employ the following parametrized model:
\begin{align}
\avrg{N_{\rm s}}\!(M)&\equiv \avrg{N_{\rm c}}\!(M)\lambda_{\rm s}\!(M)=
\avrg{N_{\rm c}}\!(M)\left[\frac{M-M_{\rm sat}}{M_1}\right]^{\alpha_{\rm sat}},
\label{eq:Ns}
\end{align}
where $M_{\rm sat}$, $M_1$ and $\alpha_{\rm sat}$ are model parameters, and we have introduced the notation $\lambda_{\rm s}(M)\equiv
[(M-M_{\rm sat})/M_1]^{\alpha_{\rm sat}}$ for convenience of our following discussion.
For our default prescription, we assume that satellite galaxies reside only in a halo that already hosts a central galaxy.
Hence, in the above equation, $N_{\rm c}=1$ for a halo where satellite galaxy(ies) can reside.
Then we assume that the number distribution of satellite galaxies in a given host halo follows the Poisson distribution with mean $\lambda_{\rm s}$:
$P(N_{\rm s}|N_{\rm c}=1) 
=(\lambda_{\rm s})^{N_{\rm s}}\exp(-\lambda_{\rm s})/(N_{\rm s}!)$ and $P(N_{\rm s}|N_{\rm c}=0) = \delta^\mathrm{K}_{N_{\rm s},0}$, where $\delta^\mathrm{K}_{i,j}$ stands for the Kronecker delta.
Thus a host halo of satellite galaxies has in total $1+N_{\rm s}$ galaxies.
Under this setting, the mean number of galaxy pairs living in the same halo with mass $M$, which is relevant for the 1-halo term calculation, can be computed as
%
\begin{align}
\avrg{N(N-1)}&= P(N_{\rm c}=1) \avrg{N(N-1)|N_{\rm c}=1} + P(N_{\rm c}=0) \avrg{N(N-1)|N_{\rm c}=0}\nonumber\\
&=\avrg{N_{\rm c}}\sum_{N_{{\rm s}}=0}^\infty~ 
\frac{(\lambda_{\rm s})^N_{\rm s}}{N_{\rm s}!}\exp(-\lambda_{\rm s})~
N_{\rm s}(1+N_{\rm s})\nonumber\\
&= \avrg{N_{\rm c}}\!(M)\left[2\lambda_{\rm s}(M)+\lambda_{\rm s}(M)^2\right].
\end{align}
Note that the second term in the first line is zero following our assumption that no satellite galaxies reside in a halo without a central galaxy.
This treatment is the same as in Ref.~\cite{2005ApJ...633..791Z}.

%
The central and satellite HOD models we use in this paper are specified by 5 parameters
$\{M_{\rm min},\sigma_{\log M},M_1,M_{\rm sat},\alpha_{\rm sat}\}$.

Once the HOD model is given, the mean number density of galaxies under consideration is given as
\begin{align}
\bar{n}_{\rm g}&=\int\!{\rm d}M~\frac{{\rm d}n}{{\rm d}M}\left[\avrg{N_{\rm c}}\!(M)+\avrg{N_{\rm s}}\!(M)\right],
\end{align}
where ${\rm d}n/{\rm d}M$ is the halo mass function which gives the mean number density of halos in the mass range $[M,M+{\rm d}M]$.
The redshift-space power spectrum of galaxies can be, without loss of generality, decomposed into two contributions within the halo model framework:
\begin{align}
P^{\rm S}_{\rm gg}\!(\bk)=P^{\rm S, 1h}_{\rm gg}\!(\bk)+P^{\rm S, 2h}_{\rm gg}\!(\bk),
\end{align}
where $P^{\rm S, 1h}_{\rm gg}$ and $P^{\rm S,2h}_{\rm gg}$ are the 1- and 2-halo terms, respectively.
The 1-halo term arises from pairs of galaxies that reside in the same halo, while the 2-halo term arises from those in different halos.
The two terms can be expressed as
\begin{align}
P^{\rm S, 1h}_{\rm gg}(\bk)&=\frac{1}{\bar{n}_{\rm g}^2}
\int\!\!{\rm d}M~\frac{{\rm d}n}{{\rm d}M}
\avrg{N_{\rm c}}\!(M)
\left[
2\lambda_{\rm s}(M){\cal H}(\bk; M, c, \sigma_{{\rm vir}, M})
+\lambda_{\rm s}(M)^2{\cal H}(\bk; M, c,\sigma_{{\rm vir}, M})^2
\right],
\label{eq:PS1h}
\end{align}
and
\begin{align}
P^{\rm S,2h}_{\rm gg}(\bk)&=
\frac{1}{\bar{n}_{\rm g}^2}
\int\!{\rm d}M\frac{{\rm d}n}{{\rm d}M}\left[\avrg{N_{\rm c}}\!(M)+\avrg{N_{\rm s}}\!(M)
{\cal H}(\bk; M, c, \sigma_{{\rm vir},M})\right]\nonumber\\
&\times \int\!{\rm d}M'\frac{{\rm d}n}{{\rm d}M'}\left[\avrg{N_{\rm c}}\!(M')
+\avrg{N_{\rm s}}\!(M'){\cal H}(\bk; M', c', \sigma_{{\rm vir},M'}^\prime)\right]P^{\rm S}_{\rm hh}(\bk; M, M').\label{eq:PS2h}
\end{align}
Here $P_{\rm hh}^{\rm S}(\bk; M, M')$ is the redshift-space power spectrum of two halos that have masses $M$ and $M'$, respectively.
The above 1-halo term has an asymptotic behavior $P^{\rm S,1h}_{\rm g}\rightarrow 1/\bar{n}_{\rm gg}$ for the limit $k\rightarrow 0$ because ${\cal H}\rightarrow 1$.
The function ${\cal H}(\bk; M, c, \sigma_{{\rm vir},M})$ is the normalized, averaged two-dimensional distribution of satellite galaxies in redshift space for halos of mass $M$.
As described in \citet{2001MNRAS.325.1359S} and \citet{2001MNRAS.321....1W} \citep[also see Refs.][]{hikage12a,hikage:2013kx}, the redshift-space profile of galaxies is given by a convolution of the real-space number density profile and the velocity distribution function for satellite galaxies in the host halo.
From the convolution theorem, the redshift-space profile in Fourier space is given as a multiplicative form
\begin{align}
{\cal H}(\bk; M, c, \sigma_{{\rm vir}, M})&=\tilde{u}_{\rm s}(k; M,c)\tilde{{\cal F}_{\rm s}}\!(\bk; \sigma_{{\rm vir},M}),
\label{eq:Hs}
\end{align}
where $\tilde{u}_{\rm s}(k; M, c)$ is the Fourier transform of the normalized radial profile of satellite galaxies in real space centered at their host halo with mass $M$.
For this, we employ the Navarro-Frenk-White (NFW) model \citep{NFW}; the normalized profile is given by
\begin{equation}
u_{\rm s}(r; M,c)\equiv u_{\rm NFW}(r; M,c)=\frac{c^3}{4\pi r_{\rm 200}^3} \left[ \ln(1+c) - \frac{c}{1+c} \right]^{-1} \frac{1}{(cr/r_{\rm 200})(1+cr/r_{\rm 200})^2},
\end{equation}
where $c$ is the concentration parameter, $r_{\rm 200}$ is the radius corresponding to the boundary of halo, defined so as to satisfy $\int^{r_{\rm 200}}_0\!4\pi r^2{\rm d}r~ u_{\rm s}(r)=1$.
Thus the normalized NFW profile is fully specified by two parameter for which we use $(M,c)$.
The Fourier transform, $\tilde{u}_{\rm s}(k; M,c)$, is given by Eq.~17 in \citet{2003MNRAS.340..580T} \citep[also see][]{scoccimarro:2001fj}. For the concentration parameter
we employ the model in \citet{2015ApJ...799..108D} that gives the scaling relation of halo concentration with halo mass and redshift
for a given cosmological model.
However, we introduce a nuisance parameter to model a possible uncertainty in the concentration amplitude, $c(M,z)\rightarrow c_{\rm conc}c(M,z)$ ($c_{\rm conc}=1$ for the fiducial model), and then study cosmological parameter forecasts after marginalizing over the parameter $c_{\rm conc}$.

The function ${\cal F}_\mathrm{s}$ in Eq.~\ref{eq:Hs} describes an apparent displacement of satellite galaxy due to its peculiar velocity with respect to the halo center -- the so-called fingers-of-God (FoG) effect \cite{jackson72,1978IAUS...79...31T}.
To model the FoG effect, we need to assume the velocity distribution of satellite galaxies which is not well understood. In this paper we assume a Gaussian distribution for the velocity function for simplicity:
\begin{align}
{\cal F}_{\rm s}(\Delta r_\parallel; r) {\rm d}\Delta r_{\parallel} \equiv \frac{1}{\sqrt{2\pi}\sigma_{{\rm vir}, M}(r)} \exp \left[
-\frac{v_{\parallel}^2}{2\sigma_{{\rm vir},M}^2(r)} \right] {\rm d}v_{\parallel},
\end{align}
where the radial displacement is given in terms of the radial component of peculiar velocity as $\Delta r_\parallel = v_\parallel/[aH(a)]$. The velocity distribution function satisfies the normalization condition:
$\int_{-\infty}^{\infty}\!{\rm d}\Delta r_{\parallel}~{\cal F}_\mathrm{s}(\Delta r_\parallel)=1$. For the one-dimensional velocity dispersion $\sigma_{{\rm vir},M}(r)$ we assume the virial velocity dispersion following \citet{hikage12a}:
\begin{align}
\label{eq:virial}
\sigma_{{\rm vir},M}^2(r) &= \frac{GM(<r)}{2r},
\end{align}
where $M(<r)$ is the mass enclosed within the radius $r$ (in the physical length scale, not comoving scale) from the center of the halo, and can be calculated by assuming the NFW profile.
Thus we assume a radially-varying velocity dispersion depending on the position of satellite galaxy at $r$ inside the host halo.
The FoG effect causes an anisotropic modulation in galaxy clustering. In order to study the impact of an uncertainty in the FoG effect on cosmological parameters, we introduce a nuisance parameter to model the uncertainty, $\sigma_{{\rm vir}, M}(r) \rightarrow c_{\rm vel}\sigma_{{\rm vir},M}(r)$ ($c_{\rm vel}=1$ for the fiducial model).

Furthermore we include a possible effect of ``off-centered'' central galaxies in our modeling.
Since dark matter halo is not a well-defined object and experiences mergers of progenitor halos, galaxies selected by specific ways based on a spectroscopic sample (e.g. color and magnitude cuts) might be off-centered (i.e. satellite) galaxies, as indicated in the results of \citet{hikage:2013kx}.
Even if a halo contains a single target galaxy in the sample and if the galaxy is off-centered from the true halo center, the galaxy is categorized as a ``central'' galaxy in a naive HOD picture.
We explicitly include effects of these off-centered galaxies on redshift-space galaxy power spectrum.
To do this, we follow the methods in \citet{hikage:2013kx} \citep[also see][]{hikage:2013kx,oka13,masaki13,2015ApJ...806....2M};
we introduce a parameter $p_{\rm off}$ $(0\le p_{\rm off}\le 1)$, which represents the probability that each central galaxy is off-centered from the center of its host halo.
In addition we assume a Gaussian distribution for the radial distribution of the off-centered galaxy in each host halo:
\begin{align}
p(r_{\rm off})=\frac{1}{(2\pi)^{3/2}(r_{\rm s}{\cal R}_{\rm off})^3}\exp\left[-\frac{(r_{\rm off})^2}{2(r_{\rm s}{\cal R}_{\rm off})^2}\right],
\label{eq:poff_real}
\end{align}
where $r_{\rm s}$ is the scale radius of each halo with mass $M$ and we include additional dimensionless parameter ${\cal R}_{\rm off}$ which models a typical off-centering radius in units of the scale radius.
Under this formulation, in Eqs.~\ref{eq:PS1h},~\ref{eq:PS2h} we can replace the central HOD as
\begin{align}
\avrg{N_{\rm c}}\!(M) \rightarrow \left[(1-p_{\rm off})+p_{\rm off}\exp\left\{-\frac{1}{2}{k^2(r_{\rm s}{\cal R}_{\rm off})^2}\right\}\tilde{\cal F}(\bk;\sigma_{{\rm vir},M})\right]\avrg{N_{\rm c}}\!(M).
\end{align}

Thus the 1-halo term of the redshift-space galaxy power spectrum involves complicated model ingredients; the spatial and velocity distributions of galaxies in their host halos,  which are difficult to theoretically predict from the first principles for a given cosmological model.
For this reason we will treat the 1-halo term as a nuisance term, and study how cosmological parameter forecasts are degraded when marginalizing over nuisance parameters to model variations in the 1-halo term.

As long as we use adequate $N$-body simulation outputs, we can accurately model the redshift-space power spectrum of halos that carries the underlying cosmological information over a range of  scales covered by the simulations.
For given initial conditions of the primordial fluctuations and a given background cosmological model such as the $\Lambda$CDM model, the power spectrum of halos of masses $M$ and $M'$ can be represented as a function of wavenumber, cosmological parameters, redshift and halo masses:
\begin{align}
P^{\rm S}_{\rm hh}(\bk; z, M,M',\Om,\sigma_8,\dots).
\end{align}
Here the redshift-space halo power spectrum is a two-dimensional function, given as a function of $\bk=(k_\parallel, k_\perp)$ due to the RSD effect where halo's positions (their centers) are modulated in the large-scale structure according to the line-of-sight component of their bulk motions that depend on cosmology.
Hence, the spatial clustering and RSD effects of halos carry cosmological information, and we address how these can be extracted even after marginalizing over uncertainties in the halo-galaxy connection parameters.

\subsection{Alcock-Paczynski effect}
\label{sec:AP}

In an actual galaxy redshift survey, another useful information can be extracted
from an apparent geometrical distortion in the observed pattern of galaxy clustering, the so-called Alcock-Paczynski (AP) effect \cite{alcock79}.
This effect arises from discrepancy between the underlying true cosmology and the ``reference'' cosmological model, where the latter needs to be assumed in a clustering analysis when mapping direct observables of galaxy positions, i.e. its spectroscopic redshift and angular positions, to the comoving coordinates.
The AP effect can be described by the coordinate transformation, and in Fourier space it is given as
\begin{align}
k_{\perp,{\rm ref}}&=\frac{D_{\rm A}(z)}{D_{\rm A, ref}(z)}k_\perp, \hspace{1em}k_{\parallel,{\rm ref}}=\frac{H_{\rm ref}(z)}{H(z)}k_{\parallel},
\end{align}
where $D_{\rm A}(z)$ denotes the angular diameter distance to redshift $z$ and $H(z)$ the Hubble expansion rate at $z$,
and quantities with and without subscript ``ref'' hereafter denote those in the reference (assumed) cosmological model and the true (unknown) cosmology, respectively.
For convenience of our discussion, we define two parameters to characterize the AP distortion, following Ref.~\cite{padmanabhan08}, as
\begin{align}
\alpha_\perp \equiv \frac{D_{\rm A}(z)}{D_{\rm A, ref}(z)}, \hspace{1em} \alpha_\parallel \equiv \frac{H_{\rm ref}(z)}{H(z)}.
\end{align}
Hence the observed redshift-space power spectrum can be expressed in terms of the underlying true power spectrum as
\begin{align}
P^{\rm S}_{\rm gg, ref}(k_{\parallel,{\rm ref}},k_{\perp,{\rm ref}})&=
\frac{1}{\alpha_\perp^2\alpha_\parallel}
P^{\rm S}_{\rm gg}(k_{\parallel},k_\perp)+P_{\rm sn},
\label{eq:PS_gg_AP}
\end{align}
where we introduced a constant parameter to model a possible residual shot noise, $P_{\rm sn}$, in the observed power spectrum
following
\citep{2003ApJ...598..720S}. In the following, we will often omit the subscript ``ref'' for notational simplicity.
The baryon acoustic oscillation (BAO) features in the galaxy power spectrum makes the AP effect very powerful to constrain the angular diameter distance and the Hubble expansion rate at the redshift of the galaxy survey, $D_{\rm A}(z)$ and $H(z)$ \citep{2003ApJ...598..720S,hu03,eisenstein05} (here note that the BAO peak location is almost unaffected by the RSD effect).
The AP effect can also be utilized to constrain the cosmological parameters which are relevant to the cosmic expansion (see e.g. \cite{Ramanah:2018eed}).

\subsection{Multipole power spectra}
\label{subsec:multipole}

As we described, the observed redshift-space power spectrum of galaxies (Eq.~\ref{eq:PS_gg_AP}) is given as a function of two wavenumbers perpendicular and parallel to the line-of-sight direction, $(k_\parallel,k_\perp)$, and contains the full information at two-point statistics level.
The multipole expansion of the observed power spectrum is a useful way of data compression as often used in the literature and can characterize anisotropies in the power spectrum, although it is not optimal unless all the multipole spectra up to infinite orders are included (which is infeasible in practice).
In addition, the covariance of the multipole power spectrum is easier to estimate, e.g. using a less number of mock catalogs, due to the dimension reduction of the data vector. The multipole power spectrum is defined from the observed spectrum as
\begin{align}
P^{\rm S}_{{\rm gg, ref},\ell}(k_{\rm ref}) = \frac{2\ell+1}{2\alpha^2_\perp\alpha_\parallel } \int_{-1}^1 {\rm d}\mu_{\rm ref}~
\left\{P^{\rm S}_{\rm gg}\left [k(k_{\rm ref}, \mu_{\rm ref}), \mu(\mu_{\rm ref}) \right ]+P_{\rm sn}\right\}
{\cal L}_\ell(\mu_{\rm ref}),
\label{eq:PS_l_AP}
\end{align}
where ${\cal L}_\ell(x)$ is the $\ell$-th order Legendre polynomial and
\begin{align}
&k(k_{\rm ref}, \mu_{\rm ref}) \equiv \sqrt{k_\parallel^2+k_\perp^2}= k_{\rm ref} \frac{1}{\alpha_\perp} \left [1+\mu_{\rm ref}^2 \left (\frac{\alpha_\perp^2}{\alpha_\parallel^2}-1 \right ) \right ]^{1/2}, \\
&\mu(\mu_{\rm ref}) \equiv \frac{k_\parallel}{k}=  \mu_{\rm ref} \frac{\alpha_\perp}{\alpha_\parallel}  \left [1+\mu_{\rm ref}^2 \left (\frac{\alpha_\perp^2}{\alpha_\parallel^2}-1 \right ) \right ]^{-1/2}
\label{eq:kmu_AP}
\end{align}
As we will show below explicitly, the monopole power spectrum is sensitive to the ``dilation'' parameter,
$\alpha_\perp^2\alpha_\parallel$ whose variation causes an isotropic shift of the BAO peak locations and a change in the power spectrum amplitudes. On the other hand, the quadrupole power spectrum is sensitive to the ``warping'' parameter $\alpha_\perp/\alpha_\parallel$, or $F_{\rm AP}$ in
\citet{2017MNRAS.470.2617A} whose variation causes an anisotropic distortion in the redshift-space power spectrum (therefore a change in the quadrupole power spectrum).

\section{$N$-body Simulations, Halo catalogs and Mocks of Galaxy Power spectra}
\label{sec:simulation}

In this paper we compute the Fisher matrix for the model parameters including the halo-galaxy connection parameters as well as the cosmological parameters, based on the derivative factors evaluated from the $N$-body simulations.
In this section we describe details of our procedures: $N$-body simulations, halo catalogs, the mock catalogs of SDSS-like galaxies, the way to compute the sensitivity of the redshift-space galaxy power spectrum to each model parameter, and the way to compute the covariance matrix.

\subsection{\texorpdfstring{$N$-Body}{N-body} Simulations and Halo Catalogs}

\begin{table}
\begin{center}
\begin{tabular}{l|ccccc} \hline \hline
	Model & realizations of sim. & $\Om$ & $\ln(10^{10}A_{\rm s})$ & $\sigma_8$ & Response ($\partial P^{\rm S}_\ell/\partial p_\alpha$) w.r.t. \\ \hline
	{\it Planck}      & $16$ & $0.3156$ & $3.094$ & $0.831$ & AP/HOD/galaxy \\
	$\Omega_{\rm m+}$ & $10$ & $0.350$  & -       & $0.820$  & $\Om,\,\sigma_8$ \\
	$\Omega_{\rm m-}$ & $10$ & $0.281$  & -       & $0.841$  & $\Om,\,\sigma_8$ \\
	$A_{\rm s+}$      & $10$ & -        & $3.249$ & $0.897$ & $\sigma_8$ \\
	$A_{\rm s-}$      & $10$ & -        & $2.939$ & $0.769$ & $\sigma_8$ \\ \hline
\end{tabular}
\caption{
Summary of $N$-body simulations used in this study. Each simulation employs a box size of $2~h^{-1}{\rm Gpc}$ and $2048^3$ $N$-body particles (see text for details).
We use $16$ realizations for the fiducial {\it Planck} cosmology to compute the response functions
of redshift-space galaxy power spectrum, $\partial P^{\rm S}_{{\rm gg},\ell}(k)/\partial p_\alpha$, with respect to each of the halo-galaxy connection parameters (see text) and the cosmological
distances at a given redshift, $D_{\rm A}(z)$ and $H(z)$, via a hypothetical clustering analysis taking into account an apparent geometrical distortion due to the Alcock-Paczynski effect.
In addition, we use $10$ realizations for each of the varied cosmological models where either of $\Omega_{\rm m}$ or the parameter of primordial curvature perturbations, $\ln(10^{10}A_{\rm s})$, is shifted by about $\pm 11\%$ or $\pm5\%$, respectively, but other cosmological parameters are kept fixed.
We use these varied cosmology simulations to compute the response functions with respect to either of $\Om$ or $\sigma_8$.
The element denoted as ``-'' means the same parameter value as the fiducial {\it Planck} cosmology.
}
\label{tab:simu}
\end{center}
\end{table}

In this subsection we briefly review details of the $N$-body simulations and the halo catalogs we use in this paper, as summarized in Table~\ref{tab:simu}.
All the $N$-body simulations are performed with $2048^3$ particles in a comoving cubic box
with the side length of $2~h^{-1}{\rm Gpc}$.
These are the \texttt{LR} simulation suite performed as a part of \texttt{Dark Quest} simulation campaign \citep[see][for details]{Nishimichi:2018etk}.
We simulate the time evolution of particle distribution using the parallel Tree-Particle Mesh code {\tt Gadget2} \citep[][]{gadget2}.
We employ the second-order Lagrangian perturbation theory \citep{scoccimarro98,crocce06a,crocce06b,nishimichi09,Valageas11a,valageas11b} to set up the initial displacement vector and the initial velocity of each $N$-body particle, where we use the publicly-available code {\tt CAMB} \citep{camb} to compute the linear matter power spectrum.
As for the fiducial cosmological model, we employ the best-fit flat-geometry $\Lambda$CDM model that is supported by the {\it Planck} CMB data \citep{planck-collaboration:2015fj}:
$(\omega_{\rm b},\omega_{\rm c},\Omega_{\rm \Lambda},\ln(10^{10}A_{\rm s}),n_{\rm s})=(0.02225,0.1198,0.6844,3.094,0.9645)$.
We set $\omega_\nu=0.00064$ for the density parameter of massive neutrinos to set up the initial condition, but we use a single $N$-body particle species to represent the ``total matter'' distribution and follow the subsequent time evolution of their distribution (i.e. we do not consider the distinct evolution of the massive neutrinos).
The {\it Planck} model has, as derived parameters, $\Omega_{\rm m}=0.3156$
(the present-day matter density parameter), $\sigma_8=0.831$ (the present-day rms linear mass density fluctuations within a top-hat sphere of radius $8~h^{-1}{\rm Mpc}$) and $h=0.672$ for the Hubble parameter.
The particle mass is $8.16\times 10^{10}~h^{-1}M_\odot$ for the fiducial {\it Planck} cosmology.

In this paper we use different sets of $N$-body simulations that are generated using the above simulation specifications.
The first set is $16$ $N$-body simulation realizations for the fiducial {\it Planck} cosmology that are presented in \citet{Nishimichi:2018etk} using different seeds of the initial conditions.
These correspond to the total volume of $16 \times 8=128~({\rm Gpc}/h)^3$ which is much larger than the volume of SDSS survey with area $8000~{\rm deg}^2$ and over the redshift range we consider in this paper, $V_{\rm SDSS}\simeq 6.2~(h^{-1}{\rm Gpc})^3$ (see Table~\ref{tab:surveyparam}).
Hence the {\it Planck} realizations allow for a precision measurement of redshift-space galaxy power spectra due to the large statistics.

For each simulation realization we identify halos in the post-processing computation, using the friends-of-friends halo finder in $6$D phase space, {\tt Rockstar} \cite{Behroozi:2013}
\citep[also see][for details]{Nishimichi:2018etk}.
Throughout this paper we adopt $M\equiv M_{200}=(4\pi/3)(r_{\rm 200})^3(200\bar{\rho}_{\rm m0})$ for the halo mass definition, where $r_{\rm 200}$ is the spherical halo boundary radius within which the mean mass density is $200\times \bar{\rho}_{\rm m0}$.
Note that the use of the present-day mean mass density
$\bar{\rho}_{\rm m0}$
is due to our use of comoving coordinates, and therefore $r_{\rm 200}$ is in the comoving length unit.
We employ the default setting of the {\tt Rockstar} finder, and define the center of each halo from the center-of-mass location of a subset of member particles in the inner part of halo, which is considered as a proxy of the gravitational potential minimum.
Our definition of halo mass includes all the $N$-body particles within the boundary $r_{\rm 200}$ around the halo center (i.e. including particles even if those are not gravitationally bound by the halo).
After identifying halos, we classify those into central and satellite halos; when the separation of two halos (between their centers) is closer than $r_{\rm 200}$ of the more massive one, we mark the less massive one as a satellite halo.
We keep only central halos with mass above $10^{12}\,h^{-1}M_\odot$ in the halo catalog.
In this paper we use the halo catalogs at output redshifts $z=0.251, 0.484$ and $0.617$ that represent the redshifts of the LOWZ galaxy sample and low-$z$ and high-$z$ sides of the CMASS galaxies in the SDSS survey, respectively.
Table~\ref{tab:surveyparam} gives a summary of these survey parameters.

The purpose of this paper is to evaluate cosmological information contents in the galaxy power spectrum after marginalization over various nuisance parameters that model halo-galaxy connection.
To do this, we numerically evaluate a response function of the redshift-space galaxy power spectrum to each model parameter, $\partial P^{\rm S}_{{\rm gg},\ell}(k)/\partial p_\alpha$, which quantifies how a change in the $\alpha$-th parameter, $p_\alpha$,
alters the multipole power spectrum $P^{\rm S}_{{\rm gg},\ell}(k)$.
More precisely we use the two-sided finite difference to numerically evaluate the partial derivative:
\begin{align}
\frac{\partial P^{\rm S}_{{\rm gg},\ell}(k)}{\partial p_\alpha}&\equiv
\frac{P^{\rm S}_{{\rm gg},\ell}(k;p_{\alpha,{\rm fid}}+\Delta p_{\alpha})
-P^{\rm S}_{{\rm gg},\ell}(k;p_{\alpha,{\rm fid}}-\Delta p_{\alpha})
}{2\Delta p_{\alpha}}
\label{eq:dPdp_alpha}
\end{align}
where  $P^{\rm S}_{{\rm gg},\ell}(k;p_{\alpha,{\rm fid}} \pm \Delta p_{\alpha})$ is
the $\ell$-th multipole power spectrum of galaxies where the parameter is shifted to $p_{\alpha,{\rm fid}} \pm \Delta p_{\alpha}$ by
a small amount $\Delta p_{\alpha}$ (see below), but other parameters are fixed to their fiducial values.
To compute this we use the mock catalog of galaxies to evaluate the power spectrum $P^{\rm S}_{{\rm gg},\ell}(k)$ for each of varied models.
For the derivatives $\partial P^{\rm S}_{{\rm gg},\ell}(k)/\partial p_\alpha$, with respect to the halo-galaxy connection parameters, we use the mock catalogs of SDSS galaxies that are constructed from 16 $N$-body simulation realizations at target redshifts
for the fiducial {\it Planck} cosmology (see below for details).

To compute the other derivatives with respect to cosmological parameters we run different sets of $N$-body simulations to estimate the dependence of the galaxy power spectrum.
In this paper we focus on $\Omega_{\rm m}$ and $\sigma_8$ to which the galaxy power spectrum is sensitive within the flat $\Lambda$CDM model.
Other cosmological parameters are fixed to their values for the fiducial {\it Planck} model.
To do this we run 10 paired realizations of $N$-body simulations by varying either of $\Omega_{\rm m}$ or $A_{\rm s}$ on either positive or negative side from their fiducial value, with other parameters being kept to the fiducial values (actually we use $\ln (10^{10} A_{\rm s})$ rather than $A_{\rm s}$ itself, as the parameter for which the numerical derivatives are evaluated, but we simply refer to as $A_{\rm s}$ to avoid the complexity of the description).
For each paired realizations, we use the same initial seed of initial conditions to run the $N$-body simulations.
Table~\ref{tab:simu} summarizes details of the $N$-body simulations we use in this paper.
The reason we use $A_{\rm s}$ instead of $\sigma_8$ is that we use $A_{\rm s}$ for the normalization parameter of the initial power spectrum for the fiducial {\it Planck} model in Ref.~\cite{Nishimichi:2018etk}.
Using the chain rule we compute the numerical derivative with respect to $\sigma_8$ as
\begin{align}
\left.\frac{\partial P^{\rm S}_{{\rm gg},\ell}(k)}{\partial \Omega_{\rm m}}\right|_{A_{\rm s}}
&=\left.\frac{\partial P^{\rm S}_{{\rm gg},\ell}(k)}{\partial \sigma_8}\right|_{\Omega_{\rm m}}
\left.\frac{\partial \sigma_8}{\partial \Omega_{\rm m}}\right|_{A_{\rm s}}
+\left.\frac{\partial P^{\rm S}_{{\rm gg},\ell}(k)}{\partial \Omega_{\rm m}}\right|_{\sigma_{8}},\nonumber\\
\left.\frac{\partial P^{\rm S}_{{\rm gg},\ell}(k)}{\partial A_{\rm s}}\right|_{\Omega_{\rm m}}
&=\left.\frac{\partial P^{\rm S}_{{\rm gg},\ell}(k)}{\partial \sigma_{8}}\right|_{\Omega_{\rm m}}
\left.\frac{\partial \sigma_8}{\partial A_{\rm s}}\right|_{\Omega_{\rm m}}.
\end{align}
We use the mock catalog of galaxies that are constructed from $N$-body simulations with varying either of $\Omega_{\rm m}$ or
$A_{\rm s}$ (see Table~\ref{tab:simu}), and then evaluate the numerical derivative,
$\left.\partial P^{\rm S}_{{\rm gg},\ell}(k)/\partial A_{\rm s}\right|_{\Omega_{\rm m}}$ or
$\left.\partial P^{\rm S}_{{\rm gg},\ell}(k)/\partial \Omega_{\rm m}\right|_{A_{\rm s}}$, using Eq.~(\ref{eq:dPdp_alpha}), where we fix the halo-galaxy connection parameters to their fiducial values.
Then we use the above equation to compute
$\left.\partial P^{\rm S}_{{\rm gg},\ell}/\partial \sigma_8\right|_{\Omega_{\rm m}}$ and
$\left.\partial P^{\rm S}_{{\rm gg},\ell}/\partial \Omega_{\rm m}\right|_{\sigma_{8}}$, where we use {\tt CAMB} to evaluate $\left.\partial \sigma_8/\partial \Omega_{\rm m}\right|_{A_{\rm s}}$ and $\left.\partial \sigma_8/\partial A_{\rm s}\right|_{\Omega_{\rm m}}$ around the fiducial {\it Planck} cosmology.

\subsection{Mock catalogs of SDSS LOWZ and CMASS galaxies}

\begin{table}
\begin{center}
\begin{tabular}{c|ccc|c} \hline \hline
	\multirow{2}{*}{Parameter} & \multicolumn{3}{|c|}{Fiducial value} &  \multirow{2}{*}{$\sigma_{\rm prior}$} \\
	& LOWZ & CMASS1 & CMASS2 & \\ \hline \hline
	$D_{\rm A}(z)\,$[\hiMpc] & $564.6$ & $861.5$ & $971.5$ & - \\
	$H(z)\,[h\,{\rm km} ~{\rm s}^{-1}~{\rm Mpc}^{-1}]$ & $114.1$ & $131.0$ & $142.1$ &  - \\ \hline
	$\log M_{\rm min}$ & $13.62$   & $13.94$   & $14.19$ &  $2.0$ \\
	$\sigma_{\log{M}}$ & $0.6915$ & $0.8860$ & $0.7919$ &  $1.0$ \\
	$\log M_1$         & $14.42$   & $14.46$   & $14.85$  &  $2.0$  \\
	$\log M_{\rm sat}$ & $13.33$   & $13.72$   & $13.01$  & $2.0$ \\
	$\alpha_{\rm sat}$           & $0.9168$  & $1.192$   & $0.9826$ &  $2.0$ \\ \hline
	$c_{\rm conc}$ & $1.0$ & $1.0$ & $1.0$ &  $2.0$ \\
	$c_{\rm vel}$ & $1.0$ & $1.0$ & $1.0$ &  $2.0$ \\
	$p_{\rm off}$ & $0.30$ & $0.30$ & $0.30$ & $1.0$ \\
	${\cal R}_{\rm off}$ & $2.0$ & $2.0$ & $2.0$ &  $2.0$ \\ \hline
	$P_{\rm SN}$ & $0$ & $0$ & $0$ & $0.1/\bar{n}_{\rm g}$ \\ \hline \hline
\end{tabular}
\caption{The fiducial values of model parameters in our Fisher matrix analysis for a hypothetical galaxy survey that resembles
the LOWZ galaxy sample and two subsamples of CMASS galaxies at lower and higher redshift sides in the SDSS-BOSS surveys (see Table~\ref{tab:surveyparam}).
Here we consider 12 parameters for each galaxy sample: the angular and radial distances, $D_{\rm A}(z_n)$ and $H(z_n)$, 9 parameters to model the halo-galaxy connection, and the residual shot noise parameter.
In addition we include the cosmological parameters $\Om$ and $\sigma_8$ for the background cosmological model.
Using these fiducial values and the slightly-varied value for each parameter (with other parameters being kept fixed), we generate the mock catalogs of these galaxies from the halo catalogs in $N$-body simulations for the fiducial {\it Planck} cosmology, make a hypothetical measurement of the redshift-space power spectrum from each mock and then compute the Fisher matrix elements from the measured power spectra.
The column denoted as ``$\sigma_{\rm prior}$'' is the value for a Gaussian prior of each parameter. For a prior of the residual shot noise,
we employ $10\%$ of the shot noise for the fiducial mock of each sample.
}
\label{tab:HOD_parameters}
\end{center}
\end{table}

\begin{table}
\begin{center}
\begin{tabular}{l|cccc} \hline \hline
Sample & Redshift range & $V_{\rm S}~[h^{-1}{\rm Gpc}]^3$ &  $\bar{n}_{\rm g}~[h\,{\rm Mpc}^{-1}]^3$ & linear bias $b_{\rm g}$ \\ \hline
LOWZ   & $[0.15,0.30]~(0.251)$ & $1.98$ & $2.173 \times 10^{-4}$ & $1.78$ \\
CMASS1 & $[0.47,0.55]~(0.484)$ & $2.26$ & $9.251 \times 10^{-5}$ & $2.12$ \\
CMASS2 & $[0.55,0.70]~(0.617)$ & $3.8$ & $5.336 \times 10^{-5}$   & $2.28$ \\ \hline\hline
\end{tabular}
\caption{
The parameters of galaxy samples that resemble the LOWZ, CMASS1 and CMASS2 galaxies in the BOSS survey.
Each sample is defined to be a volume-limited sample which is defined to be luminosity-limited, rather than flux-limited, for a given redshift range (see text for details).
The mean number density of galaxies ($\bar{n}_{\rm g}$) is computed from the mock catalogs using the fiducial HOD parameters for the fiducial {\it Planck} model (Tables~\ref{tab:simu} and \ref{tab:HOD_parameters}).
The linear bias ($b_{\rm g}$) is determined from comparison of the galaxy power spectrum with the matter power spectrum in the same mock realization at low-$k$ bins.
}
\label{tab:surveyparam}
\end{center}
\end{table}

We consider models of halo-galaxy connection that mimic spectroscopic galaxies in the SDSS survey \cite{2013AJ....145...10D}.
We consider the ``LOWZ'' galaxies in the redshift range $z=[0.15,0.30]$ and two subsamples of ``CMASS'' galaxies that are divided into two redshift bins of $z=[0.47,0.55]$ and $[0.55,0.70]$, respectively.
Hereafter we call the two CMASS samples in the lower and higher redshift bins, CMASS1 and CMASS2, respectively. We employ the HOD and other parameters that roughly resemble the SDSS galaxies for the {\it Planck} cosmology, as given in Table~\ref{tab:HOD_parameters}.
To generate mock catalogs of these galaxies we use the halo catalogs produced from $N$-body simulation outputs at redshifts $z=0.251, 0.484$ and $0.617$, which are close to the median redshifts of the LOWZ, CMASS1, and CMASS2, respectively (see Table~\ref{tab:surveyparam}).
We ignore the redshift evolution of the galaxy clustering within each redshift bin.
The following is the details of procedures to populate SDSS-like galaxies into halos identified in each $N$-body simulation realization:
\begin{itemize}
\item[(i)] {\it Central galaxies} -- We employ the central HOD, $\avrg{N_{\rm c}}\!(M)$,
with the form given by Eq.~\ref{eq:Nc}.
We select halos to populate central galaxies assuming 
a Bernoulli distribution with mean $\avrg{N_{\rm c}}\!(M)$.
We assume that each central galaxy has the same peculiar velocity as the bulk
velocity of its host halo measured as the center-of-mass velocity of particles in the core region.
\item[(ii)] {\it ``Off-centering'' effect of central galaxies} -- As our default model we include a population of off-centered ``central'' galaxies that could occur, e.g. as a result of mergers as discussed in \citet{masaki13}.
We select off-centered galaxies from all the central galaxies selected in the procedure (i), following the prescription described in \S \ref{subsec:halomodel}: each central galaxy to be off-centered is selected with a probability $p_{\rm off}$, and its displacement from its host halo center is drawn from the isotropic $3$D Gaussian distribution given by Eq.~\ref{eq:poff_real}.
The scale radius $r_{\rm s}$ of each halo is computed from the fitting formula of halo mass-concentration relation, $c(M,z)$, in \citet{2015ApJ...799..108D}, where we assume an NFW profile for the host halo mass $M$ and redshift $z$ for the {\it Planck} cosmology.
Once an off-centered radius $r_{\rm off}$ is given, we assign a ``sloshing'' velocity of the off-centered galaxy around the halo center, assuming it follows a Gaussian distribution with the typical velocity dispersion given by Eq.~\ref{eq:virial}.
The total velocity of each ``off-centered'' central galaxy is given as the sum of the assigned internal velocity and the bulk velocity of its host halo.
\item[(iii)] {\it Satellite galaxies} -- To populate satellite galaxies into halos we employ the satellite HOD with the form given by Eq.~\ref{eq:Ns}.
As described in \S \ref{subsec:halomodel} we assume that satellite galaxies reside in halos which already host a ``central'' galaxy inside, including the off-centered galaxy in the procedure~(ii).
For each of such host halos with mass $M$, we determine the number of satellite galaxies, randomly drawn from the Poisson distribution with mean $\lambda_{\rm s}$ defined in Eq.~\ref{eq:Ns}.
For each satellite galaxy we determine its position within $r_{200}$ of the halo, by randomly drawing from the NFW profile.
As we do for the off-centered galaxy in the procedure~(ii), we assign an internal velocity to each satellite galaxy using the interior mass $M(<r)$ at the satellite position.
Likewise the total velocity is given as a sum of the internal motion and the host halo's bulk velocity.
\end{itemize}
These are the procedures to populate central and satellite galaxies in the halos found in $N$-body simulation realizations including the RSD effect.
These are an empirical approach, and the properties of true galaxies would be more complicated because of the dependence on complicated physics inherent in galaxy formation/evolution processes as well as on non-trivial selection effects, e.g. color and magnitude cuts.
Hence these mock galaxy catalogs should be considered as a working example for the following discussion, and we will discuss the impact of other effects that we ignore on our results, e.g. the assembly bias separately.

\begin{figure}[h]
\centering
\includegraphics[width=0.99\textwidth]{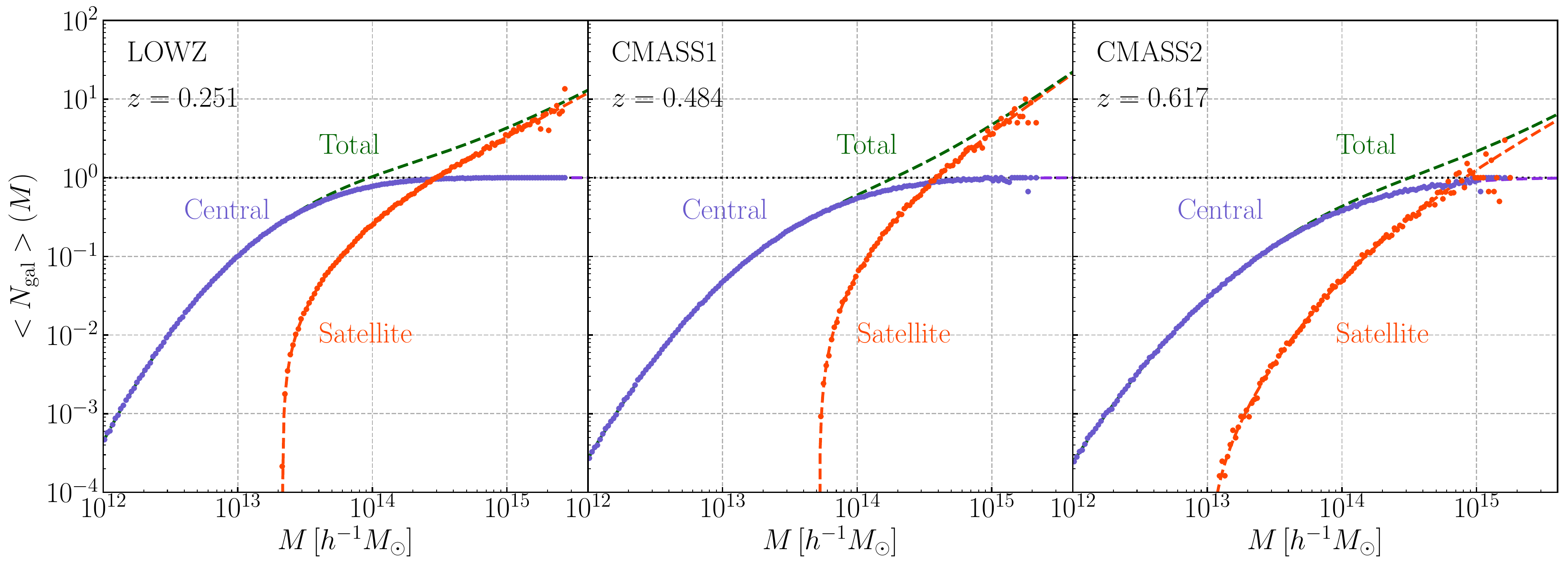}
\caption{
The halo occupation distribution (HOD), measured from one realization of the galaxy mock catalogs of LOWZ, CMASS1 and CMASS2 for the fiducial {\it Planck} cosmology (Table~\ref{tab:HOD_parameters}).
Here we show the central (blue) and satellite (orange) HODs.
Note that we use the halos with mass $M_{200}\ge 10^{12}\,h^{-1}M_\odot$ as host halos of the mock galaxies in the simulation.
For comparison the dashed curves show the input mean HODs in Table~\ref{tab:HOD_parameters}.
}
\label{fig:hod_offcenter}
\end{figure}

Fig.~\ref{fig:hod_offcenter} shows the HODs for the LOWZ, CMASS1, and CMASS2-like galaxies that we employ when building the mock catalogs in this paper.
The three dashed curves in each panel show the central and satellite HODs and the total HOD, respectively.
The points show the respective HODs that are measured from the mock catalogs for the fiducial {\it Planck} cosmology. The SDSS galaxies are  passively-evolving early-type galaxies that are selected based on color and magnitude cuts \cite{eisenstein01,2013AJ....145...10D}.
These galaxies typically reside in halos with masses $\sim 10^{13}M_\odot$, while cluster-scale halos with $\simgt 10^{14}M_\odot$ host these galaxies as satellite galaxies.
The average number density $\bar{n}_{\rm g}=2.173\times10^{-4},9.251\times10^{-5}$ or $5.336\times10^{-5}\,[h {\rm Mpc}^{-1}]^3$ for the LOWZ, CMASS1, or CMASS2 for the {\it Planck} cosmology, respectively (see Table~\ref{tab:surveyparam}).
These number densities are smaller than that of the full LOWZ and CMASS galaxies, because we mimic the luminosity-limited subsample, not the flux-limited all galaxies \citep[see][for a similar discussion]{2015ApJ...806....1M}.
The detailed shape of HOD is not essential for our purpose.
We employ these HODs as a working example.

\subsection{Measurements of redshift-space power spectra from mock galaxy catalogs}
\label{sec:measure}

\begin{figure}[h]
\centering
\includegraphics[width=0.85\textwidth]{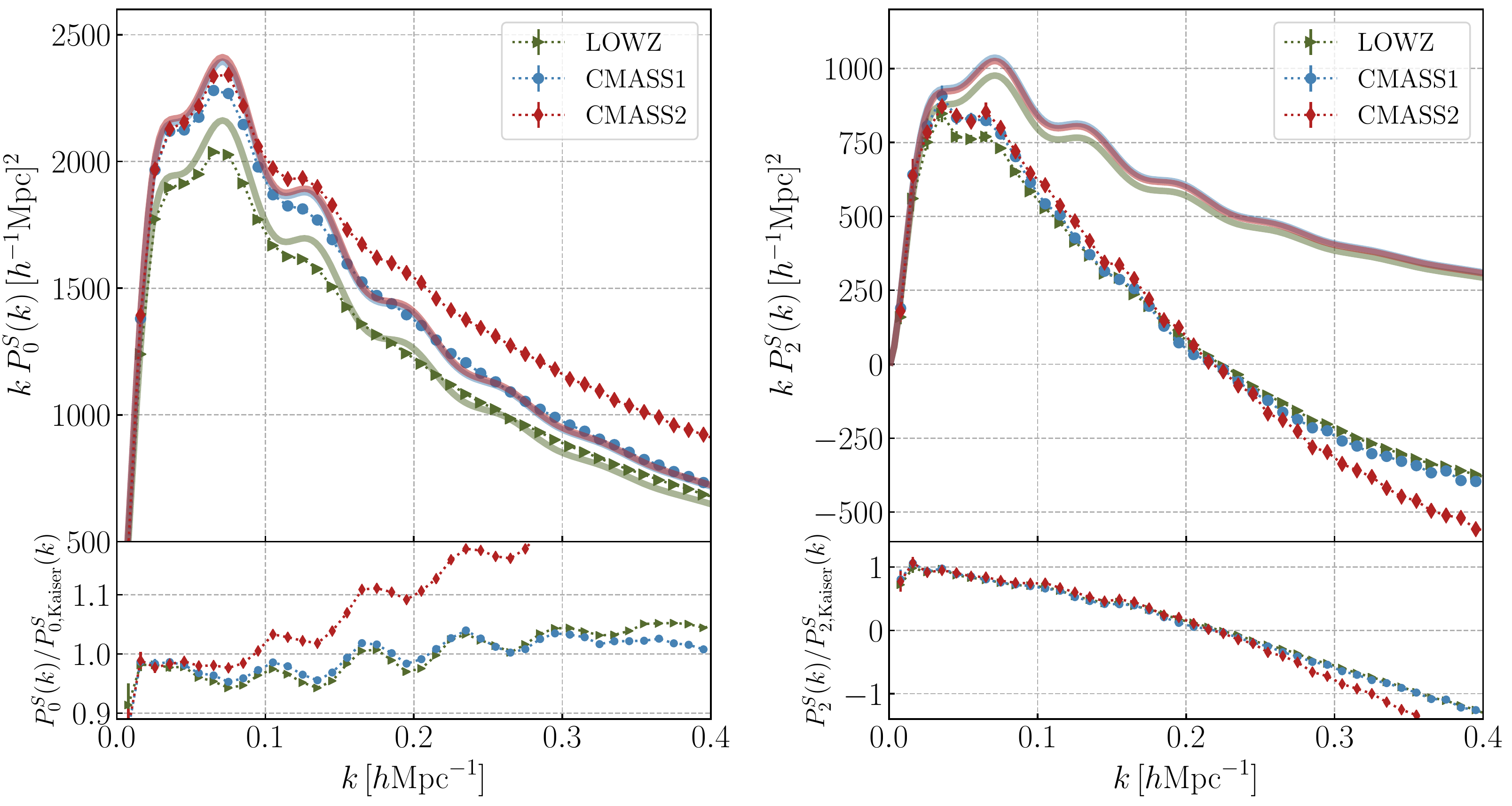}
\caption{
The symbols show the monopole (left panel) and quadrupole (right) power spectra measured from the galaxy mocks that resemble LOWZ, CMASS1 and CMASS2 galaxies, respectively.
For illustrative purpose we multiply the power spectrum by $k$ so that the power spectrum shown is in a narrow range of this amplitude (therefore $y$ axis is in a linear, rather than logarithmic, scale).
Since the power spectra are estimated from the 16 realizations of 2~$h^{-1}{\rm Gpc}$ cubic box corresponding to volume of 128~$(h^{-1}{\rm Gpc})^3$ in total, the error bar of the mean of band power measurements at each $k$ bin, although shown, is not visible.
For comparison, the solid curves show the predictions of Kaiser formula (Eq.~\ref{eq:kaiser}), where we used the linear matter power spectrum, the linear bias parameter ($b_{\rm g}$ in Table~\ref{tab:simu}), and the redshift-space distortion parameter $\beta=(1/b_{\rm g})\mathrm{d}\ln D_+/\mathrm{d}\ln a$ for the {\it Planck} model ($D_+$ is the linear growth rate).
Note that the plotting ranges of $y$-axis in the monopole and quadrupole spectra are different from each other because of their different amplitudes (see text for details).
The lower panel shows the ratio.
The Kaiser formula ceases to be accurate at $k\gtrsim 0.05~$\hMpci, especially for the quadrupole spectrum.
}
\label{fig:p0p2_fiducial}
\end{figure}
\begin{figure}[h]
\centering
\includegraphics[width=0.85\textwidth]{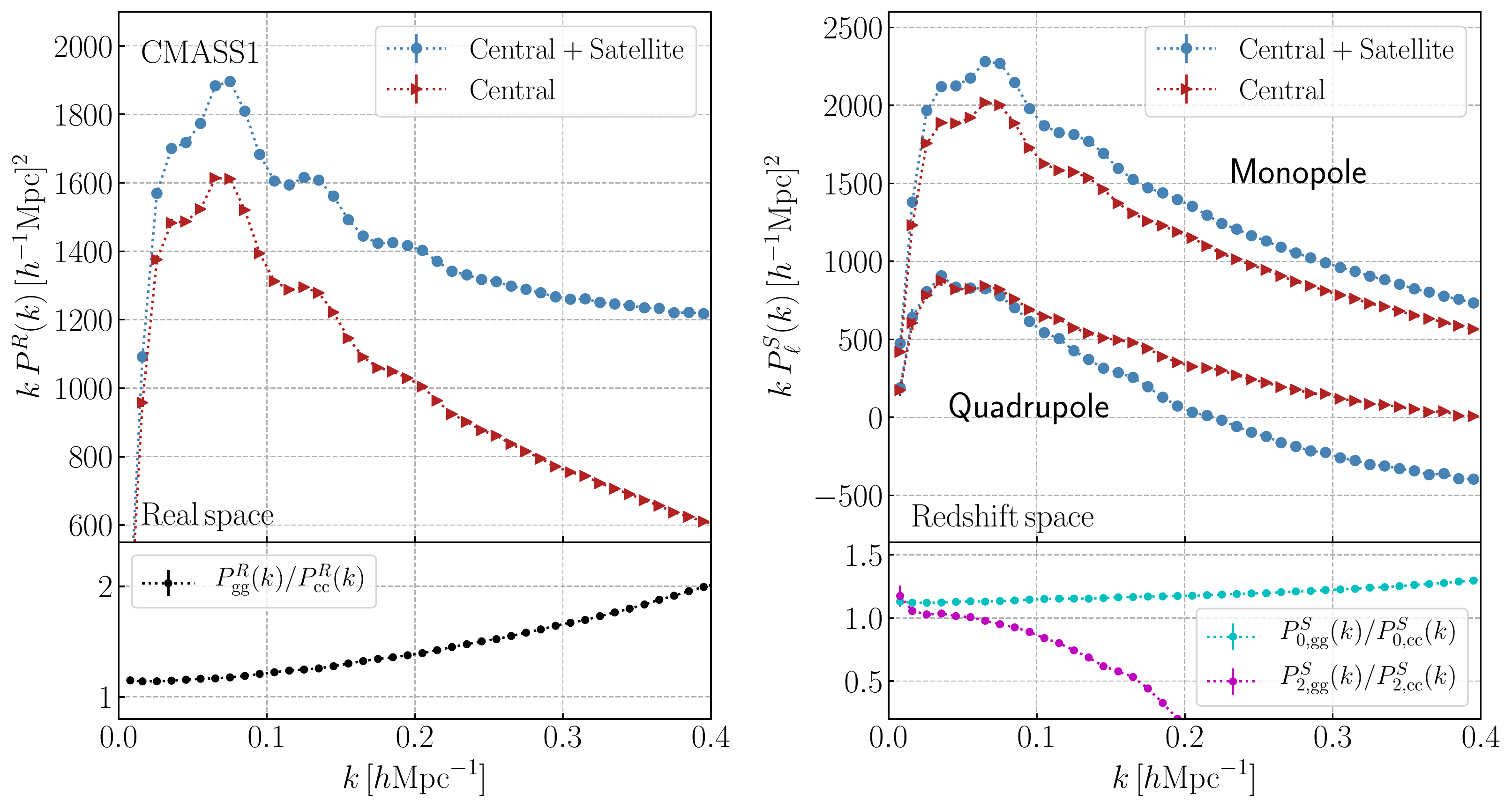}
\caption{
The impact of satellite galaxies on the real-space (left panel) and redshift-space (right) power spectra. To study this,
the triangle symbols with error bars show the power spectra when using central galaxies alone in each galaxy mock (i.e. removing satellite galaxies from the mock), while the circle symbols are the same as in Fig.~\ref{fig:p0p2_fiducial}. The lower panels show the ratio.
}
\label{fig:power_full_and_cc}
\end{figure}

We measure the power spectrum from each realization of the galaxy mocks constructed according to the method in the preceding subsection, using Fast Fourier Transform (FFT)-based method.
First we map the real-space positions of mock galaxies and their velocities to the redshift-space positions using Eq.~\ref{eq:rsd_def}, under the plane-parallel approximation where the line-of-sight is taken to be along the $z$-axis direction.
The number density field of mock galaxies is defined on $1024^3$ $3$D mesh grids using the Cloud-in-Cell (CIC) \cite{hockney81} interpolation.
Since our simulations' box size is $2000$ \hiMpc, the Nyquist wavenumber is $k_{\rm Ny} = 1.608$ \hMpci, which is sufficiently smaller than the scales we are interested in.
We then implement the FFT method to the number density field of galaxies to obtain the Fourier-transformed field, where we reduce the aliasing contamination arising from the grid interpolation, using the interlacing scheme described in \cite{Sefusatti:2015aex}.
We split each Fourier mode into linearly-spaced $k$-bins ranging $k = [0,2]$ \hMpci, and obtain the power spectrum signals by averaging the modes that enter into each bin of $k$.
We adopt the bin width $\Delta k = 0.01$ \hMpci that is sufficiently narrow compared to the BAO features in the power spectrum.
We assume the Poisson shot noise $P_{\rm Poisson} = 1/\bar{n}_{\rm g}$ and subtract it from the measured power spectrum.

Fig.~\ref{fig:p0p2_fiducial} shows the monopole and quadrupole power spectra measured from the 16 realizations of mock catalogs of LOWZ, CMASS1 and CMASS2-like galaxies for the {\it Planck} cosmology.
The error bar at each $k$ bin denotes an estimate of the error on the mean of power spectrum amplitude, estimated as $\sigma/\sqrt{16}$, where $\sigma$ is the standard deviation of band power at each $k$ bin.
Hence the error represents the statistical error when the band power at each $k$ bin is measured from a volume of $128~(h^{-1}{\rm Gpc})^3$.
For comparison we also show the linear theory predictions for each galaxy sample, which are computed using the Kaiser formula \cite{kaiser84}:
\begin{align}
\label{eq:kaiser}
  \left(
    \begin{array}{c}
      P^{\rm S}_{0,\rm Kaiser}(k) \\
      P^{\rm S}_{2,\rm Kaiser}(k)
    \end{array}
  \right) = \left(
  	\begin{array}{c}
      1+\frac{2}{3}\beta + \frac{1}{5} \beta^2 \\
      \frac{4}{3} \beta + \frac{4}{7} \beta^2
    \end{array}
  \right)
  b_{\rm g}^2 P^{\rm lin}_{\rm m}(k)
\end{align}
where $\beta = f(z) / b_{\rm g}$, $f(z) \equiv {\rm d} \ln D_+(z) / {\rm d} \ln a$ is the linear growth rate at the redshift $z$, $b_{\rm g}$ is the linear galaxy bias and $P^{\rm lin}_{\rm m}(k)$ is the linear matter power spectrum.
We determine the linear bias $b_{\rm g}$ from a comparison of the real-space power spectra of galaxies and matter measured from the galaxy mocks and $N$-body simulations, respectively,
at the lowest 5 $k$-bins (corresponding to $k\lesssim 0.05~$\hMpci) that are considered to be well in the linear regime.
The linear bias value estimated based on this method is given in Table~\ref{tab:surveyparam}.
The Kaiser formula predicts that the ratio of the quadrupole power spectrum to the monopole spectrum is given as
$P^{\rm S}_{2,{\rm Kaiser}}/P^{\rm S}_{0,{\rm Kaiser}}=(4\beta/3+4\beta^2/7)/(1+2\beta/3+\beta^2/5)$ independently of the linear matter power spectrum and the wavenumber $k$.
Galaxies with greater bias parameter has a smaller ratio of $P^{\rm S}_{2}/P^{\rm S}_{0}$. The redshift and linear bias parameter
in Table~\ref{tab:surveyparam} give about $0.43$--$0.45$ for the ratio for the LOWZ, CMASS1 and CMASS2 samples.

Fig.~\ref{fig:p0p2_fiducial} shows that the linear-theory prediction moderately matches the simulation result for the monopole power spectrum up to $k\simeq 0.1~$\hMpci.
However, the Kaiser formula ceases to be accurate very quickly at $k\gtrsim 0.05$~\hMpci for the quadrupole power spectrum, indicating that nonlinearities in the RSD effect become significant from the relatively small $k$ compared to the nonlinear effect on the real-space or monopole power spectrum.
In particular the nonlinear quadrupole power spectra have smaller amplitudes than the linear theory predicts at $k\gtrsim 0.05~$\hMpci because of the smearing effect due to streaming motions of halos or virial motions of galaxies \citep{scoccimarro04,taruya09,taruya10}.
Hence it is of critical importance to properly take into account the nonlinear effects in the galaxy power spectra in order to accurately estimate the power of redshift-space galaxy power spectra for constraining cosmological parameters.

In the halo model picture, the underlying halo power spectrum contains the cosmological information, while the effects arising from the halo-galaxy connection are considered as a source of systematic effects.
To study this, in Fig.~\ref{fig:power_full_and_cc} we plot the power spectra of the central galaxies alone; 
i.e., we repeat the measurement after we remove the satellite galaxies 
from each mock realization.
The galaxy power spectrum including the satellite galaxies is quite different from the power spectrum of central galaxies.
The satellite galaxies alter the galaxy clustering in several ways \citep[also see][for a similar discussion]{Okumura:2016mrt}.
First, the inclusion of the satellite galaxies leads to a boost in the large-scale (small-$k$) amplitude of galaxy power spectrum, because the satellite galaxies tend to reside in massive halos, which have a greater linear bias, and including the satellite galaxies upweights the contribution of massive halos leading to the greater band power at small $k$ bins in the 2-halo term regime.
Secondly, as is clear from the real-space power spectrum, the satellite galaxies boost the power spectrum amplitude at small scales due to the 1-halo term contribution, which arises from clustering of (central-satellite or satellite-satellite) galaxies in the same host halo.
Thirdly, the satellite galaxies cause a suppression in the power spectrum amplitude at small scales due to the FoG effect, as can be found from the redshift-space power spectra in the right panel.
Hence the 1-halo term contribution and the FoG effect are somewhat compensated.
Thus, the satellite galaxies, more generally the halo-galaxy connection parameters, cause complex, scale-dependent changes in the galaxy power spectrum at small scales compared to the power spectrum of the host halos or the central galaxies.

\subsection{Covariance matrix of redshift-space power spectra}
\label{sec:covariance}

\begin{figure}[h]
\centering
\includegraphics[width=0.6\textwidth]{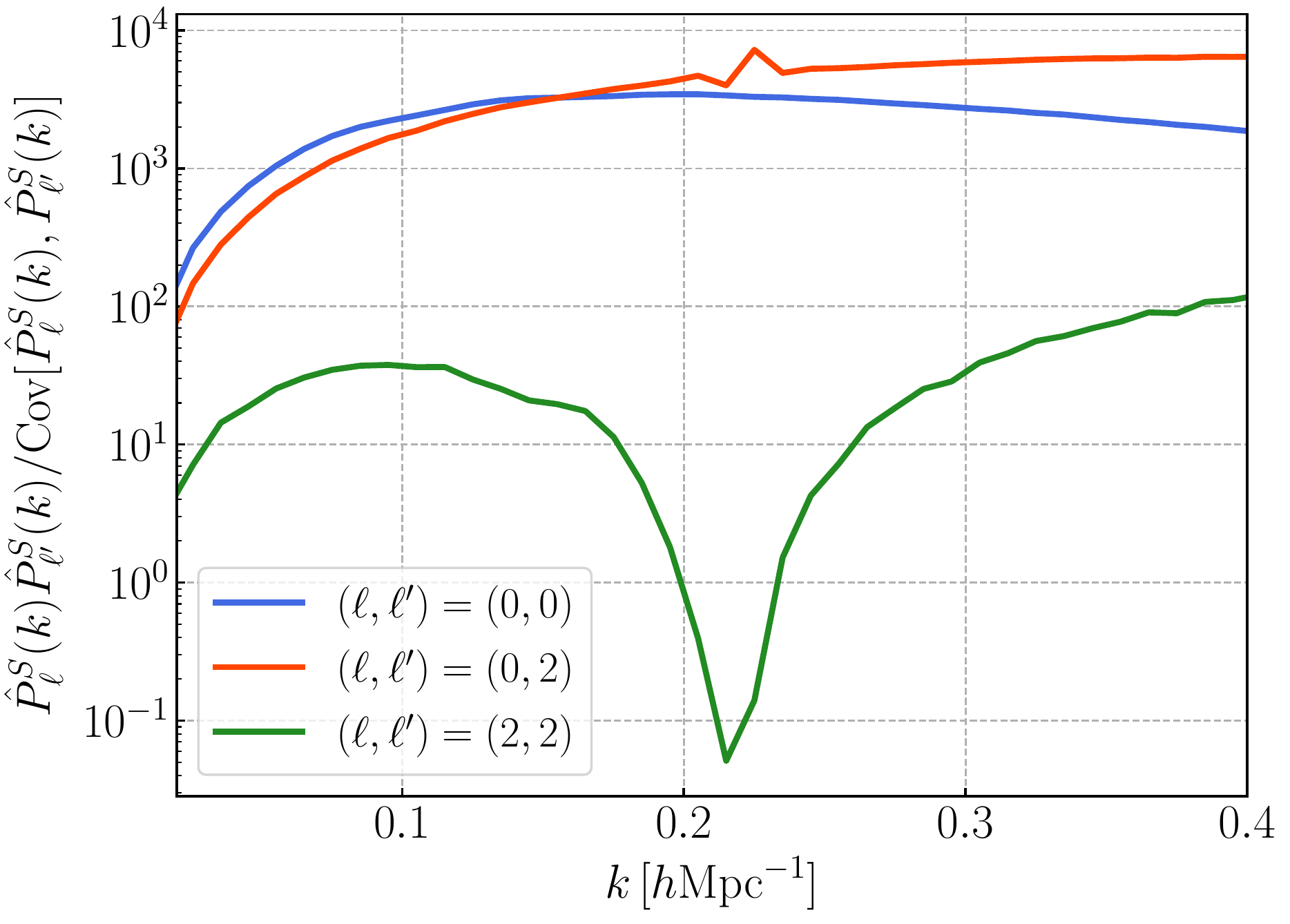}
\caption{
The Gaussian covariance of the monopole and quadrupole galaxy power spectra.
Note that, to estimate the covariance matrices, we used the redshift-space galaxy power spectrum,
$P^{\rm S}_{\rm gg}(k,\mu)$, measured from the 16 realizations of CMASS1 mocks for the {\it Planck} model (see Eq.~\ref{eq:covar}).
For illustrative purpose, we show the ratio
$P^{\rm S}_{{\rm gg},\ell}(k)P^{\rm S}_{{\rm gg},\ell'}(k)/{\rm Cov}[P^{\rm S}_{{\rm gg},\ell}(k),P^{\rm S}_{{\rm gg},\ell'}(k)]$
as a function of $k$.
Since the quadrupole power spectrum amplitude has a zero crossing at $k\simeq 0.2~$\hMpci, the curves involving the quadrupole power spectrum displays a discontinuity around the wavenumber.
}
\label{fig:cov}
\end{figure}

To compute the Fisher matrix, we need to estimate the covariance matrix for the multipole power spectra of galaxies, which is defined as
\begin{align}
\label{eq:covar}
{\rm Cov}\left[\hat{P}^{\rm S}_{{\rm gg},\ell}(k_i),\hat{P}^{\rm S}_{{\rm gg},\ell'}(k_j)\right]
\equiv \left \langle \hat{P}^{\rm S}_{{\rm gg},\ell}(k_i) \hat{P}^{\rm S}_{{\rm gg},\ell'}(k_j) \right \rangle -
\left \langle \hat{P}^{\rm S}_{{\rm gg},\ell}(k_i) \right \rangle \left \langle \hat{P}^{\rm S}_{{\rm gg},\ell'}(k_j) \right \rangle .
\end{align}
In this paper we employ the Gaussian covariance matrix as given in Appendix~A in Ref.~\cite{Guzik:2009cm} \citep[also see][]{2003ApJ...598..720S,2017PhRvD..95h3522A}, which is given by
\begin{align}
\label{eq:covar_gaussian}
{\rm Cov}\left[\hat{P}^{\rm S}_{{\rm gg},\ell}(k_i),\hat{P}^{\rm S}_{{\rm gg},\ell'}(k_j)\right]
&=\delta^K_{ij}\frac{1}{N_{\rm mode}(k_i)}
\int_{-1}^1 {\rm d}\mu \,(2\ell+1) \L_\ell(\mu) (2\ell'+1) \L_{\ell'}(\mu)
\left[P^{\rm S}_{\rm gg}(k_i,\mu) + \frac{1}{\bar{n}_{\rm g}} \right]^2,
\end{align}
where $\delta_{ij}^K$ is the Kronecker delta function, defined so as to satisfy $\delta^{K}_{ij}=1$ if $i=j$, otherwise $\delta^K_{ij}=0$;
$N_{\rm mode}(k_i)$ is the number of independent Fourier modes, 
determined by the fundamental Fourier mode for a given survey, in the spherical
shell of radius $k_i$ with width $\Delta k$; $N_{\rm mode}(k_i)\simeq 4\pi k_i^2 \Delta k/[(2\pi)^3/V_{\rm S}]$ for $k_i\gg 2\pi/(V_{\rm S})^{1/3}$, and $V_{\rm S}$ is the survey volume; $[P_{\rm gg}^{\rm S}(k_i,\mu)+1/\bar{n}_{\rm g}]$ is the ``observed'' redshift-space power spectrum given as a function of $(k_i,\mu)$, including the shot noise contamination.
For the survey volume, we adopt $V_{\rm S} = 1.98, 2.26$ or $3.80\,[h^{-1}{\rm Gpc}]^3$ for the redshift slice of LOWZ, CMASS1 or CMASS2-like galaxies, respectively (see Table~\ref{tab:surveyparam}).
For the redshift-space power spectrum $P^{\rm S}(k,\mu)$ (including the shot noise term) in the above equation, we use the power spectrum directly measured from the 16 realizations of mock catalogs for the fiducial {\it Planck} cosmology.
Note that, for the real-space power spectrum $P_{\rm gg}(k)$ (i.e. no $\mu$-dependence), the above formula reproduces the standard covariance
formula, ${\rm Cov}[P_{\rm gg}(k_i),P_{\rm gg}(k_j)]=[2\delta^K_{ij}/N_{\rm mode}(k_i)]\times [P_{\rm gg}(k_i)+1/\bar{n}_{\rm g}]^2$.

In Fig.~\ref{fig:cov} we show the Gaussian covariance matrices for the monopole and quadrupole power spectra for the CMASS1-like galaxies. For illustrative purpose, we plot
$P^{\rm S}_{{\rm gg},\ell}(k_i)P^{\rm S}_{{\rm gg},\ell'}(k_i)/{\rm Cov}[P^{\rm S}_{{\rm gg},\ell}(k_i),P^{\rm S}_{{\rm gg},\ell'}(k_i)]$, which roughly gives a signal-to-noise ratio for the band power measurement at each $k_i$ bin, $(S/N)^2|_{k_i}$. The figure clearly shows that the signal-to-noise ratio for the quadrupole spectrum is much smaller than that of the monopole spectrum, by up to a factor of 100. This large-factor reduction is due to the two facts. First, the quadrupole power spectrum has a smaller band power than the monopole spectrum by a factor of 4 (see Fig.~\ref{fig:p0p2_fiducial}), which explains a factor of 16 in $(S/N)^2|_{k_i}$ that 
scales as the square of band power. Second, the covariance of the quadrupole power spectrum also arises from the monopole. This explains the remaining factor in the factor of 100. Nevertheless, as we will show below, the quadrupole spectrum is quite powerful to constrain cosmological parameters, as it purely arises from the RSD effect and cosmological AP distortion both of which are cosmological effects. The figure also shows a significant cross-covariance between the monopole and quadrupole spectra.
Note that, in this study we consider only the diagonal, Gaussian contribution, but in reality there should be the off-diagonal covariance.
As has been recently shown in \cite{Wadekar:2019rdu}, such non-Gaussian contribution is relatively small for SDSS galaxies 
because of the relatively large shot noise contribution to the total covariance matrix.
We postpone the evaluation of this impact on the cosmological parameter inference in realistic galaxy surveys and we will further investigate this point using the SDSS galaxy clustering data in the future work.

\section{Results}
\label{sec:results}

In this section we show the main results of this paper; we show parameter forecasts that are obtained from measurements of the redshift-space power spectra of SDSS-like galaxies.

\subsection{Fisher information matrix}

To perform the parameter forecast, we employ the Fisher information formalism \cite{Tegmark:1996bz,Tegmark:1997rp}.
We include, as observables, the monopole and quadrupole power spectra of LOWZ, CMASS1 and CMASS2-like galaxies.
The Fisher matrix for the $\ell$- and $\ell'$-th multipole power spectra is defined as
\begin{align}
\label{eq:fisher}
F^{\ell \ell'}_{\alpha \beta} &= \sum_{z_{n}}
\sum_{k_i,k_j} \frac{\partial P^{\rm S}_{{\rm gg},\ell}(k_i; z_n)}{\partial p_\alpha}
{\rm Cov}^{-1}\!\left[\hat{P}^{\rm S}_{{\rm gg},\ell}(k_i; z_n),\hat{P}^{\rm S}_{{\rm gg},\ell'}(k_j; z_n)\right]
\frac{\partial P^{\rm S}_{{\rm gg},\ell'}(k_j; z_n)}{\partial p_\beta}
\end{align}
where the ${\rm Cov}^{-1}\!\left[\hat{P}^{\rm S}_{{\rm gg},\ell}(k_i; z_n),\hat{P}^{\rm S}_{{\rm gg},\ell'} (k_j; z_n)\right]$ is the inverse of the covariance matrix (Eq.~\ref{eq:covar_gaussian}), and $p_\alpha$ is the $\alpha$-th parameter.
Here we use the Gaussian covariance assumption, and hence the covariance matrix is diagonal in wavenumber bins; the summation over $k_i$ and $k_j$ reduces to a single summation $\sum_{k_i}$.
To combine the Fisher information for different samples of the LOWZ, CMASS1 and CMASS2-like galaxies, we simply sum the Fisher matrices for the three galaxy samples.
In other words, we ignore a possible cross-covariance between the different galaxy samples at different redshift slices.
When computing the Fisher information for a combined measurement of the monopole and quadrupole power spectra, we properly take into account the cross-covariance.
When computing the Fisher matrix, we set $k_{\rm min}=0.02~$\hMpci for the minimum wavenumber in the $k$-bin summation $\sum_{k_i}$, and then study how the Fisher information content varies with a maximum wavenumber, $k_{\rm max}$, for which we mainly consider $k_{\rm max}=0.1$, 0.2 or 0.3~\hMpci, respectively.
We consider the following set of model parameters:
\begin{align}
{\bf p}\equiv \left\{
\Omega_{\rm m}, \sigma_8, D_{\rm A}(z_n), H(z_n), \log M_{\rm min}(z_n), \sigma_{\log M}(z_n), \log M_1(z_n), \log M_{\rm sat}(z_n), \alpha_{\rm sat}(z_n), c_{\rm conc}(z_n), c_{\rm vel}(z_n), p_{\rm off}(z_n), {\cal R}_{\rm off}(z_n), P_{\rm SN}(z_n)
\right\},
\end{align}
where quantities including $z_n$ in the argument indicate that 
they can take different values for the three galaxy samples at different redshifts 
(either of LOWZ, CMASS1 or CMASS2).
When we include the information from the galaxy samples of three redshift bins, we consider
38 model parameters in total. Among those, 8 cosmological parameters; $(\Omega_{\rm m}, \sigma_8)$ and
$\{D_{\rm A}(z_n),H(z_n)\}$ for each of the three redshifts, where $D_{\rm A}(z_i)$ are $H(z_i)$ are the parameters to model the AP effect (see Section~\ref{sec:AP}).
On the other hand, we include 10 nuisance parameters that model the halo-galaxy connection and the residual shot noise for each of the three redshifts.
Table~\ref{tab:HOD_parameters} gives the fiducial values as well as the prior for the nuisance parameters. Note that the prior we employ is a very weak one, and we use it only to have a stable calculation of the Fisher matrix (e.g. its inversion).

\begin{figure}[h]
\centering
\includegraphics[width=0.9\textwidth]{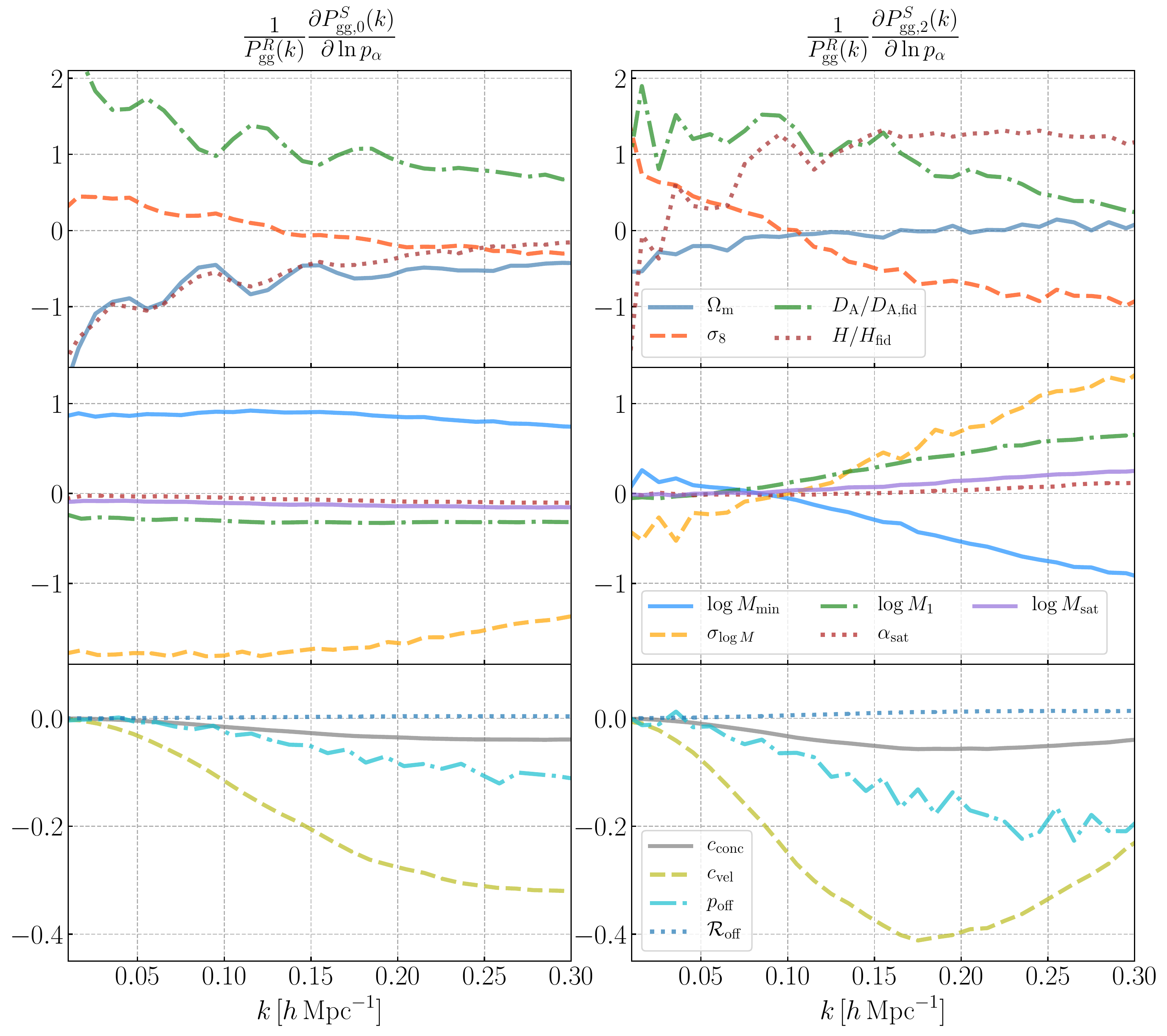}
\caption{
The partial derivative of the monopole (left panels) or quadrupole (right) power spectrum with respect to each model parameter ($p_\alpha$), which we refer to as the ``response'' function, for the CMASS1 sample.
To compute the response function, we first generate the mock catalogs where we slightly varied a parameter but fixed other parameters to their fiducial values, and then numerically compute the partial derivative.
Note that we compute the response functions for $D_{\rm A}$ and $H$ from the AP distortion effect in a hypothetical clustering analysis; we assume a slightly shifted value in $D_{\rm A}$ or $H$ from the fiducial value when mapping the real-space positions of galaxies to the comoving coordinates in each mock for the fiducial {\it Planck} model, and then measure the monopole and quadrupole spectra from the mock.
For illustrative purpose, we show the fractional response function relative to the real-space power spectrum for the fiducial mocks, $[1/P^{\rm R}_{{\rm gg}}(k)]\partial P^{\rm S}_{{\rm gg},\ell}(k)/\partial \ln p_\alpha$.
Note that we employ the parameters $\log M_{\rm min}$, $\log M_1$ and $\log M_{\rm sat}$ rather than the physical quantities $M_{\rm min}$, $M_1$ and $M_{\rm sat}$ in units of $h^{-1}M_\odot$
for numerical convenience.
We multiply the response functions for these parameters by $1/20$ so that the functions are in the similar range of $y$-axis.
}
\label{fig:param_diff_pl}
\end{figure}
\begin{figure}[h]
\centering
\includegraphics[width=0.9\textwidth]{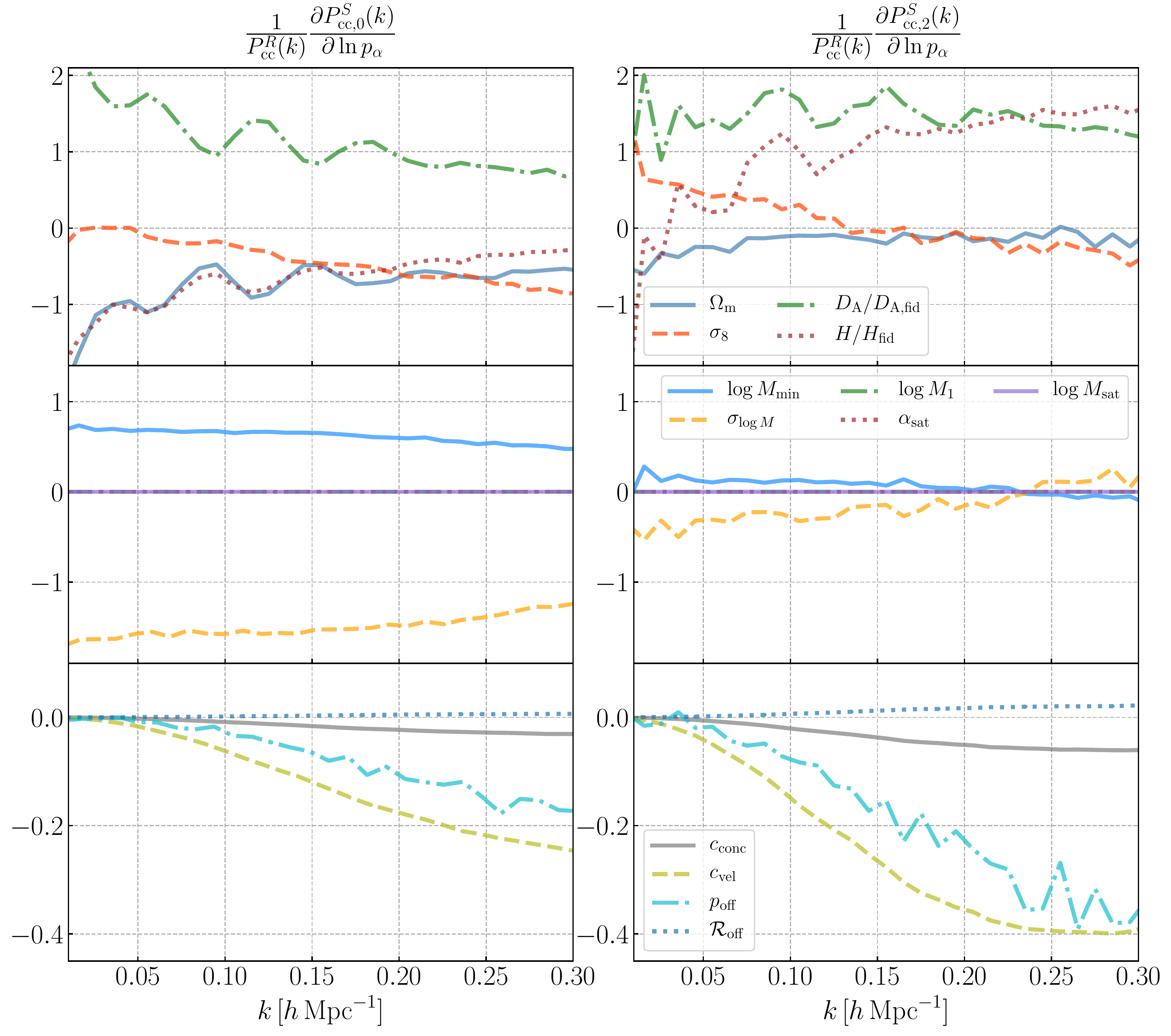}
\caption{
The similar to Fig.~\ref{fig:param_diff_pl}, but the results for the power spectra of central galaxies that are computed from the mocks including only the central galaxies as in Fig.~\ref{fig:power_full_and_cc}.
}
\label{fig:param_diff_pl_cc}
\end{figure}

The most important model ingredient for the Fisher calculation is the response function
$\partial P^{\rm S}_{{\rm gg},\ell}(k)/\partial p_\alpha$, which quantifies how a change in the $\alpha$-th model parameter alters the $\ell$-th multipole moment of the power spectrum, or equivalently a sensitivity of the $\ell$-th multipole moment to the $\alpha$-th parameter.
Fig.~\ref{fig:param_diff_pl} shows the response functions for the monopole and quadrupole spectra.
As described we use the realizations of galaxy mock catalogs to numerically evaluate the response function for each parameter up to the nonlinear scales (high $k$) (see Table~\ref{tab:simu} and Section~\ref{sec:simulation}).
We also discuss a validation of the measured responses using the linear theory prediction in Appendix~\ref{app:linear}.
As shown in the appendix, the $k$-dependence of the response function at relatively small $k$ bins is fairly well captured by variations in the underlying linear matter power spectrum, the linear galaxy bias (via the small-$k$ behavior of 2-halo term) and the linear RSD effect.
These are all useful cosmological information, while the 1-halo term causes a complex $k$-dependent effect on the response functions.

We first focus on the results for cosmological parameters $\Omega_{\rm m}$ and
$\sigma_8$ to which large-scale structure probes are most sensitive within the flat $\Lambda$CDM model.
As expected, the cosmological parameter, $\Omega_{\rm m}$ or $\sigma_8$, causes characteristic changes in
the amplitude and shape of the redshift-space power spectra over all the scales we consider.
Note that when we change $\Om$ the baryon and CDM density parameters, $\Omega_{\rm b} h^2$ and $\Omega_{c} h^2$, are kept unchanged, since these parameters are well constrained from the CMB experiments.
Also note all the responses are measured at every $k$ bin in units of \hMpci, although a change of $\Omega_{\rm m}$ leads to a change of $h$.

The responses to $\Om$ display oscillatory behaviors of the BAO, which however is not due to the shift of BAO peak itself (in physical unit) but the change in the Hubble constant $h$, since 
$\Omega_{\rm b} h^2$ and $\Omega_{c} h^2$ are kept unchanged.
For fixed $\sigma_8$ and HOD parameters, increasing $\Omega_{\rm m}$ leads to a decrease in the power spectrum amplitude at the redshifts of the galaxy samples. The change in $\Om$ also alters the RSD strength.
Such a kind of behavior is also seen on the linear matter power spectrum, and is further confirmed by the linear theory prediction using the halo bias model in Appendix~\ref{app:linear}.
%

Next let us consider the responses to $\sigma_8$.
First, increasing $\sigma_8$, for a fixed $\Omega_{\rm m}$, leads to a greater normalization of the linear matter power spectrum by definition.
In addition, increasing $\sigma_8$ leads to an increase in the abundance of massive halos, which boosts a clustering contribution from satellite galaxies in massive halos that have greater linear bias.
This yields the greater amplitude of galaxy power spectrum, even at small $k$ bins in the linear regime \citep[also see discussion around Fig.~1 in Ref.~][]{Okumura:2016mrt}.
However these boosts are to some extent compensated by a smaller halo bias due to the increased $\sigma_8$, because halos of a fixed mass scale become less biased and therefore have smaller linear bias compared to the fiducial {\it Planck} model \citep[again see Fig.~1 in Ref.][]{2015ApJ...806....2M}. Furthermore, an increase of $\sigma_8$ causes larger random or streaming motions of halos, yielding a greater suppression in the clustering amplitude at larger $k$ in the nonlinear regime.
Thus increasing $\sigma_8$ causes a scale-dependent response for the monopole and quadrupole power spectra, causing positive to negative responses from small to large $k$.
We will below study how changes in $\Omega_{\rm m}$ and $\sigma_8$ alter the power spectra of central galaxies, i.e. removing satellite galaxies, which helps to gain a clearer understanding of these behaviors.

In the same panels we also show the responses to the AP parameters, i.e. $D_{\rm A}(z)$ and $H(z)$. Note that we fix the background cosmological model to the {\it Planck} model; the real-space power spectra are the same, and the response functions are purely from a mapping of the real-space distribution of galaxies to the comoving coordinates in a hypothetical clustering analysis when the assumed cosmological model is different from the ground truth.
The AP effect distorts the BAO peak locations, and consequently the response functions display oscillatory behaviors.
As described in \S\ref{subsec:multipole} and later in this text, since the redshift-space power spectrum includes two different combinations of the AP distortion, the effects of $D_{\rm A}$ and $H$ are  distinguishable.
Note that the responses of the monopole power spectrum to $D_{\rm A}$ and $H$ have behaviors quite similar to that to $\Om$, which is not true of the quadrupole power spectrum.
This indicates that the effects of $\Om$ and distance parameters on the monopole spectrum
are highly degenerated, and hence combining the monopole and quadrupole spectra is of critical importance to break the degeneracies.

The middle-row panels of Fig.~\ref{fig:param_diff_pl} show the responses to the five HOD parameters.
As can be found from Eq.~(\ref{eq:Nc}), the parameters $\log M_{\rm min}$ and $\sigma_{\log M}$ determine the cutoff and shape of the mean central HOD at the low-mass end.
When $\log M_{\rm min}$ is decreased or when $\sigma_{\log M}$ is increased, less massive halos, which are below the cutoff mass scale in the fiducial HOD model, become to be included in the sample.
Such halos have smaller bias, leading to the smaller amplitudes of the power spectra.
Hence the response functions of the monopole power spectrum with respect to $\log M_{\rm min}$ or $\sigma_{\log M}$ are positive or negative, respectively.
On the other hand, such less massive halos have smaller random motions, and therefore cause a less suppression at small scales.
For this reason, the response of the quadrupole power spectrum changes its sign at $k \simeq 0.1~$\hMpci.

The HOD of satellite galaxies is characterized by the parameters, $\log M_1$, $\log M_{\rm sat}$ and $\alpha_{\rm sat}$.
Here $\log M_1$ determines the normalization of the mean satellite HOD; e.g., decreasing the mass scale $M_1$ leads to a higher normalization, thereby yielding more satellites to be in the sample.
Increasing $M_1$ leads to a smaller clustering amplitude at small $k$ bins due to a less contribution from massive halos, while it leads to a smaller FoG effect. For this reason, the monopole response to $M_1$ is negative, while the quadrupole response is negative at very small $k$ bins and then becomes positive at larger $k$.
The parameter $M_{\rm sat}$ determines a cut-off mass scale; $\avrg{N_{\rm s}}\rightarrow 0$ at $M\rightarrow M_{\rm sat}$.
The increase of $M_{\rm sat}$ confines the mass scale of the host halos to higher mass, which results in the decrease of the large-scale galaxy bias.
The parameter $\alpha_{\rm sat}$ determines the slope of halo mass dependence of the mean satellite HOD.
Its increase leads to two different effects on the power spectrum amplitude at large scales:
one is the enhancement of the number of satellite galaxies hosted by high-mass halos, and another is the suppression of the number of those living in low-mass halos.
The typical mass at this enhancement/suppression transition is $M-M_{\rm sat} \sim M_1$.
In addition, at small scales, an increase of the satellite population causes an enhancement of the FoG smearing, so the behavior of the response to $\alpha_{\rm sat}$ would be subtle.

Other nuisance parameters, $c_{\rm conc}$, $c_{\rm vel}$, $p_{\rm off}$ and ${\cal R}_{\rm off}$, model distributions of galaxies inside their host halos, and therefore affect the redshift-space power spectrum at $k$ bins of our interest due to the FoG effect (because the real-space changes due to variations of galaxy distribution in the same halo are all at scales below a few \hiMpc, which are outside the $k$-range we consider).
The bottom-row panels of Fig.~\ref{fig:param_diff_pl} display similar shapes for the responses to these parameters due to the changes in the amount of FoG effect.
These responses would be approximately described by a $k^2$ dependence at small $k$ bins \citep{hikage12a,hikage:2013kx}.
Hence, as long as a FoG function, which has a $k^2$-dependent term with free amplitude parameter, is introduced, one might be able to take into account variations in the FoG contamination.

For completeness, in Fig.~\ref{fig:param_diff_pl_cc} we study the responses of the power spectra of the central galaxies to the model parameters.
We numerically evaluate the response functions from the power spectra of the central galaxies.
These power spectra are measured from the central galaxies in the same mock catalogs as used in the mocks in Fig.~\ref{fig:param_diff_pl}.
Comparison of Figs.~\ref{fig:param_diff_pl} and \ref{fig:param_diff_pl_cc} manifests the role of satellite galaxies in these response functions.
For example, the $\sigma_8$ response for the monopole power spectrum of the central galaxies changes its sign at small $k$ bins, compared to Fig.~\ref{fig:param_diff_pl}, and is negative over all the $k$-scales.
Since the multipole power spectra are expressed as
$P_0(k)\sim (b^2+ 2bf/3+f^2/5)(\sigma_8)^2$ and $P_2(k)\sim (4bf/3+4f^2/7)(\sigma_8)^2$ (see Eq.~\ref{eq:kaiser}),
the sign change at small $k$ is due to the fact that a decrease in the linear bias overcomes an increase in $\sigma_8$ in the monopole response; the leading term $b^2 (\sigma_8)^2$ has a negative response to an increase in $\sigma_8$.
On the other hand, the $\sigma_8$-response of the quadrupole power spectrum is still positive at small $k$ bins, because of the weaker dependence on $b$ as the leading term has a dependence of
$b(\sigma_8)^2$.
All these complicated dependences of the power spectra on cosmological parameters are contained in the halo power spectrum.

\subsection{Cosmological parameter forecasts}
\label{sec:cosmoloigcal_parameters}

\subsubsection{Signal-to-noise ratio of multipole power spectra}
\label{sec:sn}

\begin{figure}[h]
\centering
\includegraphics[width=0.6\textwidth]{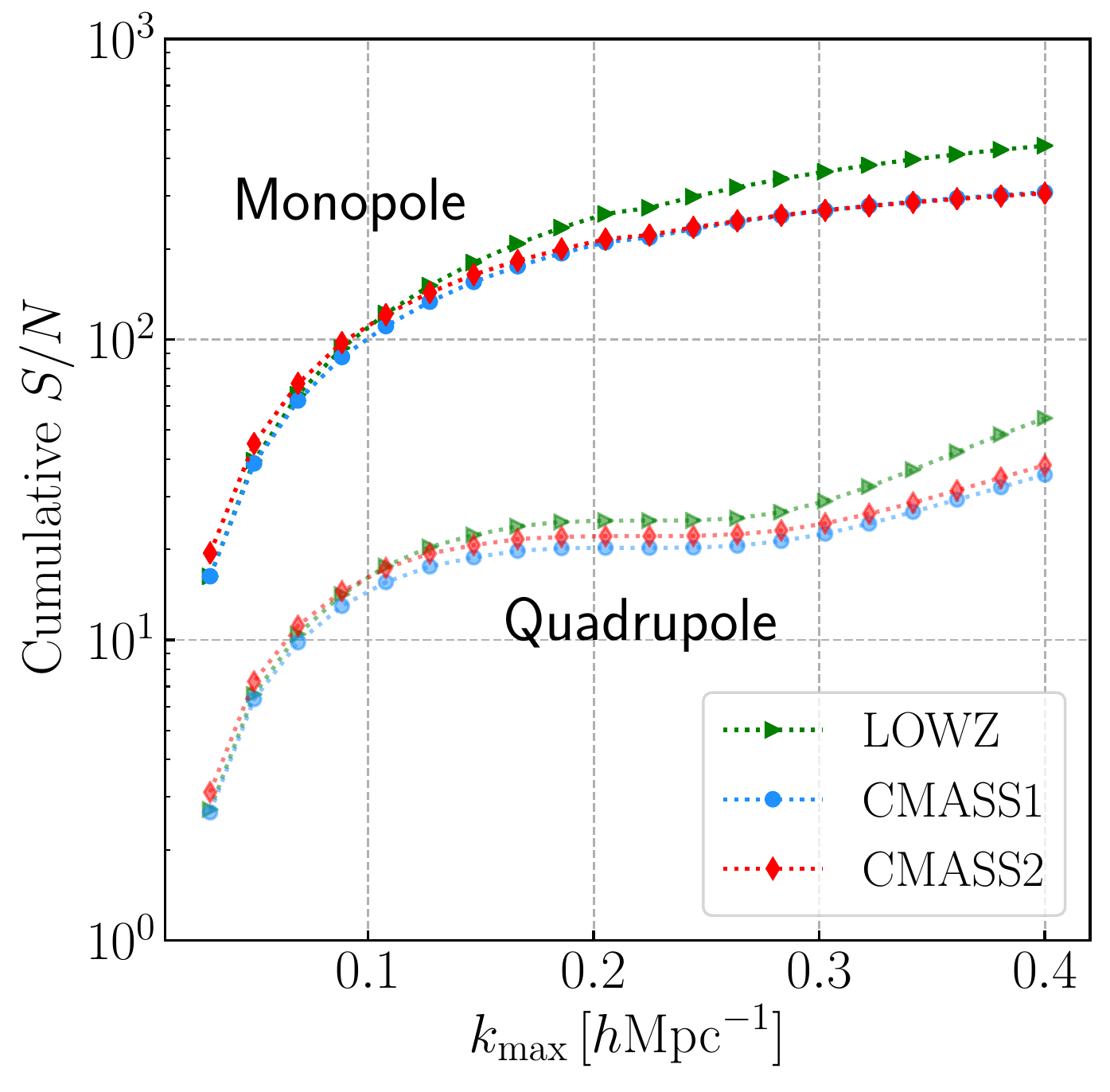}
\caption{
The cumulative signal-to-noise ratio, $S/N$ (defined by Eq.~\ref{eq:sn}) for a hypothetical measurement of the monopole and quadrupole power spectra for each of the LOWZ, CMASS1 and CMASS2 samples, respectively.
The cumulative $S/N$ is the information content of the power spectrum amplitude, obtained by integrating the signal-to-noise at each $k$ bin from the minimum wavenumber, set to $k_{\rm min}=0.02$~\hMpci here, up to a given maximum wavenumber $k_{\rm max}$ shown in the $x$-axis.
}
\label{fig:cumulativeSN}
\end{figure}

Before showing cosmological parameter forecasts, we first study, in Fig.~\ref{fig:cumulativeSN},
the cumulative signal-to-noise ratio $(S/N)^2$ expected for a measurement of the monopole and quadrupole power spectrum for
the SDSS-like survey (Table~\ref{tab:surveyparam}), which is defined as
\begin{align}
\left(\frac{S}{N}\right)^2 = \sum_{i,j}^{k_{\rm max}} {P}^{\rm S}_\ell(k_i) \, {\rm Cov}^{-1}\left[{P}^{\rm S}_\ell(k_i),{P}^{\rm S}_{\ell} (k_j)\right] {P}^{\rm S}_\ell(k_j),
\label{eq:sn}
\end{align}
The inverse of $(S/N)^2$ gives a precision in estimating the amplitude parameter of  power spectrum,
if the power spectrum up to a given maximum wavenumber, $k_{\rm max}$, is measured from an assumed galaxy survey (here the SDSS-like survey), assuming that the power spectrum shape is perfectly known.
Fig.~\ref{fig:cumulativeSN} shows the results of $S/N$ for the monopole and quadrupole power spectra as a function of $k_{\rm max}$.
Here we consider the LOWZ, CMASS1, and CMASS2-like samples.
If the power spectrum measurement is in the sample-variance-limited regime and the Gaussian covariance is valid, $S/N\propto k_{\rm max}^{3/2}$.
The figure shows that $(S/N)$ keep increasing with increasing $k_{\rm max}$ up to $k_{\rm max}\sim 0.2~$\hMpci.
However, at the larger $k_{\rm max}$ the covariance is dominated by the shot noise for the SDSS-like galaxies which have the number density of about $10^{-4}~(h{\rm Mpc}^{-1})^3$
(see Table~\ref{tab:surveyparam}), and  the $S/N$ is saturated in the shot noise regime.
The figure also shows that $S/N$ for the quadrupole has a smaller amplitude than the monopole by a factor of $10$.
Nevertheless the quadrupole spectrum carries a significant information on cosmological parameters thanks to its sensitivity to the RSD effect and the AP effect, and therefore helps break the parameter degeneracies when combined with the monopole spectrum, as we will show below.

\subsubsection{Geometrical constraints and cosmological parameter forecasts}

\begin{figure}[h]
\centering
	\centering
	\includegraphics[width=0.95\textwidth]{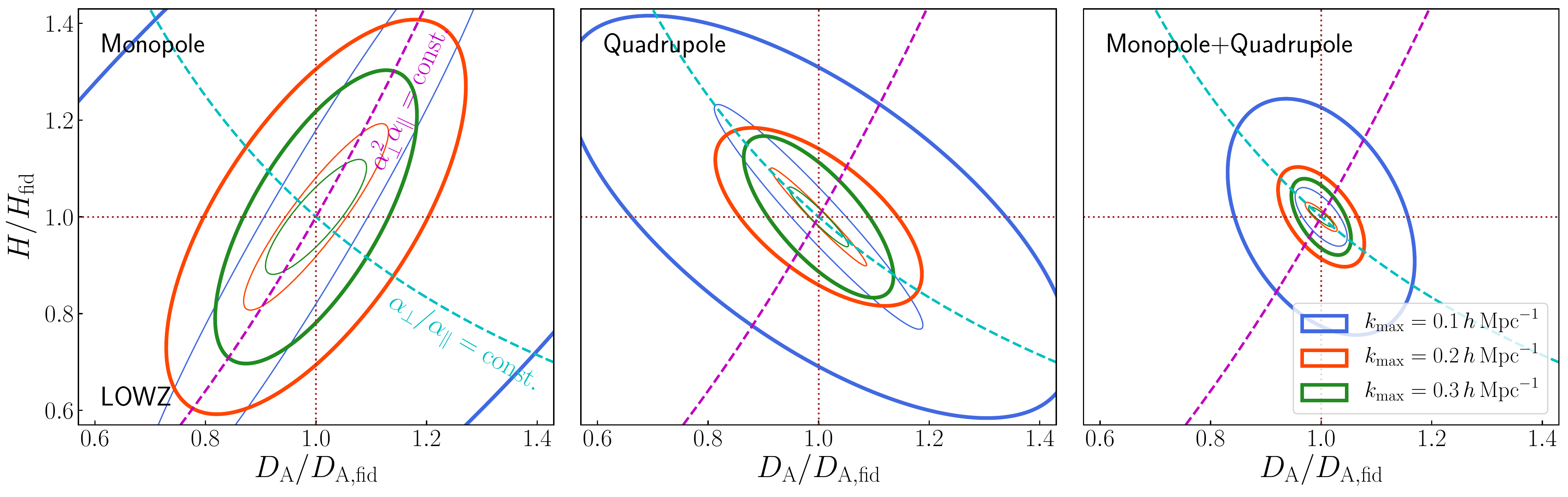}
	\includegraphics[width=0.95\textwidth]{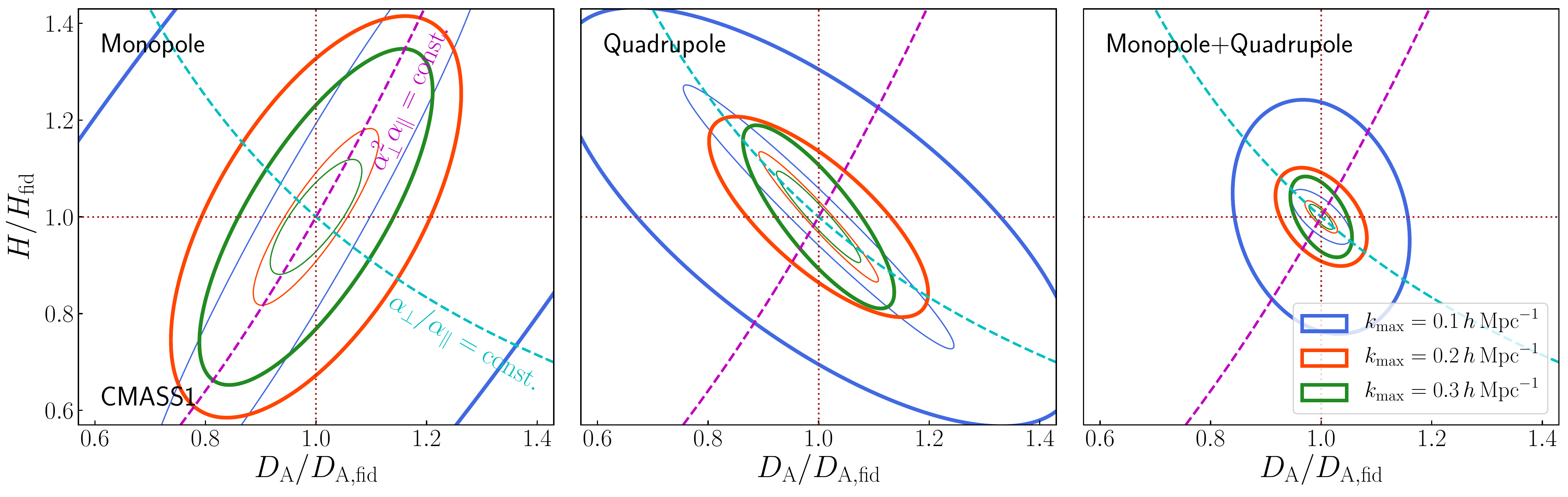}
	\includegraphics[width=0.95\textwidth]{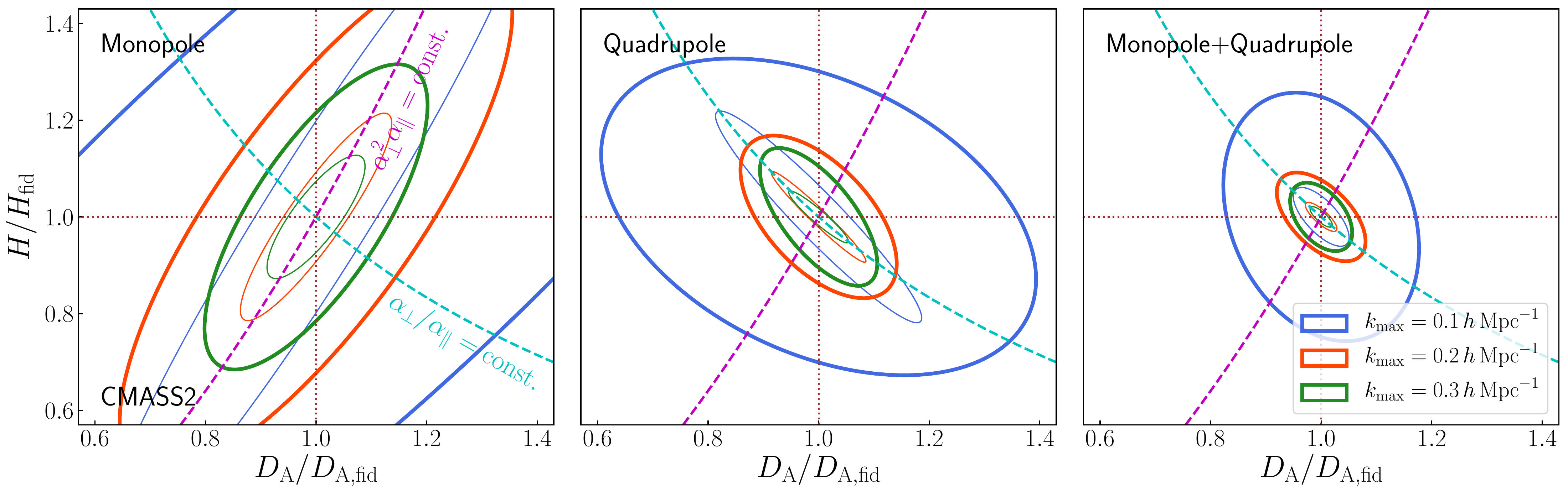}
\caption{
The Fisher-forecasted 68\% CL error ellipse in the sub-space of angular diameter distance $D_{\rm A}(z_n)$ and
the Hubble expansion rate $H(z_n)$ at each redshift of the LOWZ (top-row panels), CMASS1 (middle-row) and CMASS2 (bottom-row) samples, respectively, including marginalization over other parameters.
The left, middle or right panel in each row shows the result when using either of the monopole or quadrupole power spectrum alone or using the joint measurement of monopole and quadrupole spectra, respectively.
The three contours in thick lines in each panel show the results when using the power spectrum information up to $k_{\rm max}=0.1$, 0.2 or $0.3~$\hMpci, respectively.
These constraints are from measurements of the geometrical AP distortion of the BAO peak locations and
broad-band shape in the redshift-space power spectra.
For comparison, the magenta or cyan dashed curve shows the direction where the isotropic ``dilation'' parameter, $\alpha_\perp^2\alpha_\parallel$ or the anisotropic ``warping'' parameter $\alpha_\perp/\alpha_\parallel$ in the geometrical AP distortion is kept constant; $\alpha_\perp^2\alpha_\parallel={\rm const.}$ or $\alpha_\perp/\alpha_\parallel={\rm const}$, respectively.
Here $\alpha_\perp$ and $\alpha_\parallel$ are the ratio of the assumed cosmological distances (i.e. the assumed values) to the true values (i.e. the values for the {\it Planck} model as given in Table~\ref{tab:HOD_parameters}):
$\alpha_\perp\equiv D_{\rm A, fid}/D_{\rm A}$ and $\alpha_\parallel=H/H_{\rm fid}$ that are shown in the $x$- and $y$-axes, respectively.
We also put the thin solid-line contours, which denote the error ellipses in the case where the halo-galaxy connection parameters are fixed to their fiducial values.
}
\label{fig:fisher_boss_DA_H_margin}
\end{figure}

\begin{figure}[h]
\centering
\includegraphics[width=0.9\textwidth]{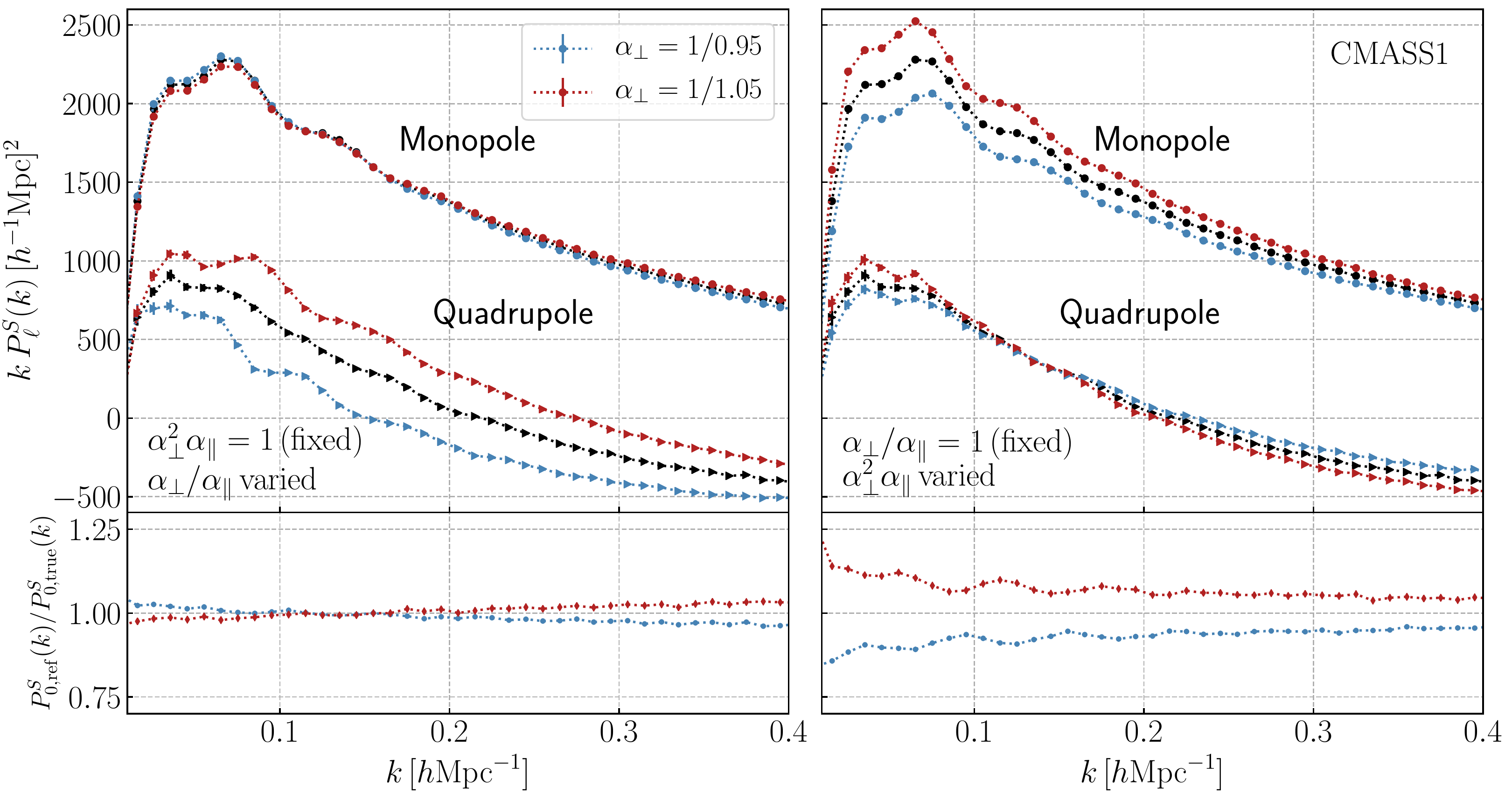}
\caption{
The geometrical AP effect on the monopole and quadrupole power spectra for the fiducial CMASS1 mocks.
To study this, we focus on the two components of AP effect: the isotropic dilation effect ``$\alpha_\perp^2\alpha_\parallel$'' and the anisotropic warping effect ``$\alpha_\perp/\alpha_\parallel$'' (see the caption in the previous figure).
The left panel shows the spectra when the warping parameter $\alpha_\perp/\alpha_\parallel$ is varied by $\alpha_\perp=1/0.95$ or $1/1.05$, while the dilation parameter $\alpha_\perp^2\alpha_\parallel$ is kept fixed.
While the monopole power spectrum amplitude and the BAO peak locations are almost unchanged, this variation alters the relative amplitude of the quadrupole power spectrum.
The right panel shows the spectra when the dilation parameter $\alpha_\perp^2\alpha_\parallel$ is varied by the same amount of $\alpha_\perp$ as in the left panel, while the warping parameter $\alpha_\perp/\alpha_\parallel$ is kept fixed.
In this case, while the monopole power spectrum is largely changed, the change of the quadrupole is relatively moderate.
}
\label{fig:Pl_cmass1_AP_2comb}
\end{figure}
\begin{figure}[h]
\centering
\includegraphics[width=0.98\textwidth]{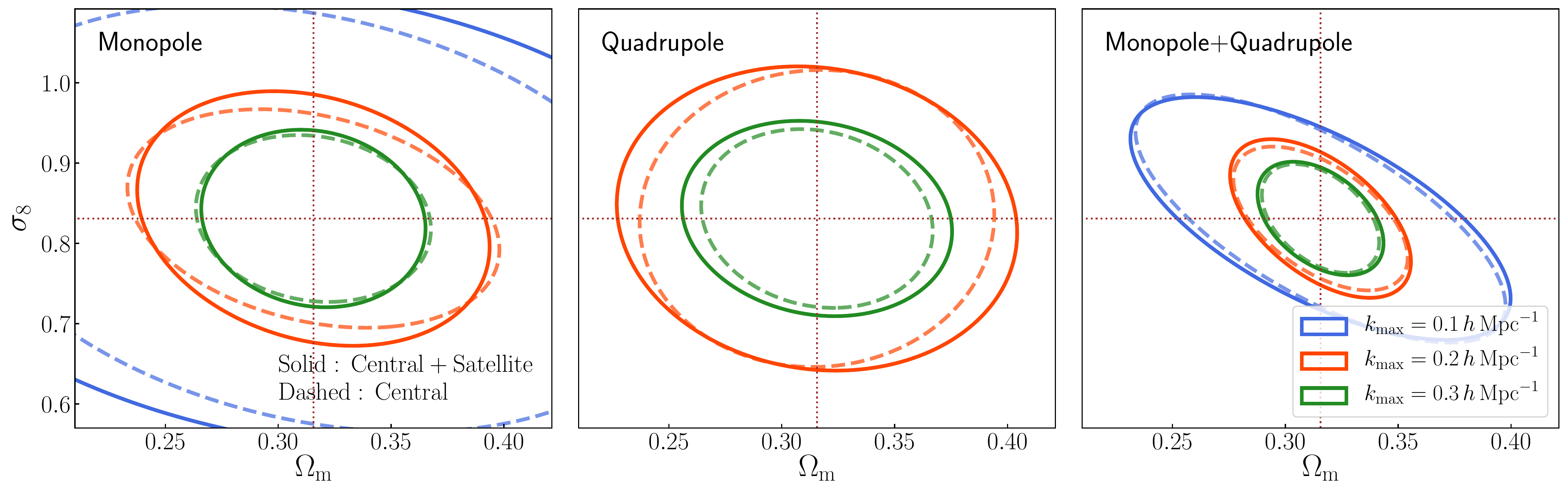}
\caption{
The marginalized error ellipses in the sub-space of parameters, $\Omega_{\rm m}$ and $\sigma_8$, as in Fig.~\ref{fig:fisher_boss_Om_s8_fixed}.
To derive this, we used the Fisher matrix jointly combining the power spectrum information for the LOWZ, CMASS1, and CMASS2 samples.
Since we treated the geometrical information in the galaxy power spectrum by $D_{\rm A}(z_n)$ and $H(z_n)$ at each redshift, the constraints on $\Omega_{\rm m}$ and $\sigma_8$ are mainly from the RSD effect and the amplitude information in the redshift-space galaxy power spectrum.
The dashed-line contours represent the constraints from the power spectra that are measured from the mocks including central galaxies
alone (i.e. removing satellite galaxies from each mock).
Comparison of the solid and dashed contours shows that the constraining power of $\Omega_{\rm m}$ and $\sigma_8$ is mainly from the redshift-space power spectra of central galaxies.
}
\label{fig:fisher_boss_Om_s8_margin}
\end{figure}
\begin{figure}[h]
\centering
\includegraphics[width=0.98\textwidth]{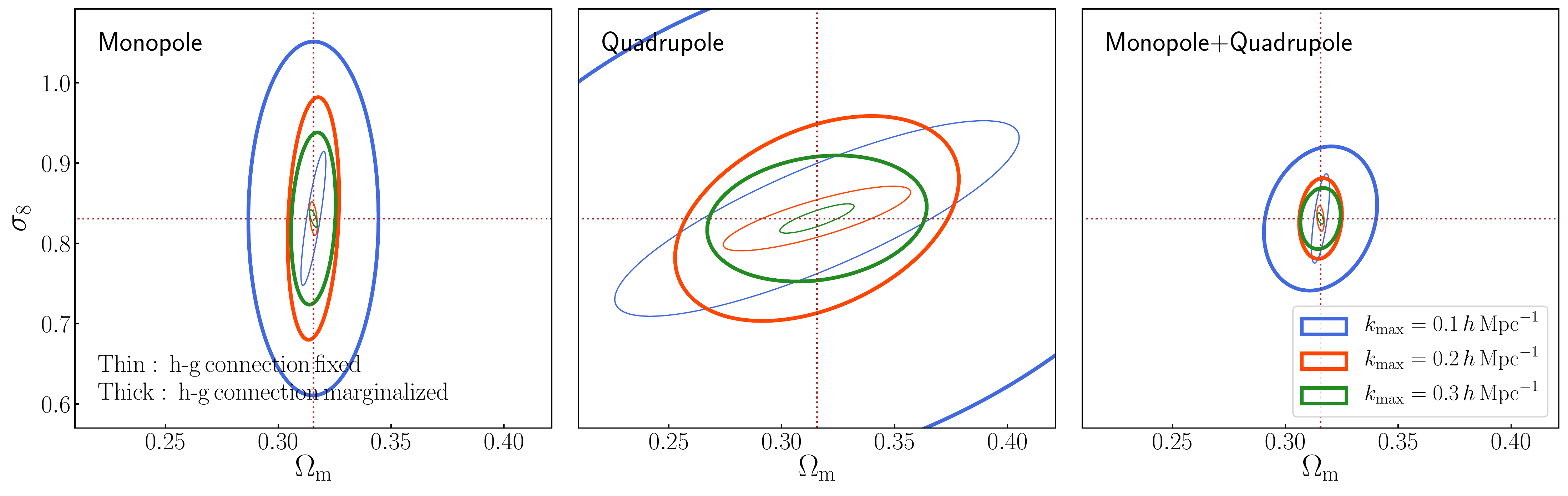}
\caption{
Similar results to the previous figure, but show the expected errors on $\Omega_{\rm m}$ and $\sigma_8$ for the flat $\Lambda$CDM framework.
Here we include the $\Omega_{\rm m}$ information in $D_{\rm A}(z_n)$ and $H(z_n)$ at each redshift of the LOWZ, CMASS1 and CMASS2 samples, since these quantities at each redshift is specified by an assumed $\Om$ if a flat $\Lambda$CDM model is {\it a priori} assumed.
Consequently the error of $\Om$ is significantly tightened.
The thin solid-line contours are the error ellipses when we fix the halo-galaxy connection parameters to their fiducial values, i.e. unmarginalized errors.
}
\label{fig:fisher_boss_Om_s8_lcdm}
\end{figure}
\begin{figure}[h]
\centering
\includegraphics[width=0.98\textwidth]{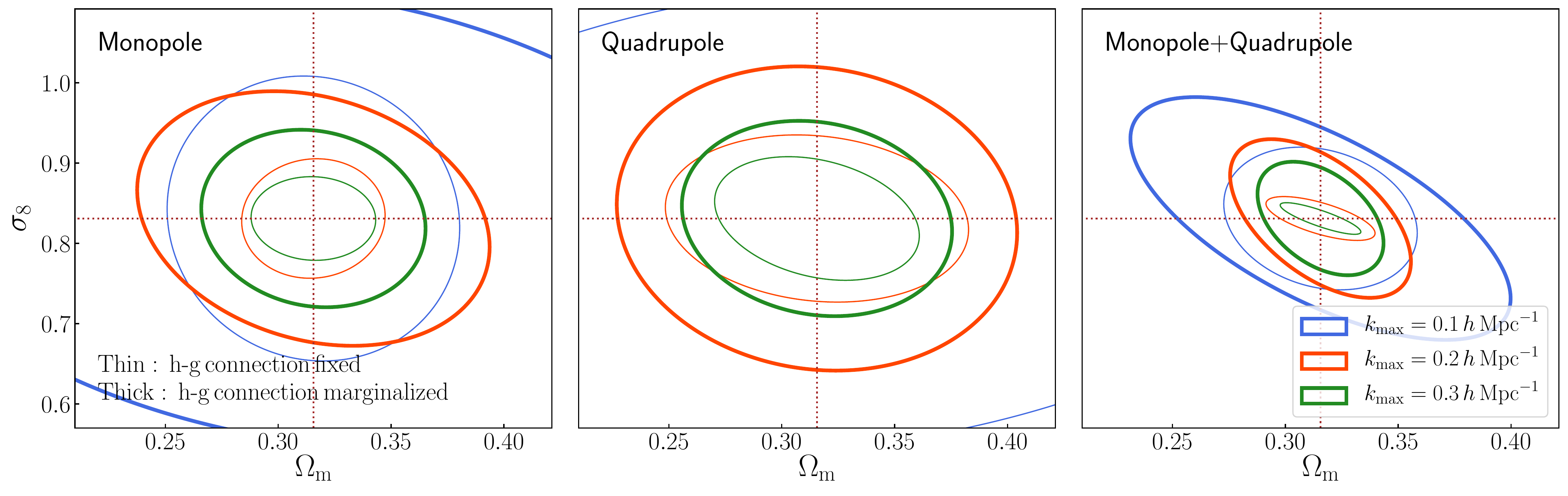}
\caption{
The impact of uncertainties in halo-galaxy connection on the errors of $\Om$ and $\sigma_8$.
Similar plot to Fig.~\ref{fig:fisher_boss_Om_s8_margin}, but shows the results when fixing the halo-galaxy connection parameters to their fiducial values as the thin solid-line contours.
To compute this, we use $(8\times 8)$ sub-matrix elements of the full Fisher matrix that contain only $\Om$ and $\sigma_8$ and the two distances, $D_{\rm A}$ and $H$, for each of the three redshift bins.
}
\label{fig:fisher_boss_Om_s8_fixed}
\end{figure}

We now show the forecasts for cosmological constraints, which are the main results of this paper.
A notable advantage of the redshift-space galaxy clustering method, compared to
other large-scale structure probes, is that it enables us to simultaneously constrain the growth of matter clustering (usually $f\sigma_8$) as well as the cosmological distances via the AP effect.
As we have described above, in this study we treat these parameters independently,
$\Omega_{\rm m}$, $\sigma_8$, $D_{\rm A}(z_n)$, and $H(z_n)$ in our parameter forecasts, even though the cosmological distances can be specified for a given $\Om$ at a given redshift for a flat $\Lambda$CDM model.
Thus the redshift-space galaxy clustering method enables a stringent geometrical test of the cosmological distances free of uncertainties in the large-scale structure growth or galaxy bias \citep{Eisenstein:2005su}
\citep[also see][for a similar discussion]{2015PhRvD..92l3518T}.
We keep this advantage for our parameter forecasts, and then will discuss how the $\Om$ constraint is improved if we use the $\Om$-dependence of distances for a flat $\Lambda$CDM model.

The BAO peak position and shape in the galaxy power spectrum is measured anisotropically if a reference cosmological model, which needs to be assumed in converting the observed angular scales and redshifts to the comoving coordinates in the analysis, differs from the underlying true cosmology.
Fig.~\ref{fig:fisher_boss_DA_H_margin} shows the 68\% CL error ellipses in determination of $D_{\rm A}(z_n)$ and $H(z_n)$, including marginalization over other 32 parameters including nuisance parameters to model the halo-galaxy connection.
The left, middle and right panels in each row show the results when using either of the monopole or quadrupole power spectrum alone or when combining the two spectra, respectively.
The top-, middle- and bottom-row panels are the results for the LOWZ, CMASS1 and CMASS2-like galaxy samples, respectively.
First of all, each panel shows that the geometrical constraints are improved as the information up to the higher $k_{\rm max}$ is included.
The monopole power spectrum gives a tighter constraint on the combination of ``$D_{\rm A}(z)^2/H(z)(\propto \alpha_\perp^2\alpha_\parallel)$'', where $\alpha_\parallel
=H/H_{\rm fid}$ and $\alpha_\perp=D_{\rm A, fid}/D_{\rm A}$ ($D_{\rm A}$ and $H$ are model parameters, and the quantities with ``fid'' are the values for the {\it Planck} cosmology, i.e. the true values).
The value of  ``$\alpha_\perp^2\alpha_\parallel$'' is varying along the direction perpendicular to
``$\alpha_\perp^2\alpha_\parallel={\rm const.}$'' shown in each panel, which is close to the direction  of
the constant warping parameter ($\alpha_\parallel/\alpha_\perp={\rm const.})$.
A change in $H(z)$ and $D_{\rm A}(z)$ while keeping $\alpha_\parallel/\alpha_\perp$ fixed, causes an ``isotropic'' distortion in the monopole spectrum, or causes an isotropic shift in the BAO peak locations and the broad-band $k$-dependent shape as a function of $k$.
It means that the distance constraints in the monopole power spectrum are from the isotropic distortion (sometimes called the ``dilation'' effect).
On the other hand, the quadrupole spectrum gives a tighter constraint on the combination of $\alpha_\perp/\alpha_\parallel$.
Again the direction of varying the value of $\alpha_\perp/\alpha_\parallel$ is close to the direction of $\alpha_\perp^2\alpha_\perp={\rm constant.}$
While the monopole power spectrum amplitude is kept fixed along this direction of $\alpha_\perp^2\alpha_\perp={\rm const.}$, it causes an anisotropic distortion in the redshift-space power spectrum and therefore alters the quadrupole power spectrum.
Fig.~\ref{fig:Pl_cmass1_AP_2comb} explicitly shows how a change in either of $\alpha_\parallel/\alpha_\perp$ or $\alpha_\perp^2\alpha_\parallel$, keeping the other fixed, alters either of the monopole or quadrupole power spectra, while the other is almost unchanged.
Interestingly, even though the signal-to-noise ($S/N$) in the quadrupole power spectrum is smaller than that of the monopole spectrum by a factor of 10 (see Fig.~\ref{fig:cumulativeSN}), the quadrupole carries a better precision of the distance measurements than that of the monopole.
This would be  explained by the fact that a change of $\Om$ has a strong degeneracy with that of $D_{\rm A}(z)$ and $H(z)$ in the monopole power spectrum (see Fig.~\ref{fig:param_diff_pl}).
On the contrary, the quadrupole has distinct behaviors in the responses to $\Om$ and AP parameters, as well as to another cosmological parameters $\sigma_8$.

Fig.~\ref{fig:fisher_boss_Om_s8_margin} shows the 68\% CL error ellipses in the ($\Omega_{\rm m}, \sigma_8$)-subspace, including marginalization over other 36 parameters (the cosmological distances and the halo-galaxy connection parameters).
Here we include the information for the three redshift slices of the LOWZ-, CMASS1- and CMASS2-like surveys (Table~\ref{tab:surveyparam}).
For comparison, we here show the results when using either of the monopole or quadrupole power spectrum alone,
or combining the monopole and quadrupole signals up to a given maximum wavenumber, $k_{\rm max}=0.1, 0.2$ or $0.3$~\hMpci, respectively.
As we have shown in Fig.~\ref{fig:fisher_boss_DA_H_margin}, variations in the BAO peak locations and anisotropic features in the 
power spectrum are captured by the cosmological distance parameters, $D_{\rm A}(z)$ and $H(z)$.
The constraints on $\Omega_{\rm m}$ and $\sigma_8$ shown here are from the RSD effect and the amplitude information in the power spectrum.
Similarly to the previous figure, this figure shows that the quadrupole power spectrum carries a similar-level information on these cosmological parameters to that of the monopole power spectrum.
Thus the RSD effect to which the quadrupole power spectrum is sensitive carries useful cosmological information.
Hence, combining the monopole and quadrupole power spectra improves the cosmological constraints compared to the constraints from either of the two power spectra alone.
The figure also shows that including the power spectrum information up to the higher $k_{\rm max}$, from 0.1 to 0.3~\hMpci, continues to improve the cosmological constraints, even after marginalization over other nuisance parameters.
This implies that changes in $\Om$ and $\sigma_8$ cause quite different scale-dependent changes in the redshift-space power spectra from the dependences of other parameters, as quantified by the response functions in Fig.~\ref{fig:param_diff_pl}.
Therefore the cosmological parameters are distinguishable from other parameters such as the halo-galaxy connection parameters in the measured redshift-space power spectrum.
Table~\ref{tab:fisher_table} summarizes the marginalized 68\% CL error of $\Omega_{\rm m}$ or $\sigma_8$.
The table shows that including the information up to $k_{\rm max}=0.3$~\hMpci from $0.1$~\hMpci leads to an improvement in the constraint on $\Omega_{\rm m}$ or $\sigma_8$ by a factor of $3.15$ or $3.63$, respectively.
This would be compared to a factor of 5.2($=27^{1/2}$), which corresponds to a naive improvement if the parameter error scales with the inverse of Fourier volume, $1/k_{\rm max}^{3/2}$ in a case that
all the parameters are totally independent and the information is in the sample-variance-limited regime.
The results in Fig.~\ref{fig:fisher_boss_Om_s8_margin} and Table~\ref{tab:fisher_table} are quite encouraging.

\begin{table}
\begin{center}
\begin{tabular}{l|ccc|ccc} \hline\hline
 & \multicolumn{3}{c|}{$\sigma(\Om) / \Om$} & \multicolumn{3}{|c}{$\sigma(\sigma_8) / \sigma_8$} \\ \hline
$k_{\rm max}$~[\hMpci] & $0.1~$ & $0.2~$ & $0.3~$ & $0.1~$ & $0.2~$ & $0.3~$ \\ \hline
$P^{\rm S}_0$		& $0.40~(0.060)$ & $0.16~(0.023)$ & $0.10~(0.020)$ & $0.23~(0.17)$ & $0.13~(0.12)$ & $0.087~(0.085)$ \\
$P^{\rm S}_2$		& $0.71~(0.29)$  & $0.18~(0.11)$  & $0.12~(0.085)$ & $0.51~(0.27)$ & $0.15~(0.10)$ & $0.096~(0.062)$ \\
$P^{\rm S}_0 + P^{\rm S}_2$ & $0.18~(0.052)$ & $0.083~(0.020)$ & $0.058~(0.018)$ & $0.12~(0.071)$ & $0.078~(0.040)$ & $0.056~(0.030)$ \\ \hline\hline
\end{tabular}
\caption{
The $68\%$ fractional error of $\Om$ or $\sigma_8$, including marginalization over other parameters.
Shown here is the fractional error when using either of the monopole ($P^{\rm S}_0)$ or quadrupole ($P^{\rm S}_2$) power spectrum information or using the joint measurement ($P^{\rm S}_0+P^{\rm S}_2$) up to a given maximum wavenumber, $k_{\rm max}=0.1$, 0.2 or $0.3~$\hMpci, respectively.
The number in the parentheses gives the error if including the $\Om$ information in the angular diameter distance $D_{\rm A}$ and the Hubble expansion rate $H$ for the flat $\Lambda$CDM model (see Fig.~\ref{fig:fisher_boss_Om_s8_lcdm} for details).
}
\label{tab:fisher_table}
\end{center}
\end{table}

An alternative approach one might want to employ is to use the measured redshift-space galaxy power spectrum to make a most stringent test or even falsify the flat $\Lambda$CDM paradigm.
Within the flat $\Lambda$CDM model, the cosmological distances, $D_{\rm A}(z)$ and $H(z)$, are specified by a given $\Omega_{\rm m}$, and hence one could obtain an even tighter constraint on $\Omega_{\rm m}$ than the case where $D_{\rm A}(z)$ and $H(z)$ are treated as free parameters.
Fig.~\ref{fig:fisher_boss_Om_s8_lcdm} shows the marginalized errors of $\Omega_{\rm m}$ and $\sigma_8$ based on this approach, where we used 32 parameters in the Fisher analysis; 38 minus 6 parameters ($D_{\rm A}(z)$ and $H(z)$ for each of the three redshift slices).
The figure clearly shows that including the geometrical information significantly improves the constraint on $\Omega_{\rm m}$, while the constraint on $\sigma_8$ is almost unchanged.
This implies that most of the information in $\Omega_{\rm m}$ is from the geometrical constraints, the BAO peak locations and the AP effect, if assuming the flat $\Lambda$CDM model, while the constraint on $\sigma_8$ is mainly from the RSD and amplitude information.
Table~\ref{tab:fisher_table} gives a summary of the marginalized 1D error in $\Omega_{\rm m}$ and $\sigma_8$, and shows that we could achieve about 2\% or 3\% accuracy in $\Omega_{\rm m}$ or $\sigma_8$, respectively, if we can use the information up to $k_{\rm max}=0.3$ \hMpci, even after including marginalization over the halo-galaxy connection parameters.

What is the impact of nuisance parameters such as the halo-galaxy connection parameters on the cosmological parameter inference?
To address this question, in Fig.~\ref{fig:fisher_boss_Om_s8_fixed}, we show the results if we fix the halo-galaxy connection parameters to their fiducial values by the thin lines on top of the results after marginalization (thick), which were already shown in Fig.~\ref{fig:fisher_boss_Om_s8_margin}.
Here we used only 8 parameters in the Fisher analysis; $\Omega_{\rm m}$, $\sigma_8$ and 6 distance parameters (3 $D_{\rm A}$'s and $H$'s parameters).
First of all, if we use the information up to $k_{\rm max}=0.1~$\hMpci, we cannot obtain any meaningful constraint on either of $\Om$ or $\sigma_8$ due to strong degeneracies between the power spectrum amplitudes and the halo-galaxy connection parameters, where the latter controls the overall galaxy bias at small $k$ limit. If we use either of the monopole or quadrupole information up to $k_{\rm max}=0.2$ or 0.3~\hMpci, the halo-galaxy connection parameters degrade the accuracy of $\Om$ or $\sigma_8$ by a similar amount.
In the case of the joint constraints using both of the monopole and quadrupole, the error of $\Omega_{\rm m}$ is not largely changed when $k_{\rm max} \geq 0.2$ \hMpci, because $\Om$ is mainly determined by the AP effect as we discussed.
On the other hand, the $\sigma_8$ constraint is largely degraded by the halo-galaxy connection parameters.
Hence, an accuracy of $\sigma_8$ estimation from the redshift-space galaxy power spectrum largely depends on a level of our understanding of the halo-galaxy connection in the nonlinear regime.

\begin{figure}[h]
\centering
\includegraphics[width=0.6\textwidth]{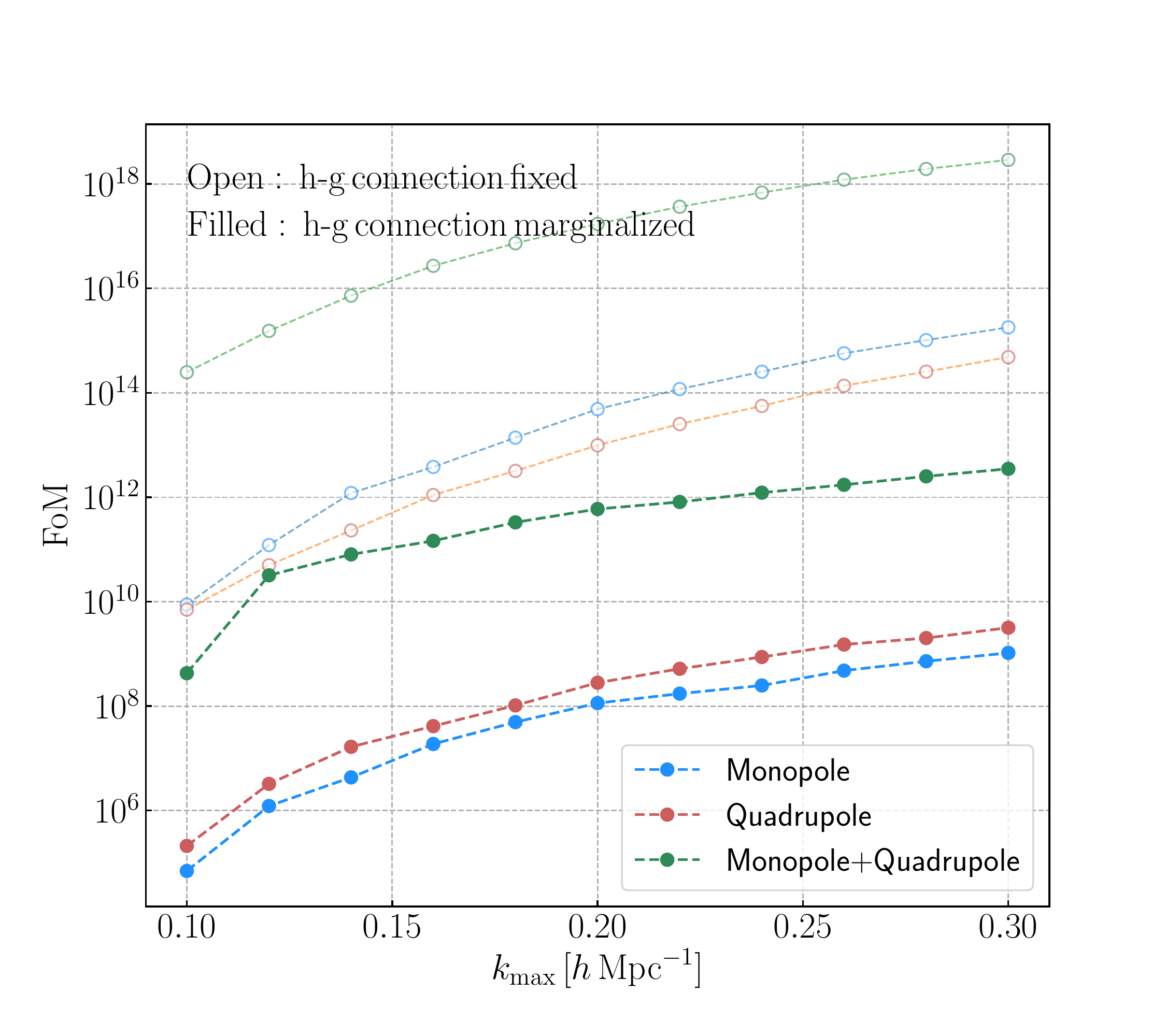}
\caption{
The figure-of-merit (FoM) of cosmological information, $\{\Om,\sigma_8,D_{\rm A}(z_n)/D_{\rm A,fid},H(z_n)/H_{\rm fid}\}$ (8 parameters in total), in the redshift-space galaxy power spectra for the LOWZ, CMASS1 and CMASS2 samples.
The FoM quantifies a volume of the error ellipsoid in the eight-dimensional space of the cosmological parameters.
The filled symbols show the results when including the power spectrum information up to a given maximum wavenumber $k_{\rm max}$ shown in the $x$-axis.
Here we show the results for the monopole or quadrupole power spectrum alone or when combining measurements of the two spectra.
The open symbols show the results when other halo-galaxy connection parameters are kept fixed to their fiducial values.
}
\label{fig:FoM_boss_margin}
\end{figure}

To quantify the constraining power of the BOSS-like galaxy survey on the cosmological parameters, we compute the following figure of merit (FoM), defined in terms of the Fisher matrix as
\begin{align}
{\rm FoM} = \frac{1}{\sqrt{|{\rm det} \, ({\bf F}^{-1})_{\rm sub}|}},
\end{align}
where ${\bf F}^{-1}$ is the inverse of Fisher matrix, ``sub'' in $({\bf F}^{-1})_{\rm sub}$ means $(8\times 8)$ sub-matrix elements containing only $\Omega_{\rm m}$, $\sigma_8$ and 6 distance parameters, and ``det'' denotes the determinant of the sub-matrix.
The FoM quantifies a volume of the marginalized ellipsoid in the 8-dimensional space of cosmological parameters.

Fig.~\ref{fig:FoM_boss_margin} shows that the FoM of the quadrupole power spectrum is comparable to that of the monopole power spectrum.
What is remarkable is that the quadrupole gets a larger FoM than the monopole when the HOD and other halo-galaxy connection parameters are marginalized.
This is ascribed to huge degeneracies between the cosmological parameters and the HOD parameters in the monopole power spectrum (see Fig.~\ref{fig:param_diff_pl}).
By combining the two information, the impact of halo-galaxy connection parameters, i.e. uncertainties in galaxy physics or small-scale structures, is mitigated, which boosts the FoM by a factor of over 1000 compared to either of the two alone.
In addition, the power spectrum information in the quasi nonlinear regime is useful; including the information up to $k_{\rm max}=0.3~$\hMpci leads to about factor of 6 gain in the information content compared to $k_{\rm max}=0.2~$\hMpci. This gain is larger than a naive expectation, $(0.3/0.2)^{3/2}\simeq 1.8$, and is ascribed to the fact that the nonlinear information helps efficiently break the parameter degeneracies.
This result clearly demonstrates that the redshift-space power spectrum in mildly nonlinear regime is quite powerful to constrain the cosmological parameters.

\section{Discussion}
\label{sec:discussion}

\subsection{A model-independent measurement of the RSD effect}

\begin{figure}[h]
\centering
\includegraphics[width=0.6\textwidth]{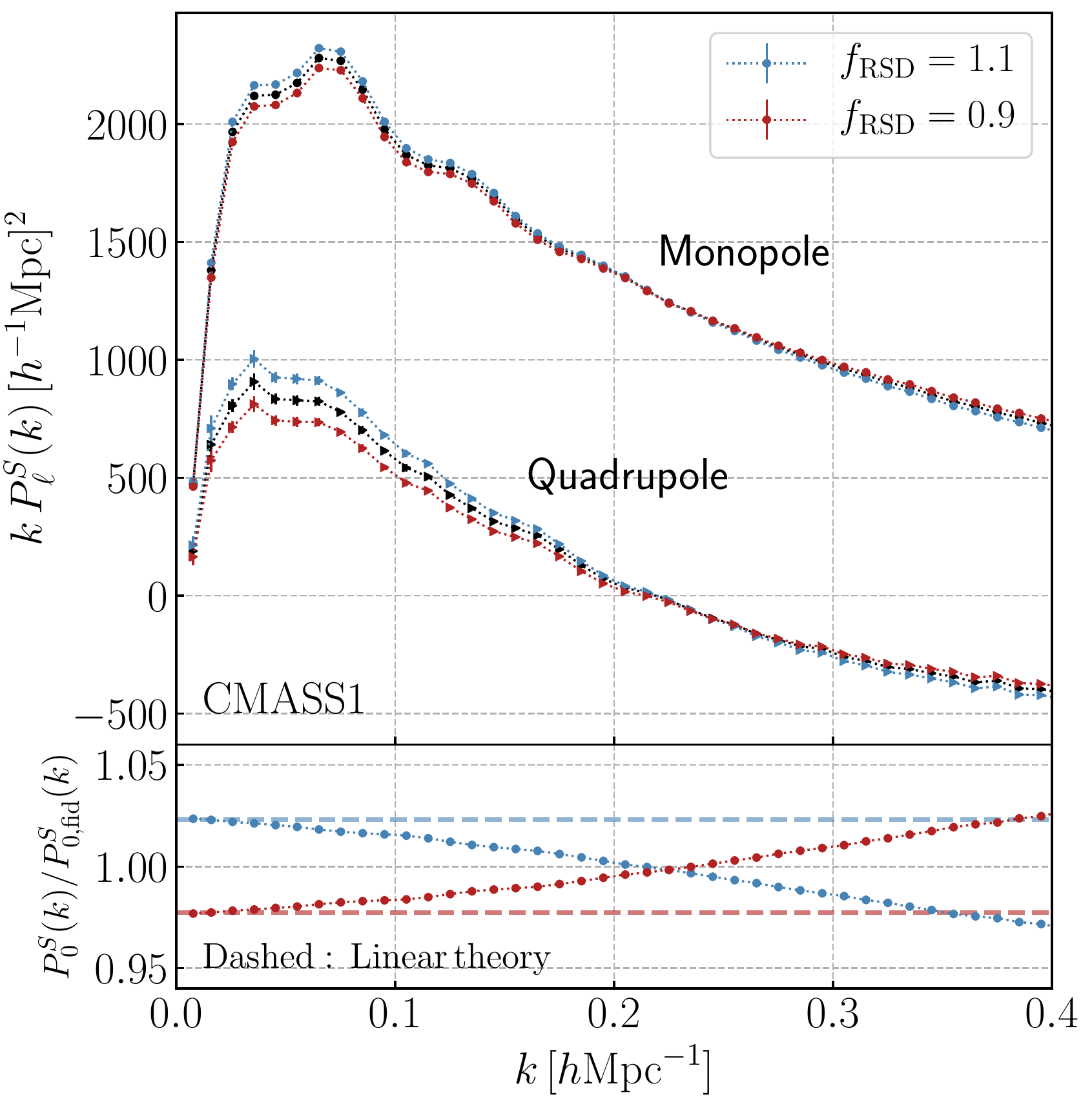}
\caption{
The effect of RSD control parameter, $f_{\rm RSD}$, on the monopole and quadrupole power spectra; when generating the mocks of of CMASS1 galaxies in redshift space, we modified the RSD displacement of host halos by the amount of $f_{\rm RSD}$, and then computed these multipole power spectra from the mocks (we did not change the RSD effects due to internal motions of galaxies inside the host halo).
Note that $f_{\rm RSD}=1$ corresponds to the $\Lambda$CDM model, and all the models shown here have the same real-space power spectrum. A change in the RSD parameter $f_{\rm RSD}$ alters these power spectra depending on $k$. The lower panel shows the fractional ratio relative to the fiducial model. The dashed curve shows the Kaiser formula prediction (Eq.~\ref{eq:kaiser}) that is computed by replacing $\beta_{\Lambda {\rm CDM}}$ in the formula with $f_{\rm RSD}\beta_{\Lambda{\rm CDM}}$, and fairly well reproduce the mock results at small $k$ limit.
}
\label{fig:p0p2_fRSD}
\end{figure}
\begin{figure}[h]
\centering
\includegraphics[width=0.7\textwidth]{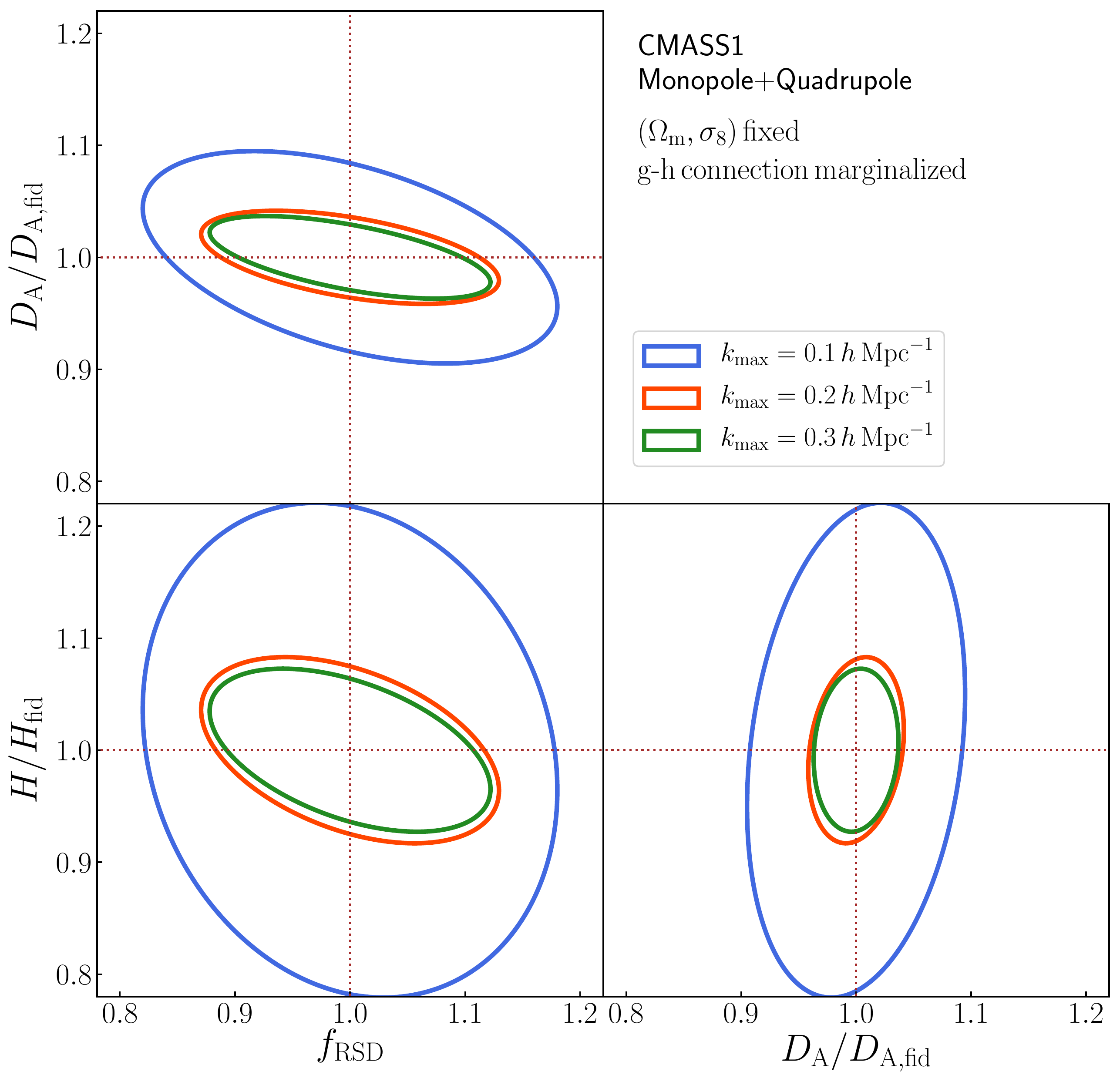}
\caption{The marginalized error ellipse in each of the two-dimensional subspaces of $D_A$, $H$ or $f_{\rm RSD}$, where we consider
the distance constraints for the CMASS1 sample and the constraint on $f_{\rm RSD}$ is from the information of all the three redshift slices (i.e. we use the same parameter for the RSD modification of all the three slices by the same amount $f_{\rm RSD}$). To obtain these forecasts, we use $f_{\rm RSD}$, and 3 $D_A$ and $H$ parameters for the three redshift slices, instead of $\Omega_{\rm m}$ and $\sigma_8$, and include other 30 nuisance parameters of the halo-galaxy connection. These constraints are purely from the RSD, the BAO peak locations and the AP effects. The constraint $f_{\rm RSD}$ (its fiducial value $f_{\rm RSD}=1$) corresponds to a fractional error of $f \sigma_8$ in the linear regime.
}
\label{fig:fisher_fRSD}
\end{figure}
The uniqueness of the redshift-space power spectrum method is it enables to measure the cosmological distances ($D_{\rm A}$ and $H$) and the strength of peculiar velocities via the RSD effect.
Because peculiar velocities of galaxies arise from the gravitational field in large-scale structure, it can be used to test gravity theory on cosmological scales.
Here we more explicitly address the power of the redshift-space galaxy power spectrum for making a ``model-independent'' test of the RSD effect
\citep[also see][for similiar discussion]{2011ApJ...726....5O,2014MNRAS.444.1400N}.
To do this, when generating a mock catalog of galaxies in redshift space, we modify the radial displacement of each halo by an amount of model parameter $f_{\rm RSD}$, from Eq.~(\ref{eq:rsd_def}), as
\begin{align}
&\Delta{\bf s}= \frac{v'_{\rm LoS}(\bx)}{aH(z)} {\bf \hat{e}}_{\rm LoS}, \nonumber\\
&v'_{\rm LoS} = f_{\rm RSD} v_{\rm h, LoS} + v_{\rm vir, LoS},
\end{align}
%
where $v_{\rm h,LoS}$ and $v_{\rm vir, LoS}$ are the line-of-sight components of the halo bulk velocity and of the virial velocity of galaxies inside the halo, respectively,
and $f_{\rm RSD}$ is a model parameter; if $f_{\rm RSD} = 1$, the RSD displacement is the same as that for the fiducial {\it Planck} $\Lambda$CDM cosmology.
If we assume $f_{\rm RSD}\ne 1$, the amount of RSD effect is artificially modified. When we consider LOWZ, CMASS1, and CMASS2, we employ the same constant factor $f_{\rm RSD}$ in generating the mock catalogs of the galaxies. Note that the RSD effect in the linear regime is proportional to $f\sigma_8$. The standard RSD analyses to constrain the parameter combination $f\sigma_8$ to test gravity theories are usually based on the nonlinear templates of the spectra constructed for $\Lambda$CDM cosmology within the GR framework but float $f\sigma_8$ as an independent parameter. Our test here is exactly along this line.
Assuming a constant $f_{\rm RSD}$ across the three redshifts of galaxies, it gives a simplest model of the RSD modification.
However, due to large uncertainties in internal random motions of galaxies inside the host halo, we do not modify the RSD displacement due to the relative motion of galaxies to the halo bulk velocity, i.e., the FoG effect.
Thus the following forecast on an estimation of $f_{\rm RSD}$ is purely from the effect on the halo power spectrum.
However note that velocities of these random motions are actually modified according to the details of the modified gravity model, e.g. \cite{Gronke:2014gaa}, which shows using simulations that the Hu-Sawicki $f(R)$ \cite{hu:2007lr} and the Symmetron \cite{Hinterbichler:2010es} gravity model affect the velocity profiles inside halos in a model-dependent way.

Fig.~\ref{fig:p0p2_fRSD} shows how a change in $f_{\rm RSD}$ alters the monopole and quadrupole power spectra as a function of $k$.
Note that the real-space power spectra are independent of $f_{\rm RSD}$ and the same for all these cases. As expected, the parameter $f_{\rm RSD}$ alters the monopole and quadrupole spectra, and is different from the AP distortion (Fig.~\ref{fig:Pl_cmass1_AP_2comb}).
The effect at small $k$ is as expected by the Kaiser formula, but the effect at large $k$ changes its sign. This is because the RSD effect at large $k$ is dominated by the smearing effect due to streaming motions of different halos. This effect is enhanced or reduced respectively by changing $f_{\rm RSD}>1$ or $<1$ from the fiducial value of $f_{\rm RSD}=1$, which explains the sign change at large $k$ in Fig.~\ref{fig:p0p2_fRSD}.

In Fig.~\ref{fig:fisher_fRSD}, we show the marginalized error ellipses in the two-dimensional sub-space of $D_{\rm A}$, $H$ or $f_{\rm RSD}$.
To do this, we fix $\Omega_{\rm m}$ and $\sigma_8$ to their fiducial values; that is, we do not include these parameters in the Fisher matrix. Hence we use 37 model parameters in the Fisher analysis; $f_{\rm RSD}$ and 3 $D_{\rm A}$'s and $H$'s parameters for three redshift slices, and 30 nuisance parameters of the halo-galaxy connection.
Because we fix $\Omega_{\rm m}$ and $\sigma_8$ to their fiducial values, all the real-space power spectra are the same for models varying either of $f_{\rm RSD}, D_{\rm A}$ or $H$, and the changes in the redshift-space power spectrum are due to variations in the RSD effect (via $f_{\rm RSD}$) or the AP distortion.
Therefore this forecast for $f_{\rm RSD}$ assesses the power of the redshift-space power spectrum for constraining the RSD effect strength, or more generally a deviation of the RSD effect from the $\Lambda$CDM prediction.
If we can find $f_{\rm RSD}\ne 1$ from this kinds of analysis, that would be a smoking gun evidence of non-GR gravity. The figure shows that the SDSS-like galaxy survey would allow for a 10\% accuracy of $f_{\rm RSD}$ determination, even after marginalization over the AP effects and other parameters.

\subsection{Galaxy assembly bias}

We have so far employed the HOD method to model the biased relation between distributions of galaxies and matter, i.e. galaxy bias, in an $N$-body simulation realization.
One critical assumption employed in the HOD method is it assumes that a probability of populating galaxies in halos is solely described by halo masses, $\avrg{N}\!(M)$. Even if halos have the same mass, the halos have various assembly histories due to the nature of hierarchical structure formation in a CDM scenario. For this reason halos, even with the same mass, could have a different large-scale bias. That is, the large-scale bias of halos could depend on another parameter besides the halo mass -- this effect is usually referred to as an assembly bias \cite{gao:2005fk,2006ApJ...652...71W,Gao:2006qz,2008ApJ...687...12D,faltenbacher:2010lr}.

Therefore we here study a possible impact of assembly bias on the cosmological parameter forecast.
To include the assembly bias effect in the mock galaxy catalog, we used the following method.
First, using the member $N$-body particles of each halo in a given $N$-body simulation realization, we compute the mass enclosed within the sphere of radius of $0.5r_{\rm 200}$ and use the mass fraction to the whole halo mass $M_{200}$ as a proxy of the mass concentration.
We then make a ranked list of all the halos in the ascending order of the inner-mass fraction at each halo mass bin ranging $M = [10^{12},10^{16}] h^{-1}M_\odot$ with width $\Delta\log_{10}M=0.02$, and populate central galaxies into halos from the top of the list (from the lowest-concentration halo), according to the number fraction specified by the central HOD $\avrg{N_{\rm c}}\!(M)$ (Eq.~\ref{eq:Nc}).
Finally we populate satellite galaxies into the halos which already host a central galaxy using the fiducial satellite HOD.
Thus this method does not change the mean HOD as a function of mass, but would change clustering properties of galaxies by preferentially selecting host halos with low concentration.
We then measure the redshift-space power spectrum from these modified mocks, in the same way as we do for the fiducial mocks.

\begin{figure}[h]
\centering
\includegraphics[width=0.85\textwidth]{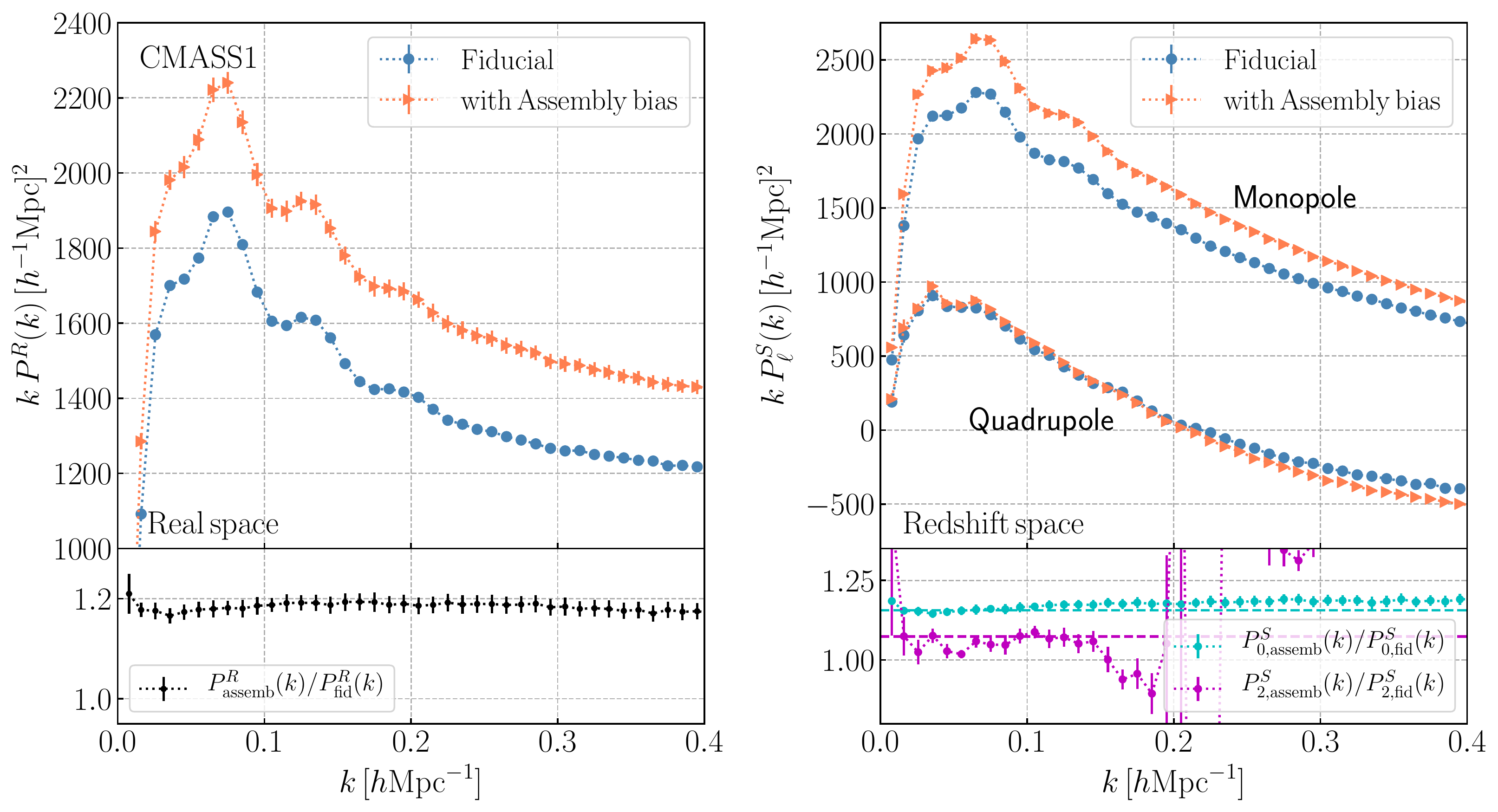}
\caption{
Comparison of the fiducial signals with those that included the effect of halo assembly bias (see text for details).
The result for CMASS1-like galaxies is shown.
Here we implemented the assembly bias effect by populating galaxies into halos according to the ranked list of mass concentration at each halo mass bin (from the lowest mass concentration).
The left panel shows that
the real-space power spectrum is amplified by almost constant factor, 1.2, over all the scales we consider.
The right panel shows the redshift-space power spectrum, while the lower panel shows the comparison with the Kaiser formula, where the linear bias is modified by the factor that matches the real-space power spectrum amplitude.
The modified Kaiser formula fairly well reproduces the monopole and quadrupole spectra. Thus the assembly bias appears to be mainly from
the density field around mock galaxies, while
the velocity field is not largely changed by the assembly bias.
}
\label{fig:power_assembly}
\end{figure}

Fig.~\ref{fig:power_assembly} shows the power spectra measured from the mocks including the assembly bias effect.
The real-space power spectrum, measured from the mocks with assembly bias, displays greater amplitudes than that in the fiducial mocks by 20\%, over all the scales we consider. Thus, halos with low mass concentrations
have greater clustering amplitudes than the average, displaying a physical connection between inner-structure of halos and the large-scale environments. This is consistent with the claims in the previous work \citep{2006ApJ...652...71W}, for relatively massive halos with $M\gtrsim M_\ast$, where $M_\ast$ is a typical nonlinear halo mass satisfying $\delta_{\rm collap}/\sigma(M_\ast)\simeq 1$ ($\delta_{\rm collap}=1.67$), because BOSS galaxies reside in such massive halos.
This result should be considered as an extreme case, because we completely followed the ranked list of the halo mass concentration when populating central galaxies in a fully deterministic manner, and did not include any scatter in the galaxy-halo concentration.
Hence a more realistic case would be in between the fiducial model and the assembly bias result even if the assembly bias exists in the real galaxy catalog.
The right panel shows the effect of assembly bias on the monopole and quadrupole power spectra.
The assembly bias amplifies the monopole spectrum by a similar amount to that of the real-space power spectrum.
The effect on the quadrupole power spectrum is less significant.
The lower panel shows the comparison with the predictions of Kaiser formula, if we change the linear bias by an amount of the assembly bias effect (about factor of 1.1), but do not change 
$f\sigma_8$.
The figure shows that the modified Kaiser prediction fairly well reproduces the simulation results, although the quadrupole power spectrum has a zero-crossing at $k\simeq 0.2$~\hMpci, so the results look noisy around the scale.
This means that the assembly bias mainly affects the halo bias, and does not change the RSD effect. Or equivalently the peculiar velocity field of low-concentration halos is not so different from that of halos with the same mass -- i.e. little assembly bias effect on the velocity field \citep[see][for similar discussion]{2019MNRAS.487.2424M,2019MNRAS.486..582P}.

\begin{figure}[h]
\centering
\includegraphics[width=5in]{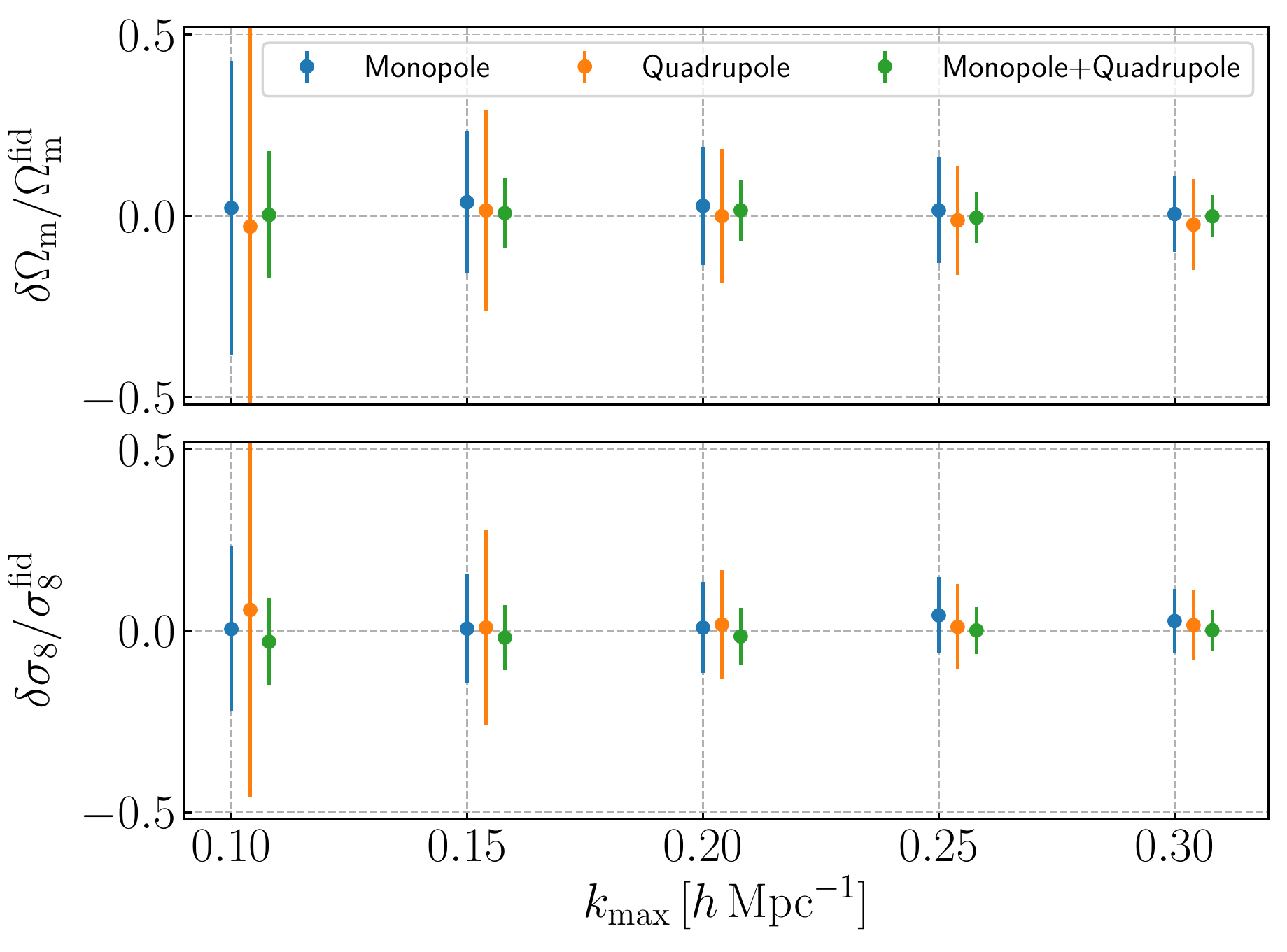}
\caption{
An estimation of a possible bias in $\Om$ (upper panels) or $\sigma_8$ (lower panels) due to the effect of assembly bias on the redshift-space power spectrum, if the effect is ignored in the model prediction.
We here assume that the assembly bias, which is implemented in the same way as in Fig.~\ref{fig:power_assembly}, affects the redshift-space power spectra for all the LOWZ, CMASS1, and CMASS2-like galaxies, and then estimate the parameter bias using the Fisher method (Eq.\ref{eq:fisher_bias}).
We show the results, relative to the fiducial value (true value) of each parameter, as a function of the maximum wavenumber $k_{\rm max}$, and also show the marginalized error for comparison.
}
\label{fig:bias_assemb}
\end{figure}

To quantify the impact of assembly bias on cosmological parameter estimation, we estimate a possible bias on the cosmological parameters due to the ignorance of the dependence on the concentration in the model, using the formula \cite{Huterer:2004tr,Joachimi:2009fr}
\begin{align}
 \delta p_\alpha = \sum_{\beta} \left(F^{\ell \ell'} \right)^{-1}_{\alpha \beta} \sum_{ij} \left[ P^{\rm S+AB}_\ell(k_i) - {P}^{\rm S}_\ell(k_i) \right] {\rm Cov}^{-1}\left[\hat{{P}}^{\rm S}_\ell(k_i),\hat{{P}}^{\rm S}_{\ell'}(k_j)\right] \frac{\partial {P}^{\rm S}_{\ell'}(k_j)}{\partial p_\beta},
 \label{eq:fisher_bias}
\end{align}
where $P^{\rm S+AB}_{\ell}(k_i)$ is the power spectrum measured from the mocks with assembly bias, and $P^{\rm S}_\ell(k)$ without
superscript ``+AB'' is the fiducial power spectrum.
The quantity $\delta p_\alpha$ estimated from the above equation quantifies a bias in the parameter $p_\alpha$ due to the assembly bias effect, if the model predictions do not include the effect.
Fig.~\ref{fig:bias_assemb} shows the results for a possible bias in $\Om$ or $\sigma_8$ if the assembly bias affects the redshift-space power spectra for all the LOWZ, CMASS1, CMASS2-like galaxies.
Since the assembly bias we consider is an extreme case and the assembly bias, even if exists, unlikely affects all the SDSS galaxies at different redshifts, this can be considered as a worst case scenario.
The figure shows that the assembly bias does not cause an amount of bias in these parameter greater than the marginalized error.
This is because the assembly bias changes only the overall amplitudes of the real-space power spectrum, i.e. a change in the apparent galaxy bias, and does not affect the RSD effect,
as shown in Fig.~\ref{fig:power_assembly}.
We have actually confirmed that this assembly bias effect causes a significant bias in the forecast on $\Om$ and $\sigma_8$ when we do not include any uncertainty in the halo-galaxy connection, but can be absorbed by the changes in HOD parameters, even if all the other nuisance parameters, $c_{\rm conc}$, $c_{\rm vel}$, $p_{\rm off}$, ${\cal R}_{\rm off}$ and $P_{\rm SN}$, are fixed to their fiducial values.

\section{Conclusion}
\label{sec:conclusion}

In this paper we have studied the cosmological information content in the redshift-space power spectrum of galaxies over a range of wavenumber scales from the linear to quasi-nonlinear regimes.
For this purpose we made use of a suite of the halo catalogs constructed from high-resolution $N$-body simulations to build mock catalogs of SDSS-like galaxies by populating galaxies into halos based on a prescription of the halo occupation distribution (HOD) model.
Using these mocks we studied how changes in cosmological parameters and the halo-galaxy connection parameters alter the redshift-space power spectrum of galaxies.
Assuming that the redshift-space power spectrum of halos hosting the galaxies carries the cosmological information, we performed a Fisher information analysis on the redshift-space galaxy power spectrum.
We studied how the power spectrum of SDSS-like galaxies can be used to constrain the cosmological parameters ($\Om$ and $\sigma_8$) and the cosmological distances ($D_{\rm A}$ and $H$), even after marginalization over the halo-galaxy connection parameters.

We showed that the cosmological parameters and the cosmological distances via the AP effect cause characteristic effects on the monopole and quadrupole power spectra of galaxies over the range of scales we considered (up to $k_{\rm max}=0.3$~\hMpci) that appear quite different from the effects caused by changes in the halo-galaxy connection parameters (Fig.~\ref{fig:param_diff_pl}).
In particular, we found that the BAO features are quite powerful, even in the presence of the effects of halo-galaxy connection, and allow for robust measurements of the angular and radial cosmological distances, $D_{\rm A}$ and $H$ at a redshift of the galaxy survey (Fig.~\ref{fig:fisher_boss_DA_H_margin}).
Compared to this, the parameters ($\Om,\sigma_8$), which control the amplitude of redshift-space power spectrum and the RSD strength, are more affected by the galaxy-halo connection parameters, although the constraints can be continuously improved by including the redshift-space power spectrum information from the linear regime up to the nonlinear regime, up to $k_{\rm max}=0.3$~\hMpci in our exercise.
Even if the signal-to-noise ratio of the quadrupole power spectrum amplitude is smaller than that of the monopole power spectrum by up to a factor of 100, the quadrupole power spectrum carries the sensitivity to the cosmological parameters similar to that of the monopole power spectrum, and therefore combining the two allows one to improve the cosmological constraints, mitigating the impact of halo-galaxy connection parameters.
For example, changes in the HOD parameters, more physically changes in the way to populate galaxies into halos, alters the overall amplitude of the redshift-space power spectrum because such changes alter relative contributions of different-mass halos hosting galaxies, where different-mass halos have different biases in the amplitude and scale dependence.
This effect appears to be different from the effects of $\Om$ and $\sigma_8$, and can be distinguishable if combining the monopole and quadrupole information.
Thus our results imply that combining the monopole and quadrupole power spectra allows for a {\em self-calibration} of cosmological parameters, mitigating the impacts of the halo-galaxy connection parameters, to some extent.

\if
As we have shown, the redshift-space power spectrum of halos contains a wealth of
the useful cosmological information.
First of all, halos are biased tracers of the underlying matter distribution, and the halo power spectrum in the linear or quasi-nonlinear regimes can be used to infer
the matter power spectrum that has characteristic BAO features as a standard ruler.
The AP effect enables us to use the BAO scales to constrain both the angular and radial distances, $D_{\rm A}$ and $H$.
Second, halos are biased tracers of the underlying dark matter distribution, and more massive halos are rare and have greater bias amplitudes, where the halo bias amplitude and the halo mass dependence are sensitive to the cosmological parameters ($\Om$ and $\sigma_8$).
Hence a relative contribution of different-mass halos hosting galaxies controls the overall amplitude of galaxy power spectrum.
In addition different-mass halos have scale-dependent bias, beyond linear bias, in the quasi-nonlinear regime.
Third, the RSD effect due to coherent bulk motions of halos imprint characteristic, anisotropic features in the redshift-space power spectrum of halos, as the linear-scale anisotropies are predicted by the Kaiser formula.
Furthermore, the RSD effects due to random motions of halos cause a smearing effect of the redshift-space power spectrum, and the amount of the smearing effect depends on the mass of host halos.
The redshift-space power spectrum of halos can be precisely modeled, e.g. by using a suite of $N$-body simulations as we did in this paper.
Hence if we can recover the redshift-space power spectrum of halos from the observed galaxy distribution, we can in principle use these information to obtain cosmological constraints, which is the final goal for this kind of approach.
\fi

To perform cosmological parameter inference from an actual data
such as the SDSS spectroscopic galaxies, it requires a sufficiently accurate model of the redshift-space halo power spectrum as a function of cosmological models, separation ($k$), redshift and halo masses (exactly speaking, masses of two halos taken in the power spectrum measurement).
With the advent of high-performance computational resources, it would be possible to build a database or ``emulator'' that allows for a fast, accurate computation of the redshift-space power spectrum of halos, e.g. extending our previous method for the real-space power spectrum of halos in Ref.~\cite{Nishimichi:2018etk}.
Once such a model of the redshift-space power spectrum of halos is available, we can use the HOD-type prescription to obtain the model prediction for the redshift-space power spectrum of galaxies; we can make a weighted sum of the halo power spectrum over halo masses to obtain the 2-halo term, and then add the 1-halo term taking into account a sufficient number of nuisance parameters to account for variations in the model ingredients (the spatial and velocity distributions of satellite galaxies in the host halos and a possible off-centering effect of central galaxies), based on the halo model (Section~\ref{subsec:halomodel}).
This is our ongoing project, and will be presented elsewhere.

A uniqueness of the redshift-space power spectrum, compared to other large-scale structure probes, is that the anisotropic features allow one to measure the cosmological distances (also via the BAO peaks) as well as the RSD effect.
Here the RSD effect is expected to be a powerful probe of the gravity theory on cosmological scales.
We have also addressed this question.
By introducing a parameter $f_{\rm RSD}$ to control the amplitude of the RSD effect of halos hosting galaxies in the mocks, we assessed a power of the redshift-space power spectrum for making a model-independent estimation of
the RSD parameter together with the distance parameters $D_{\rm A}$ and $H$.
With this parameterization, the constraint on $f_{\rm RSD}$ is purely from the anisotropic features in the redshift-space power spectrum, because it does not alter the real-space power spectrum.
We found that a fractional accuracy of $f_{\rm RSD}$, corresponding to a fractional error of $f\sigma_8$ in the linear regime, is about 10\% if we include the redshift-space power spectrum up to $k_{\rm max}=0.3~$\hMpci, after marginalization over the halo-galaxy connection parameters and uncertainties in the FoG effect due to virial motions of galaxies in their host halos.
Our forecast might be considered as a conservative forecast, but generally implies that there are severe degeneracies between the RSD effects and the systematic effects due to the halo-galaxy connection or the 1-halo term.
It would be worth to explore a measurement-side method of mitigating the FoG effect, e.g. a reconstruction method of halos from the observed galaxy distribution \citep{reid09,2010MNRAS.404...60R,Okumura:2016mrt}, which enables to mitigate the FoG contamination due to satellite galaxies.

Furthermore we have studied the impact of possible assembly-bias effect on the cosmological parameters from the redshift-space power spectrum.
Although we used a prescription to populate galaxies into halos depending on the mass concentration of individual halos, which gives a proxy of the assembly history of each halo, we found that the assembly bias mainly changes the overall amplitude of the redshift-space power spectrum, i.e. the bias parameter, but little alters the peculiar velocities of galaxies and therefore the RSD effect (Fig.~\ref{fig:power_assembly}).
Hence we concluded that the assembly bias unlikely causes a significant bias in cosmological parameters if the monopole and quadrupole power spectra are combined.
However, our conclusion is based on the simplified treatment, and it would require a more detailed study using a more realistic model of the assembly-bias effects such as the effects predicted by hyrdodynamical simulations.

It would be also interesting to explore a method of combining the redshift-space power spectrum with the galaxy-galaxy weak lensing information.
The galaxy-galaxy lensing signals for the same spectroscopic galaxies used in the clustering analysis are available if a deeper imaging data, which allows one to select background galaxies behind each spectroscopic galaxy, is available for the same region of the sky.
This is the case, for example for the SDSS BOSS spectroscopic survey and the Subaru Hyper Suprime-Cam survey \cite{2018PASJ...70S...4A}.
The galaxy-galaxy weak lensing probes the average matter distribution around the spectroscopic galaxies, and can be used to constrain the bias relation between galaxy and matter distributions, which will in turn enable to help break degeneracies between the cosmological parameters (e.g. $\sigma_8$ or RSD parameter) and the HOD parameters. This is our future work.

\smallskip{\em Acknowledgments} --  YK thanks to the Yukawa Institute for Theoretical Physics, Kyoto University for the warm hospitality where this work was partly done.
This research was supported by the Munich Institute for Astro- and Particle Physics (MIAPP) of the DFG cluster of excellence
ORIGINS (\url{http://www.munich-iapp.de}).
We would like to thank Leonardo Senatore and Marko Simonovic for useful discussion.
This work was supported in part by World Premier International
Research Center Initiative (WPI Initiative), MEXT, Japan, and JSPS
KAKENHI Grant Numbers JP15H03654,
JP15H05887, JP15H05893, JP15K21733, JP17H01131, JP17K14273 and 19H00677, and by Japan Science and Technology Agency CREST JPMHCR1414.
YK is also supported by the Advanced Leading Graduate Course for Photon Science at the University of Tokyo.
The $N$-body simulations and subsequent halo-catalog creation for this work were carried out on Cray XC50 at Center for Computational Astrophysics, National Astronomical Observatory of Japan.

\appendix

\section{Comparison of responses with simple theoretical prescriptions}
\label{app:linear}

\begin{figure}[h]
\centering
\includegraphics[width=0.8\textwidth]{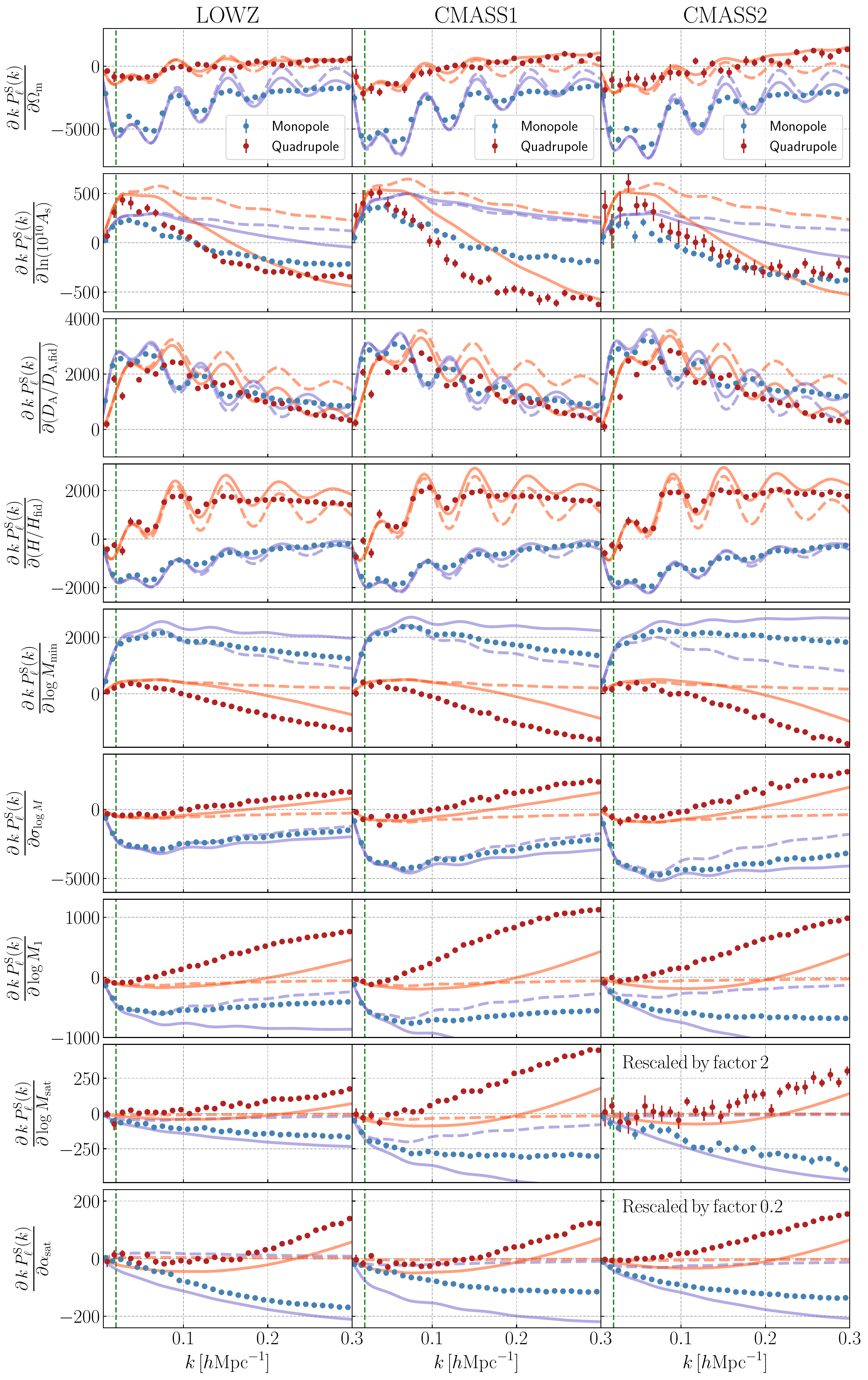}
\caption{
The comparison of the derivatives of the power spectrum computed from the SDSS-like galaxy mocks to
the predictions by two simple models (see text for details).
We compare the derivatives with respect to $\Om$, $\ln (10^{10} A_{\rm s})$, $D_{\rm A}(z_n)$, $H(z_n)$ and five HOD parameters with the corresponding theoretical predictions.
Blue and red lines are for the monopole and quadrupole moments, and the solid and dashed lines denote the model (i) and (ii), respectively.
The green, vertical dashed line indicates the scale $k = 0.02$ \hMpci, which is adopted as the minimum wavenumber $k_{\rm min}$ in all the Fisher calculations in this study.
}
\label{fig:deriv_lin}
\end{figure}
In this paper we execute a Fisher matrix analysis using the response of the redshift-space power spectrum with respect to various parameters.
For completeness, in this appendix we compare the estimation of these responses from our simulation-based galaxy mocks with their predictions based on the linear theory and the halo model formalism, which is presented in Eqs.~(\ref{eq:PS1h})
and (\ref{eq:PS2h}).
We predict the responses following two simple models.

\begin{itemize}
\item[(i)] {\it The HOD power spectrum combined with the FoG damping} -- Following Eqs.~(\ref{eq:PS1h}) and (\ref{eq:PS2h}), we compute the real-space galaxy power spectrum based on the halo power spectrum, a NFW radial profile of galaxies and the HOD 
defined by Eqs.~(\ref{eq:Nc}) and (\ref{eq:Ns}).
The halo power spectrum is computed from the linear matter power spectrum and the large-scale halo bias $b_{\rm h}(M)$.
To model the redshift-space distortions, we employ the well-known Kaiser's effect and the FoG damping factor:
\begin{align}
	P^{\rm S}_{\rm gg}(k,\mu) = \left(1 + \beta \mu^2 \right)^2 P^{\rm R}_{\rm gg}(k) D_{\rm FoG}(k \mu f \sigma_{\rm v}),
\end{align}
where $\beta = f / b_{\rm g}$ and $\sigma_{\rm v}$ is the parameter which represents the velocity dispersion of the virial velocities.
For the FoG damping we assume the broadly-used Gaussian form, i.e., $D_{\rm FoG}(k \mu f \sigma_{\rm v}) = e^{- k^2\mu^2 f^2 \sigma_{\rm v}^2}$, and we compute $\sigma_{\rm v}$ by simply using the linear theory expression:
\begin{align}
	\sigma_{{\rm v,lin}}^2 = \frac{1}{3} \int \frac{{\rm d}^3q}{(2\pi)^3} \frac{P_{\rm lin}(q)}{q^2}.
\end{align}
\item[(ii)] {\it The linear power spectrum with the linear galaxy bias} -- We compute the galaxy power spectrum following Kaiser's formula from the linear matter power spectrum and the linear galaxy bias. Note that the HOD parameters affect only on this galaxy bias in this model.
\end{itemize}
In both models above, we use the linear galaxy bias $b_{\rm g}$ given as
\begin{align}
b_{\rm g} = \int\!{\rm d}M~\frac{{\rm d}n}{{\rm d}M}\left[\avrg{N_{\rm c}}\!(M)+\avrg{N_{\rm s}}\!(M)\right] b_{\rm h}(M),
\end{align}
where $b_{\rm h}(M)$ is the linear halo bias.
To compute the halo mass function and the linear halo bias in these models for an input cosmology, we employ {\tt Dark Emulator} in Ref.~\cite{Nishimichi:2018etk}.
Since the parameters to model the spatial and velocity distributions of galaxies in their host halos, $c_{\rm conc}$, $c_{\rm vel}$, $p_{\rm off}$ and ${\cal R}_{\rm off}$, are difficult to include in the linear theory prediction, we focus only on the cosmological and HOD parameters here.

Fig.~\ref{fig:deriv_lin} shows the comparison of the parameter responses measured from mocks to these two models.
The three columns show the results for LOWZ, CMASS1 or CMASS2-like galaxies, respectively.
The filled symbols are the measured responses with respect to each parameter, and the solid and dashed lines are the predictions given by the models (i) and (ii) described above, respectively.
We see that behaviors of these responses at highly large scales are well described by both models, and the model (i) has a slightly better performance on reproducing the responses to the cosmological parameters than the simplest linear theory prediction (ii). However, the modification included in the model (i) gives little improvements on reproducing the HOD-responses. This means that the model (i) is still too naive to capture these behaviors and hence we must employ more complicated models in analyzing the galaxy power spectrum in the current galaxy surveys.

\bibliography{lssref}

\end{document}